\documentclass[aps,prl,twocolumn,floatfix,final,superscriptaddress,citeautoscript,longbibliography]{revtex4-2}
\usepackage{graphicx} 
\usepackage{xcolor}
\usepackage{subcaption}
\usepackage[justification=raggedright,singlelinecheck=false]{caption}
\usepackage[colorlinks=true,linkcolor=blue,citecolor=blue,urlcolor=blue]{hyperref}
\usepackage{amsmath}
\usepackage{braket}
\usepackage{comment}

\providecommand{\BenchSquareAshNDepthGeomean}{\ensuremath{1.24\times}}
\providecommand{\BenchSquareAshNGateGeomean}{\ensuremath{1.13\times}}
\providecommand{\BenchSquareCZDepthGeomean}{\ensuremath{1.86\times}}
\providecommand{\BenchSquareCZGateGeomean}{\ensuremath{1.65\times}}

\newcommand{\SuppRoutingBenchmarkSummaryTable}{Table~S2}

\newcommand{\GJZ}[1]{{\color{blue} {[GJZ: #1]}}}

\begin{document}
\title{Lifting connectivity bottlenecks in superconducting quantum processors \\
    via enriched native two-qubit gates}

\newcommand{\THUCS}{Department of Computer Science and Technology, Tsinghua University, Beijing 100084, P.R. China}
\newcommand{\CTQ}{China Telecom Quantum Information Technology Group Co., Ltd., Hefei 230094, Anhui, P.R. China}
\newcommand{\QRI}{Quantum Research Institute of China Telecom, Hefei 230031, Anhui, P.R. China}
\newcommand{\TQ}{TraverseQuantum Co., Ltd., Beijing, P.R. China}
\newcommand{\HKUST}{Department of Electronic and Computer Engineering, The Hong Kong University of Science and Technology, Hong Kong}
\newcommand{\ZGC}{Zhongguancun Laboratory, Beijing, P.R. China}

\affiliation{\THUCS}
\affiliation{\CTQ}
\affiliation{\QRI}
\affiliation{\TQ}
\affiliation{\HKUST}
\affiliation{\ZGC}

\author{Hanyi Wang}
\thanks{These authors have contributed equally to this work.}
\affiliation{\CTQ}
\affiliation{\QRI}
\author{Jingzhe Guo}
\thanks{These authors have contributed equally to this work.}
\affiliation{\THUCS}
\affiliation{\TQ}
\author{Lijun Sun}
\thanks{These authors have contributed equally to this work.}
\affiliation{\CTQ}
\affiliation{\QRI}
\author{Zhaohui Yang}
\affiliation{\HKUST}
\author{Weizhi Tao}
\affiliation{\CTQ}
\affiliation{\QRI}
\author{Xingye Yuan}
\affiliation{\THUCS}
\affiliation{\TQ}
\author{Qiankun Wang}
\affiliation{\CTQ}
\affiliation{\QRI}
\author{Bihao Guo}
\affiliation{\CTQ}
\affiliation{\QRI}
\author{Chunwang Liu}
\affiliation{\CTQ}
\affiliation{\QRI}
\author{Rui Yang}
\affiliation{\CTQ}
\affiliation{\QRI}
\author{Yang Li}
\affiliation{\CTQ}
\affiliation{\QRI}
\author{Yu Fan}
\affiliation{\CTQ}
\affiliation{\QRI}
\author{Jiasheng Hu}
\affiliation{\CTQ}
\affiliation{\QRI}
\author{Junhe Wang}
\affiliation{\CTQ}
\affiliation{\QRI}
\author{Shuyue Zheng}
\affiliation{\CTQ}
\affiliation{\QRI}
\author{Shengbin Wang}
\affiliation{\CTQ}
\affiliation{\QRI}
\author{Xinfang Zhang}
\affiliation{\CTQ}
\affiliation{\QRI}
\author{Feng Wu}
\email{wufeng@iqubit.org}
\affiliation{\ZGC}
\author{Hantao Sun}
\email{sunhantao@chinatelecom.cn}
\affiliation{\CTQ}
\affiliation{\QRI}
\author{Jianxin Chen}
\email{chenjianxin@tsinghua.edu.cn}
\affiliation{\THUCS}

\begin{abstract}
Limited qubit connectivity is a central architectural constraint in superconducting quantum processors, whose planar layouts require additional gates to mediate interactions between distant qubits. Here, we use the AshN control scheme, where rich two-qubit control on every nearest-neighbour pair allows a logical interaction and the required qubit routing to be merged into a single native operation, effectively transforming a sparse hardware graph into a more connected computational architecture. For the benchmark instances studied, the resulting synthesis capability enables reliable execution on constrained one- and two-dimensional lattices, with compiled two-qubit gate counts approaching those of an all-to-all-connected reference. Across seven benchmark circuits on one- and two-dimensional topologies, the AshN-based implementation achieves geometric-mean reductions of $45.2\%$ and $43.7\%$ in two-qubit gate count compared with controlled-$\mathrm{Z}$-based compilation, respectively. Using AshN gates, we prepare an eight-qubit two-excitation Dicke state with a fidelity of $0.736$ and certify its genuine multipartite entanglement using a fully positive-partial-transpose witness, whereas the same witness does not certify entanglement for the $\mathrm{CZ}$-based implementation. The state fidelity and entanglement certification remain robust across the tested lattice configurations, including those with up to three connectivity defects. Our work establishes native-gate engineering as a practical approach to mitigating connectivity constraints.
\end{abstract}

\maketitle

Superconducting quantum processors offer fast, high-fidelity control, but their two-qubit interactions are typically confined to sparse, approximately planar coupling graphs~\cite{Arute2019,Acharya2023,Acharya2025,gao2025establishing,He2025}. This connectivity bottleneck is usually framed as a shortage of physical couplers. At the circuit level, however, the graph alone does not determine the cost: routing also depends on the native operations available on each edge. Logical interactions between non-adjacent qubits must be brought onto connected pairs, typically by inserting $\mathrm{SWAP}$ operations~\cite{Li2019,tan2021optimal}. The resulting increase in gate count and depth limits the fidelity attainable at a given physical error rate. This geometric routing penalty is less severe in platforms with long-range or reconfigurable interactions~\cite{Haffner2008,Linke2017,Pino2021,Barredo2016,Bluvstein2022,Bluvstein2024}.

Adding edges to the hardware graph is a direct remedy. Long-range superconducting interactions have been demonstrated using resonators, communication channels and three-dimensional integration~\cite{Majer2007,Kurpiers2018,Rosenberg2017}, and are especially relevant to quantum error-correcting codes whose interaction graphs are difficult to embed in a planar nearest-neighbour array~\cite{cohen2022low,Bravyi2024}. Such extensions, however, introduce fabrication, frequency-allocation, control and calibration constraints~\cite{Kosen2022,Muller2019}. This motivates a complementary question: how much more connected can a processor become at the circuit level without changing its physical graph?

An alternative is to increase what each existing edge can do. Superconducting processors have increasingly moved beyond a single conventional entangler such as $\mathrm{CZ}$, adopting alternative entanglers, multiple native gates or continuously parameterized interactions, primarily to reduce logical circuit synthesis costs~\cite{Abrams2020,Google2020,huang2023quantum,IBM2026}. Broader native control can also benefit qubit routing. If a logical operation $U$ and a routing $\mathrm{SWAP}$ act consecutively on the same pair, their product can be implemented as a single native operation, up to local rotations, whenever it remains within the native gate family~\cite{krizan2025quantum}. The explicit $\mathrm{SWAP}$ is then absorbed while the logical-to-physical mapping is updated. The decisive property is therefore not the size of the gate set alone, but its closure under composition with $\mathrm{SWAP}$. Full two-qubit expressivity guarantees this property for arbitrary $U$~\cite{chen2024one}. A restricted family can also support $\mathrm{SWAP}$ absorption, but only for the subset of operations that remains within the family after composition with $\mathrm{SWAP}$. Most previously studied restricted gate families do not provide an arbitrary-$U$ guarantee\cite{Kandala2017,Arute2019}. Whenever the required closure is available, a compiler can exploit it to change the effective connectivity seen by a circuit without modifying the coupler graph~\cite{Li2019}.

The recent AshN control scheme offers an experimentally accessible route to $\mathrm{SWAP}$ absorption for arbitrary two-qubit operations~\cite{chen2024one,chen2025efficient}. By combining a tunable exchange interaction with simultaneous local microwave drives, AshN spans, up to single-qubit rotations, the nonlocal content of arbitrary two-qubit operations on a connected pair. A logical operation and an adjacent routing $\mathrm{SWAP}$ can consequently be consolidated into a single AshN operation while the logical-to-physical mapping is updated~\cite{tan2021optimal,yang2026reconfigurable}. This capability does not erase geometry: circuits with rapidly changing interaction partners may offer few absorption opportunities~\cite{supplemental}. Nor is the additional control free, because a broader gate family must be calibrated and executed with competitive error rates. The relevant test is therefore not expressivity in isolation, but whether the resulting circuit compression outweighs the added control and calibration costs.

Here we carry out this end-to-end test by connecting AshN control, hardware-aware compilation and circuit execution. We develop parallel calibration and gate-characterization procedures for the native operations used in our circuits and benchmark the resulting gate set on one- and two-dimensional superconducting-qubit arrays against controlled-$Z$-based implementations on the same coupling graphs. For the structured circuits studied, $\mathrm{SWAP}$ absorption brings compiled two-qubit gate counts closer to an all-to-all-connectivity compilation reference, with the lower counts accompanied by higher computational-basis success probabilities. We further prepare an eight-qubit two-excitation Dicke state and certify genuine multipartite entanglement across the tested connectivity configurations, including configurations in which up to three couplers are unavailable. Together with larger-scale compilation studies and random-circuit stress tests, these results delineate the reach and limits of native-gate engineering and show how usable connectivity can emerge from the co-design of the coupling graph, native control and compilation.

\begin{figure*}[htbp]
    \centering
    \includegraphics[width=0.85\textwidth]{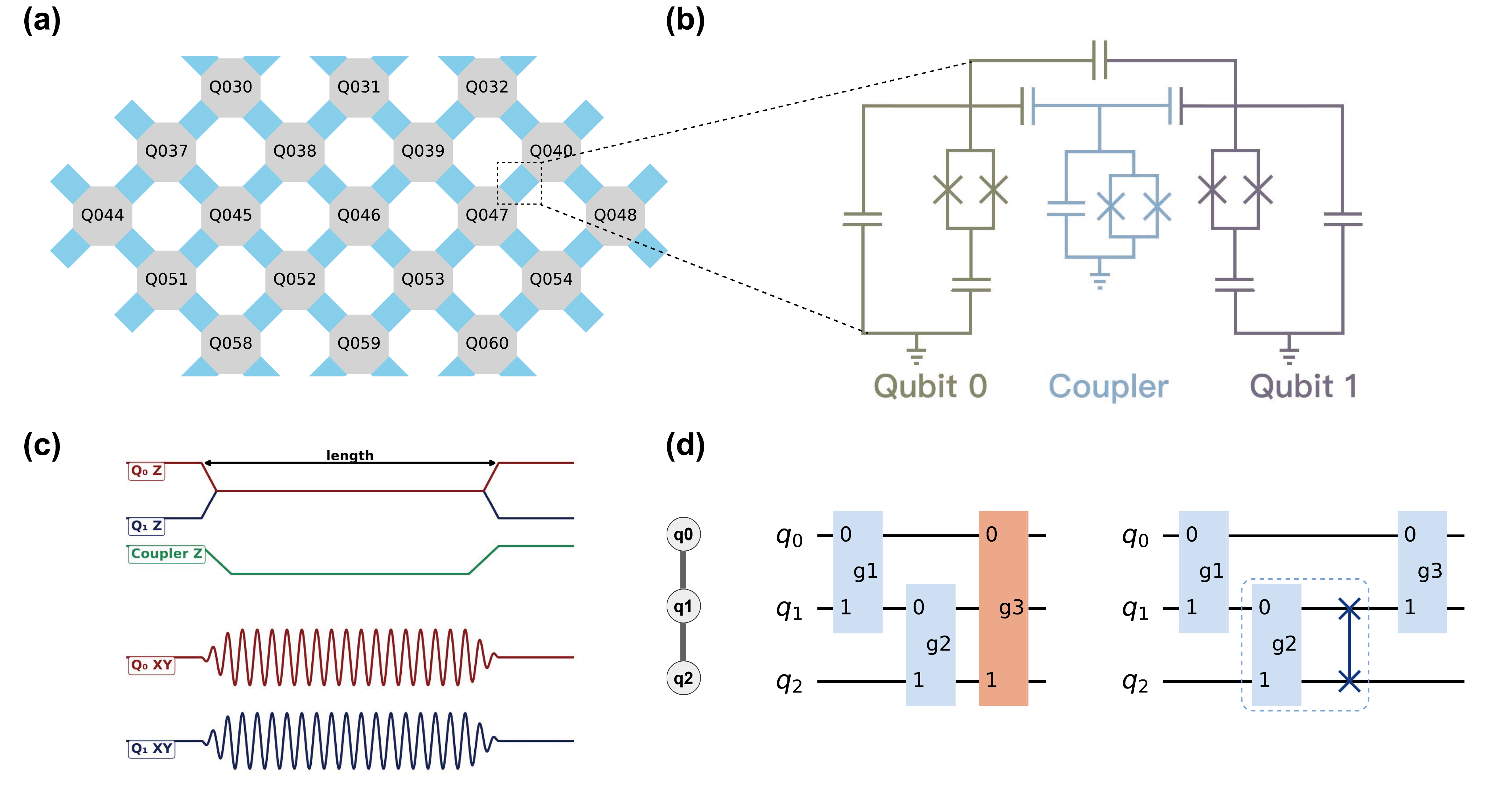}
    \caption{(a) Topological layout of qubits and couplers. Dark gray denotes qubits, and light blue denotes couplers. (b) Simplified circuit schematic of two qubits coupled via a tunable coupler. (c) Compiled pulse sequence for an arbitrary AshN two-qubit gate. From top to bottom: $Z$-control pulses on the two qubits, the $Z$-control pulse on the tunable coupler, and the XY-drive pulses applied to the two qubits. (d) The AshN gate reduces routing overhead via $\mathrm{SWAP}$ absorption. The left panel shows a three-qubit 1D chain topology ($q_0 \leftrightarrow q_1 \leftrightarrow q_2$). The middle panel illustrates that, under a trivial initial mapping, gate $g_3$ in the logical circuit is not directly executable. The right panel shows that a $\mathrm{SWAP}$ gate is inserted after $g_2$ to route qubits and enable execution of $g_3$. By consolidating $g_2$ and the $\mathrm{SWAP}$ gate into a single AshN gate (dashed box), $g_3$ becomes executable without incurring explicit routing overhead.
    }
    \label{Figure1}
\end{figure*}

\section{AshN Control, Compilation, and Calibration}

The native two-qubit operation used in this work is generated by the AshN control scheme~\cite{chen2024one,chen2025efficient}. As illustrated in Fig.~\ref{Figure1}(a)--(c), two frequency-tunable qubits interact through a tunable coupler while local microwave drives are applied simultaneously. The combination of transverse $XX/YY$ coupling and local driving generates a continuous family of two-qubit operations, parameterized by the coupling strength, drive amplitudes, phases and detuning. In the rotating frame, the effective Hamiltonian is
\begin{align}
    H_R    =\, & \frac{g}{2}(XX+YY)+\frac{\Omega_1}{2}(\cos\phi_1 XI+\sin\phi_1 YI)            \\
    +          & \frac{\Omega_2}{2}(\cos\phi_2 IX+\sin\phi_2 IY ) + \frac{\Delta}{2}(ZI + IZ), \nonumber\end{align}
where $g$ is the transverse coupling strength, $A_j$ and $\phi_j$ denote the amplitude and phase of the microwave drive applied to qubit $j$, and $\delta$ is half of the drive detuning. 

To implement a target two-qubit gate $U$, we first extract its nonlocal interaction coefficients $(a,b,c)$ using the KAK decomposition~\cite{zhang2003geometric}:
\begin{align}
    U = e^{i\phi_g}\, (O_{1}\otimes O_{2})\, e^{i(a XX+ b YY + c ZZ)}\, (O_{1}'\otimes O_{2}'),
\end{align}
Here $\phi_g$ indicates a global phase, while $O_1$, $O_2$, $O_1'$ and $O_2'$ are single-qubit rotations. The coordinates $(a,b,c)$ specify the local-equivalence class of $U$ in the Weyl chamber~\cite{zhang2003geometric}. Given $(a,b,c)$ and the available coupling strength $g$, the AshN protocol determines a gate duration and a set of local-drive parameters whose evolution realizes the target nonlocal operation up to single-qubit rotations. In experiment, these parameters must be calibrated against device imperfections, waveform distortions and model mismatch. The theoretical control construction covers the full Weyl chamber; experimentally, we calibrate and characterize the representative subset of gates required by the circuits studied below.

This expressivity is useful for routing because the product of a two-qubit operation and a $\mathrm{SWAP}$ is itself a two-qubit unitary. When an inserted $\mathrm{SWAP}$ acts on the same physical pair as an adjacent two-qubit block, their product can therefore be implemented as a single AshN operation, up to local rotations [Fig.~\ref{Figure1}(d)]~\cite{tan2021optimal,yang2026reconfigurable}. The logical-to-physical mapping is updated accordingly, without adding an explicit two-qubit gate or an additional layer. Such absorption is not generally available to more restricted native gate families and does not remove the need to route interactions between non-adjacent physical qubits.

To exploit these opportunities, we use MirrorSABRE, a gate-set-aware extension of SABRE that supplements its distance-based look-ahead heuristic with a preference for $\mathrm{SWAP}$s that can be absorbed into adjacent two-qubit blocks. The router thereby balances the proximity of upcoming interaction partners against the gate-count and depth savings enabled by absorption~\cite{Li2019,yang2026reconfigurable,yangUnifyingQubitRouting2026}.

The availability of $\mathrm{SWAP}$ absorption is workload dependent. Structured circuits may repeatedly use the same physical edges, whereas random circuits with rapidly changing interaction partners provide fewer merging opportunities; the latter regime is examined in the Supplemental Material~\cite{supplemental}. Realizing the compiled circuits experimentally further requires the requested two-qubit operations to be calibrated and characterized across multiple couplers. To this end, we develop a parallel calibration framework and demonstrate it across the coupler groups used to configure a 12-qubit chain and a $3\times3$ array. These experiments establish parallel-calibration feasibility at the hardware scale tested here; extension to larger processors will require further validation of calibration time, crosstalk, and long-term stability. The framework provides access to a substantial portion of the Weyl chamber and enables simultaneous calibration of different AshN gates across multiple couplers~\cite{supplemental}.

\begin{figure*}[htbp]
    \centering
    \includegraphics[width=0.85\textwidth]{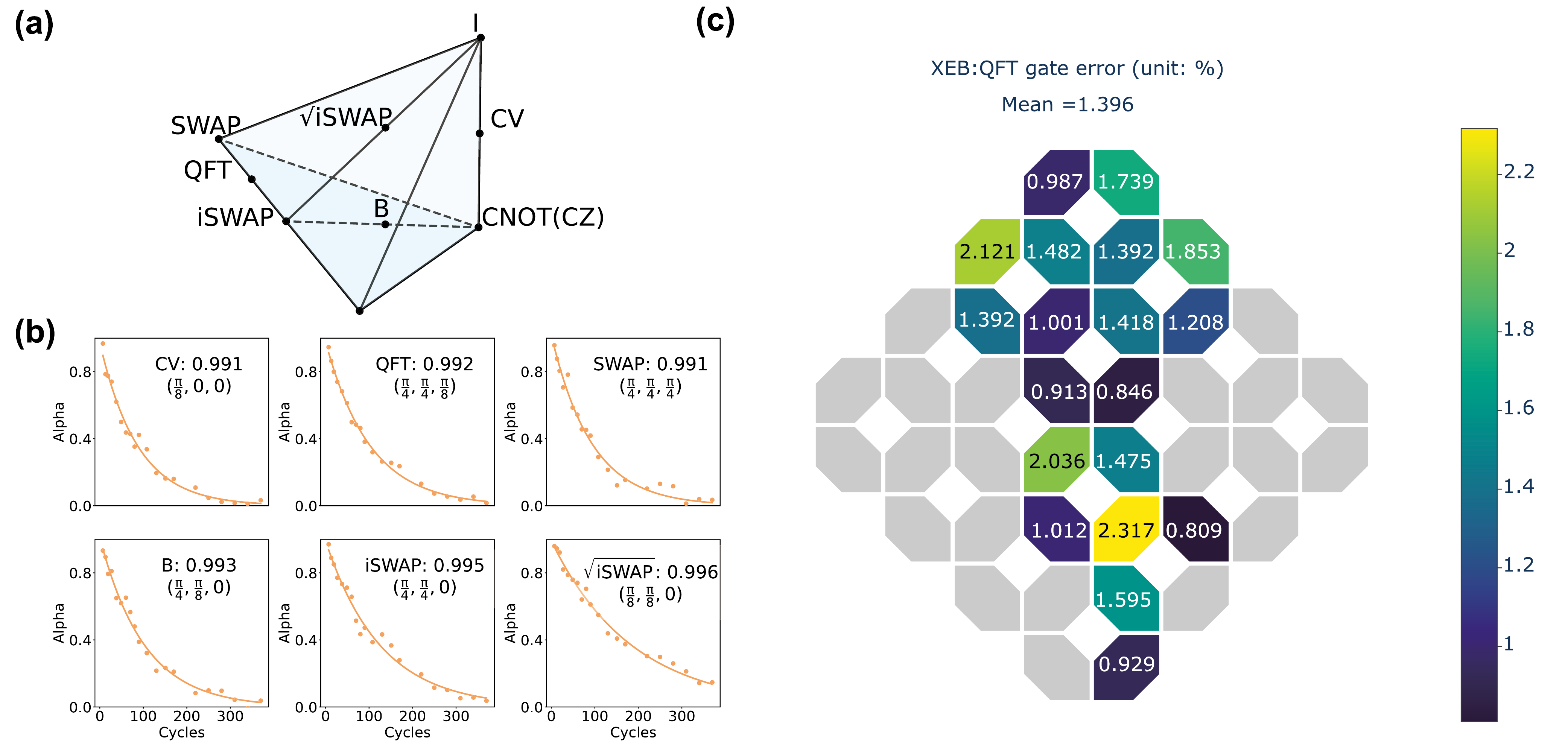}
    \caption{
        AshN gate performance of the system. (a) Illustration of the AshN two-qubit gate in the Weyl chamber calibrated in this work. Under the KAK decomposition, every two-qubit operation in $\mathrm{SU}(4)$ is locally equivalent to a point in the Weyl chamber. (b) Cross-entropy benchmarking (XEB) results for different types of AshN two-qubit gates. Here, ``cycles'' denotes the number of layers in the random circuits, and Alpha represents the XEB fidelity extracted from the cross-entropy between the experimental and ideal probability distributions. Each panel corresponds to a different type of AshN two-qubit gate; the annotations indicate the gate type, the extracted XEB fidelity, and the corresponding Weyl chamber coordinates of the gate. (c) Error of the two-qubit $\mathrm{QFT}$ gate obtained from parallel calibration experiments. The device supports a maximum twelve-qubit chain and a $3$$\times$$3$ 2D topology.
    }
    \label{Figure2}
\end{figure*}

The framework is built upon gate-specific calibration protocols that exploit the characteristic Hamiltonian dynamics of each AshN gate family. For $\mathrm{SWAP}$ gates, for example, the pulse parameters are extracted from the measured $\ket{10}\leftrightarrow\ket{01}$ population dynamics under different waveform settings. Similar dynamical signatures are used for the remaining gate families, allowing the calibration parameters to be determined directly from computational-basis measurements without iterative feedback optimization. Figure~\ref{Figure2}(c) shows the parallel calibration of the quantum Fourier transform ($\mathrm{QFT}$) gate at the Weyl coordinates $(\pi/4,\pi/4,\pi/8)$. Because this gate requires simultaneous $Z$ and XY control on both qubits, it incorporates all pulse types used in the AshN gate set. The corresponding couplers are divided into four groups and calibrated simultaneously before being configured into a 12-qubit 1D chain and a $3\times3$ 2D array~\cite{supplemental}.

For gate characterization, randomized benchmarking has long been the standard tool, but it is primarily tailored to Clifford gates~\cite{knill2008randomized,magesan2011scalable}. As non-conventional and continuously parameterized gate sets become increasingly relevant, alternative protocols have been developed, including fully randomized benchmarking~\cite{huang2023quantum} and cross-entropy benchmarking~\cite{boixo2018characterizing}. Here we adopt cross-entropy benchmarking, which has become a widely used approach for characterizing non-Clifford and hardware-native quantum operations~\cite{Arute2019,gao2025establishing}.

A further practical challenge in implementing general AshN circuits is that the local phases generated by most two-qubit gates cannot always be absorbed into virtual-$Z$ corrections and may therefore require additional physical single-qubit rotations. To mitigate this overhead, we develop a hybrid phase-compensation strategy that selectively uses virtual-$Z$ corrections whenever allowed by the underlying two-qubit interaction, while retaining physical $Z$ rotations when virtual compensation is unavailable. This strategy extends beyond conventional $\mathrm{CZ}$ and $\mathrm{iSWAP}$ gates and applies to a broad class of native two-qubit interactions. In particular, many experimentally relevant AshN gates are locally equivalent to two-qubit operations that permit $R_Z$ commutation, provided that their Weyl chamber coordinates $(a,b,c)$ satisfy either $a=b=\pi/4$ or $b=c=0$~\cite{supplemental}. Consequently, a substantial fraction of the single-qubit phase corrections can be implemented virtually, reducing the number of required physical single-qubit gates. For this class of AshN gates, each preceding single-qubit operation on either qubit can be implemented with at most two native single-qubit gates, reducing the single-qubit-gate count and improving circuit fidelity.

\section{Circuit Benchmarks on Restricted Topologies}

\begin{figure*}[htbp]
    \centering
    \includegraphics[width=0.85\textwidth]{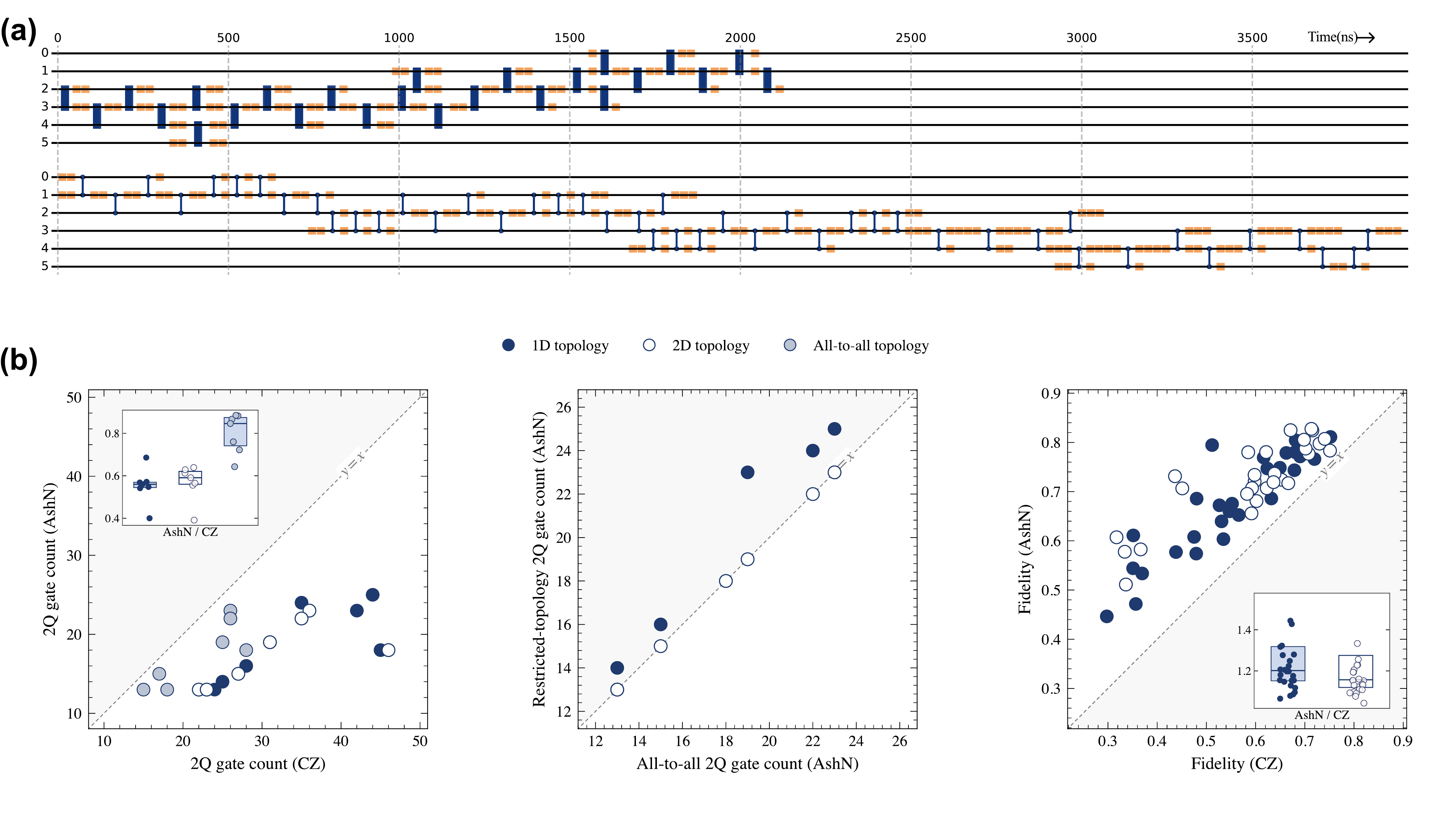}
    \caption{
        Comparison between $\mathrm{CZ}$- and AshN-based two-qubit gate schemes. (a) Representative benchmark circuit compiled using the AshN- and $\mathrm{CZ}$-based frameworks. The upper panel shows the AshN implementation, consisting of native AshN two-qubit gates and single-qubit gates, while the lower panel shows the corresponding $\mathrm{CZ}$-based implementation composed of $\mathrm{CZ}$ and single-qubit gates. The circuit shown is one of the seven benchmark circuits used in the experiments. (b) Comparison of compiled two-qubit gate counts and circuit fidelities for the AshN- and $\mathrm{CZ}$-based compilation schemes. The left panel compares the compiled two-qubit gate counts for the seven benchmark circuits using the AshN and $\mathrm{CZ}$ gate sets on one-dimensional (1D) and two-dimensional (2D) topologies, together with an all-to-all-connectivity compilation reference. The inset summarizes the median gate-count ratio (AshN/$\mathrm{CZ}$), with the box indicating the interquartile range (25th--75th percentiles). The middle panel compares the compiled two-qubit gate counts obtained with the AshN gate set on the 1D and 2D topologies against the corresponding all-to-all-connectivity compilation reference; points with identical values overlap. The all-to-all results are compilation references rather than measurements on fully connected hardware. The right panel shows the circuit fidelities for all 28 experimental instances (seven benchmark circuits, each evaluated using four computational-basis input states), defined as the computational-basis success probability. The inset summarizes the median fidelity ratio (AshN/$\mathrm{CZ}$) over all experimental instances.
    }
    \label{Figure3}
\end{figure*}

Having calibrated and characterized the representative native gates, we next test whether their additional expressivity translates into measurable circuit-level benefits under restricted connectivity. We benchmark seven representative quantum circuits on both 1D and 2D topologies, using either the AshN or $\mathrm{CZ}$ gate set; these circuits span a diverse set of two-qubit operations, including $\mathrm{iSWAP}$, $\mathrm{SWAP}$, $\mathrm{QFT}$, and controlled-$V$ ($\mathrm{CV}$) (see~\cite{supplemental} for the specific AshN gate parameters). We restrict the benchmarks to systems of up to seven qubits, since for larger practical circuits the $\mathrm{CZ}$-based implementations become too deep on current hardware, preventing a meaningful comparison between the two gate sets.


As shown in Fig.~\ref{Figure3}, AshN compilation consistently yields lower two-qubit gate counts than $\mathrm{CZ}$-based compilation on both topologies. Here, we use the two-qubit gate count as a simplified measure of circuit execution cost. A more refined estimate would additionally account for the durations of the individual native gates. As discussed in~\cite{chen2024one}, when decoherence is the dominant error source, gate duration can provide a first-order estimate of the corresponding gate error; a duration-based comparison was presented in~\cite{chen2025efficient}. For simplicity and consistency across the benchmark circuits, we therefore focus on two-qubit gate counts throughout this paper. Across the seven circuits on the 1D topology, AshN compilation reduces the two-qubit gate count to approximately $54.8\%$ of the $\mathrm{CZ}$ value based on the geometric mean, consistent with a reduction to $56.0\%$ of the CZ value based on the median shown in Fig.~\ref{Figure3}. On the 2D topology, AshN compilation achieves a similar reduction, reaching approximately $56.3\%$ of the $\mathrm{CZ}$ value based on the geometric mean and $59.1\%$ based on the median.

Beyond the overall reduction in two-qubit gate count with AshN compilation, Fig.~\ref{Figure3} also illustrates the interplay between device connectivity and the native gate set. Restricted connectivity introduces substantial routing overhead for $\mathrm{CZ}$-based compilation, creating greater opportunities for AshN compilation to reduce this overhead through $\mathrm{SWAP}$ absorption. In contrast, the denser connectivity of the 2D topology reduces the need for routing and consequently lowers the two-qubit gate count for both native gate sets.
Owing to current hardware limitations, our circuit-level experiments are restricted to seven benchmark circuits containing up to seven qubits. Notably, for these benchmark circuits, compiling onto the 2D topology with the AshN gate set introduces no additional two-qubit gates beyond those required by the corresponding logical circuits, indicating zero routing overhead in terms of two-qubit gate count. To determine whether this advantage persists beyond the experimentally accessible regime, we further benchmark larger instances from a broader range of circuit families, with up to several tens of qubits and thousands of two-qubit gates.
Across these larger benchmarks, AshN compilation reduces the geometric mean of the two-qubit gate-count and gate-depth overheads from $\BenchSquareCZGateGeomean$ and $\BenchSquareCZDepthGeomean$ for $\mathrm{CZ}$-based compilation to $\BenchSquareAshNGateGeomean$ and $\BenchSquareAshNDepthGeomean$, respectively, as summarized in \SuppRoutingBenchmarkSummaryTable{} of the Supplemental Material. The Supplemental Material further extends this analysis to even larger circuits, demonstrating that the routing advantage of AshN persists at larger circuit scales while also revealing workload-dependent cases where the benefit becomes less pronounced. In particular, the reduction in routing overhead is limited for less structured circuits, such as circuits dominated by a single long-range two-qubit interaction between distant qubits (e.g., between the top-left and bottom-right corners of a planar topology), as well as quantum volume circuits, whose intrinsic randomness limits opportunities for routing optimization. This dependence on device connectivity further motivates the study of more realistic scenarios in which regular connectivity is locally disrupted by hardware defects.

\section{Dicke-State Preparation on Defective Topologies}

\begin{figure*}[htbp]
    \centering
    \includegraphics[width=0.85\textwidth]{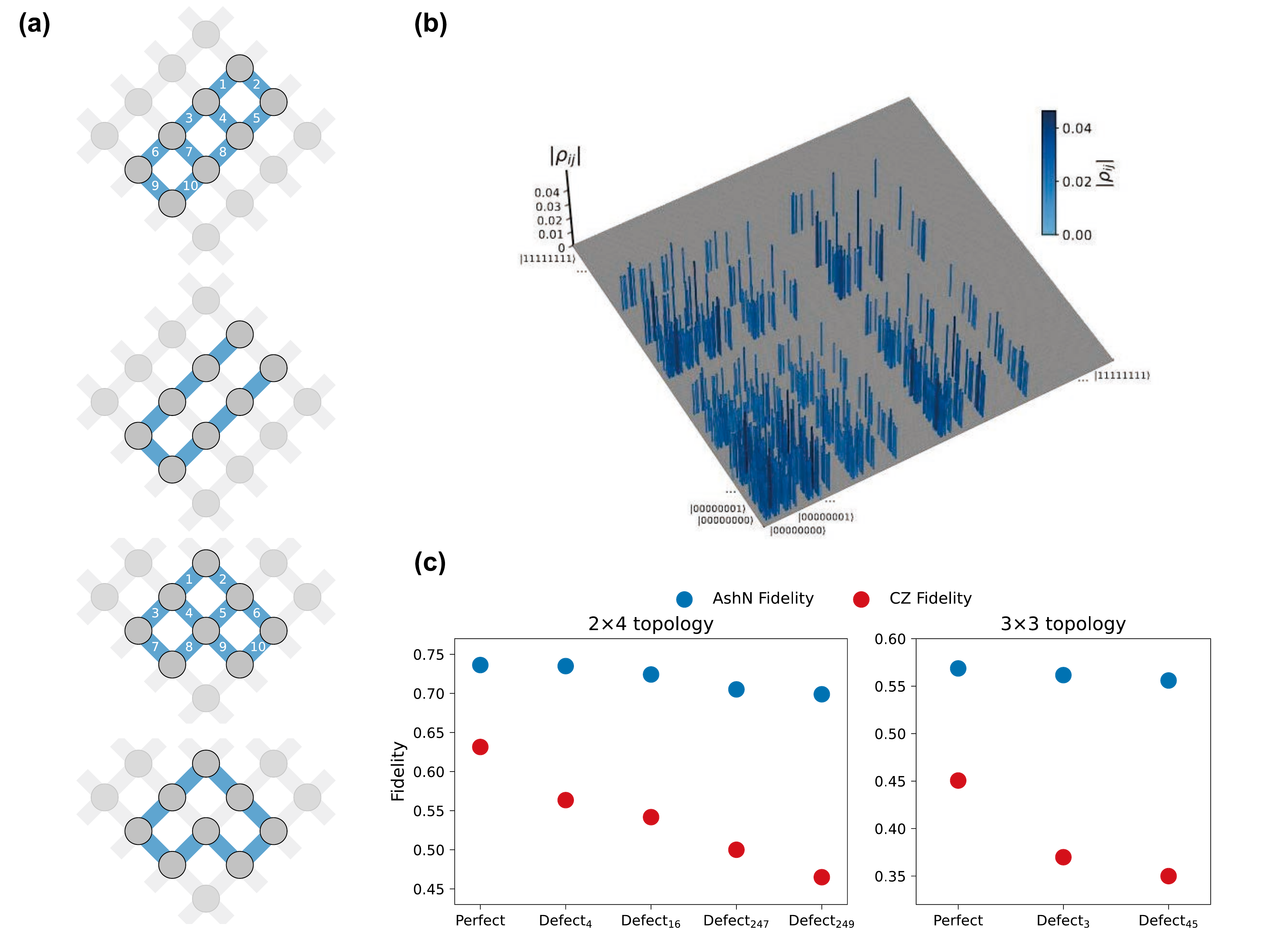}
    \caption{(a) Schematic of the qubit topologies used in this work. From top to bottom: the defect-free $2$$\times$$4$ topology, the corresponding Defect$_{247}$ configuration, the defect-free $3$$\times$$3$ topology, and the corresponding Defect$_{45}$ configuration. (b) Full quantum state tomography (QST) of the Dicke state prepared on the defect-free $2$$\times$$4$ topology using the AshN scheme, reconstructed in the computational basis. The ideal two-excitation Dicke state has support over 28 computational-basis states with equal probability $1/28$, and the reconstructed state achieves a fidelity of $F=0.736\pm0.001$ with the ideal Dicke state. Bar height and color represent the magnitude of the reconstructed density-matrix element $|\rho_{ij}|$, where $\rho_{ij}=\langle i|\rho|j\rangle$. Elements with $|\rho_{ij}|<0.01$ are omitted for clarity. (c) Comparison of the reconstructed Dicke-state fidelity under different topological configurations. The left panel corresponds to the $2$$\times$$4$ system, and the right panel corresponds to the $3$$\times$$3$ system. Error bars of state fidelity are smaller than the marker size and are not visible.
    }
    \label{Figure4}
\end{figure*}

To evaluate the resilience of the AshN gate scheme to hardware defects in a practically relevant quantum-information task, we prepared an eight-qubit, two-excitation Dicke state on both $2$$\times$$4$ and $3$$\times$$3$ topologies. The preparation of multipartite entangled states constitutes an important benchmark for quantum processors, as it requires coherent multi-qubit interactions and is highly sensitive to gate errors and connectivity constraints. In the AshN implementation, the circuit is compiled using a native gate set comprising $\mathrm{CZ}$, $\mathrm{iSWAP}$, $\sqrt{\mathrm{iSWAP}}$, $\mathrm{SWAP}$, and a B-like entangling gate. Representative compiled circuits for the AshN- and $\mathrm{CZ}$-based implementations are provided in~\cite{supplemental}. The calibrated $\mathrm{CZ}$ gates achieve an average XEB error of $0.550\%$, whereas the remaining native two-qubit gates ($\mathrm{iSWAP}$, $\sqrt{\mathrm{iSWAP}}$, $\mathrm{SWAP}$, and B-like) exhibit a comparable average XEB error of $0.615\%$~\cite{supplemental}. This slightly higher error primarily stems from the increased control complexity of these gates, including more elaborate pulse sequences and additional crosstalk introduced by simultaneous XY drives. The experimental density matrix is reconstructed via quantum state tomography, from which the state fidelity and the collective-spin observables $\langle J_z^2\rangle$ and $\langle J_x^2+J_y^2\rangle$ are evaluated to quantify the quality of the prepared Dicke state and its multipartite coherence~\cite{James2001,Lucke2014}.

        For the defect-free $2$$\times$$4$ topology, the Dicke state prepared with the AshN scheme achieves a fidelity of $0.736\pm0.001$, with corresponding collective-spin observables $\langle J_z^2 \rangle = 3.91\pm0.03$ and $\langle J_x^2 + J_y^2 \rangle = 13.64\pm0.13$. The ideal eight-qubit Dicke state with two excitations has $\langle J_z^2\rangle=4$ and $\langle J_x^2+J_y^2\rangle=16$, providing the theoretical benchmarks for comparison. Under identical conditions, the $\mathrm{CZ}$-based implementation reaches a lower fidelity of $0.631\pm0.001$, with $\langle J_z^2 \rangle = 3.79\pm0.04$ and $\langle J_x^2 + J_y^2 \rangle = 12.39\pm0.13$. The improved performance of the AshN scheme originates from its lower compilation overhead, requiring only 23 entangling gates compared with 31 for the $\mathrm{CZ}$-based realization. We further characterize the entanglement structure using the fully positive-partial-transpose (PPT) entanglement witness optimized for white-noise robustness $W_{8, 2}^{\textrm{opti}}$ proposed in~\cite{Bergmann2013}. The AshN-prepared state yields $\mathrm{Tr}(W_{8, 2}^{\textrm{opti}}\cdot\rho_{\mathrm{AshN}})=-0.0344\pm0.0008$, significantly below zero and thereby certifying genuine multipartite entanglement (GME). In contrast, the $\mathrm{CZ}$-prepared state gives $\mathrm{Tr}(W_{8, 2}^{\textrm{opti}}\cdot\rho_{\mathrm{CZ}})=0.0593\pm0.0008$, for which GME cannot be established. Thus, under the same experimental conditions, the chosen witness certifies GME for the AshN-based realization but does not certify GME for the $\mathrm{CZ}$-based realization.

        More importantly, the advantage of the AshN scheme persists in the presence of hardware defects. As shown in Fig.~\ref{Figure4}, the fidelity of the $\mathrm{CZ}$-prepared Dicke state deteriorates rapidly as defective couplers are introduced, whereas the fidelity obtained with the AshN scheme remains nearly unchanged across different defect configurations, demonstrating strong resilience to connectivity disruptions. Even with three defective couplers, the Dicke-state fidelity decreases by only about $0.04$ relative to the defect-free topology. Furthermore, the PPT-mixture entanglement witness remains negative across all defective topologies investigated, thereby certifying genuine multipartite entanglement throughout the entire set of defect configurations. These results show that, for the Dicke-state circuits and defect configurations tested here, AshN compilation mitigates the additional routing overhead associated with connectivity loss while preserving the certification of genuine multipartite entanglement.

        \section{Implications and Outlook}
        Taken together, the hardware experiments and compilation analysis provide an end-to-end test of whether richer native control can alter the effective connectivity of a sparse superconducting processor. On the hardware available here, we calibrate AshN gates across configurations extending to a 12-qubit chain and a $3\times3$ array, and use circuit experiments at smaller scales to verify that the resulting reduction in compiled overhead survives physical implementation. The larger compilation study then probes regimes beyond current experimental reach. For many application-derived circuits, the routing-overhead factor---the mapped two-qubit cost normalized by its all-to-all-connectivity compilation reference---clusters near unity, most clearly on the square lattice, where its geometric mean is $\BenchSquareAshNGateGeomean{}$ in gate count and $\BenchSquareAshNDepthGeomean$ in two-qubit depth~\cite{supplemental}. This behavior is not universal. In quantum-volume circuits, interaction partners change rapidly with circuit width, leaving few adjacent operations with which a $\mathrm{SWAP}$ can be absorbed; the routing overhead consequently continues to grow with system size. The contrast identifies circuit interaction structure, rather than size alone, as a boundary of the approach~\cite{supplemental}.

        The scaling implication is therefore conditional. Our compilation results suggest that sparse processors could approach all-to-all-connectivity compilation costs for broad classes of structured workloads, provided that larger devices can retain comparable gate quality and extend parallel AshN calibration without prohibitive growth in crosstalk, drift or calibration complexity. Establishing those conditions will require hardware experiments beyond the $12$-qubit-chain and $3\times3$ regimes studied here. Within this boundary, connectivity is usefully viewed as an operational property: the coupling graph determines where interactions occur, whereas the native control and compiler determine how strongly that geometry constrains a circuit~\cite{gokhale2020optimized,smith2022programming}.

        Crucially, this principle does not require access to arbitrary $\mathrm{SU}(4)$ operations. Full two-qubit expressivity, as realized here through AshN control, is one sufficient way to ensure that an arbitrary two-qubit operation remains implementable after composition with $\mathrm{SWAP}$. A restricted native family can provide the same routing resource whenever it is closed under the relevant $\mathrm{SWAP}$ compositions, up to local operations. The architectural quantity of interest is thus not expressivity in isolation, but the match between this closure property and the interaction structure of the target workload.

        This distinction opens a concrete route towards quantum error correction. The Louvre and Bunny-code studies exploit the same algebraic mechanism: a $\mathrm{CNOT}$ composed with $\mathrm{SWAP}$, often denoted $\mathrm{CXSWAP}$, is locally equivalent to an $\mathrm{iSWAP}$-class operation~\cite{zhou2026louvre,zhou2026bunny}. Access to $\mathrm{CNOT}$- and $\mathrm{iSWAP}$-class interactions therefore supplies the relevant $\mathrm{SWAP}$ closure without universal two-qubit control. Those studies predict that such control can reduce the nonlocal coupling requirements of particular quantum low-density parity-check code layouts. The present work does not implement encoded operations or assess logical error rates, but provides an experimental basis for testing this mechanism on superconducting hardware. Demonstrating that the same control advantage survives repeated syndrome-extraction cycles, leakage constraints and fault-tolerant scheduling is a natural next step.

\section{Acknowledgements}
We thank David Ding, Hui-Hai Zhao, Linghang Kong, Fang Zhang, and Gengyan Zhang for helpful discussions on AshN-gate calibration and many other aspects of hardware–software co-design. Part of this work was conducted while J.G. and X.Y. were interns at TraverseQuantum Co., Ltd. This work was supported by the National Science and Technology Major Project ``Quantum Science and Technology'' (Grant Nos.~2025ZD0300600 and 2025ZD0300604), Zhongguancun Laboratory and National Key Research, Development Program of China (Grant No. 2025YFE0200900) and the Research Grants Council of Hong Kong SAR (No.~16217326).

\bibliography{BeyondNN_abbr}
\end{document}


\title{Supplemental Material for ``Lifting connectivity bottlenecks in superconducting quantum processors via enriched native two-qubit gates''}

\newcommand{\THUCS}{Department of Computer Science and Technology, Tsinghua University, Beijing 100084, P.R. China}
\newcommand{\CTQ}{China Telecom Quantum Information Technology Group Co., Ltd., Hefei 230094, Anhui, P.R. China}
\newcommand{\QRI}{Quantum Research Institute of China Telecom, Hefei 230031, Anhui, P.R. China}
\newcommand{\TQ}{TraverseQuantum Co., Ltd., Beijing, P.R. China}
\newcommand{\HKUST}{Department of Electronic and Computer Engineering, The Hong Kong University of Science and Technology, Hong Kong}
\newcommand{\ZGC}{Zhongguancun Laboratory, Beijing, P.R. China}

\affiliation{\THUCS}
\affiliation{\CTQ}
\affiliation{\QRI}
\affiliation{\TQ}
\affiliation{\HKUST}
\affiliation{\ZGC}

\author{Hanyi Wang}
\thanks{These authors have contributed equally to this work.}
\affiliation{\CTQ}
\affiliation{\QRI}
\author{Jingzhe Guo}
\thanks{These authors have contributed equally to this work.}
\affiliation{\THUCS}
\affiliation{\TQ}
\author{Lijun Sun}
\thanks{These authors have contributed equally to this work.}
\affiliation{\CTQ}
\affiliation{\QRI}
\author{Zhaohui Yang}
\affiliation{\HKUST}
\author{Weizhi Tao}
\affiliation{\CTQ}
\affiliation{\QRI}
\author{Xingye Yuan}
\affiliation{\THUCS}
\affiliation{\TQ}
\author{Qiankun Wang}
\affiliation{\CTQ}
\affiliation{\QRI}
\author{Bihao Guo}
\affiliation{\CTQ}
\affiliation{\QRI}
\author{Chunwang Liu}
\affiliation{\CTQ}
\affiliation{\QRI}
\author{Rui Yang}
\affiliation{\CTQ}
\affiliation{\QRI}
\author{Yang Li}
\affiliation{\CTQ}
\affiliation{\QRI}
\author{Yu Fan}
\affiliation{\CTQ}
\affiliation{\QRI}
\author{Jiasheng Hu}
\affiliation{\CTQ}
\affiliation{\QRI}
\author{Junhe Wang}
\affiliation{\CTQ}
\affiliation{\QRI}
\author{Shuyue Zheng}
\affiliation{\CTQ}
\affiliation{\QRI}
\author{Shengbin Wang}
\affiliation{\CTQ}
\affiliation{\QRI}
\author{Xinfang Zhang}
\affiliation{\CTQ}
\affiliation{\QRI}
\author{Feng Wu}
\email{wufeng@iqubit.org}
\affiliation{\ZGC}
\author{Hantao Sun}
\email{sunhantao@chinatelecom.cn}
\affiliation{\CTQ}
\affiliation{\QRI}
\author{Jianxin Chen}
\email{chenjianxin@tsinghua.edu.cn}
\affiliation{\THUCS}

\date{\today}
\maketitle

\renewcommand\thefigure{S\arabic{figure}}
\renewcommand\theequation{S\arabic{equation}}
\renewcommand{\thetable}{S\arabic{table}}
\setcounter{secnumdepth}{1}
\renewcommand\thesection{\Roman{section}}

\section{Experimental Setup}
The experiment was performed on a superconducting quantum processor with the same hardware architecture as \textit{Zuchongzhi}~3.0, comprising 105 frequency-tunable transmon qubits and 182 tunable couplers~\cite{jiang2026one,gao2025establishing}. Detailed device and readout characteristics can be found in Ref.~\cite{group2025tianyan}, including the maximum $f_{01}$ frequencies of the qubits, the idle $f_{01}$ frequencies of the qubits, idle $T_1$ times, as well as the readout error and the single-qubit gate error.

\begin{figure*}[htbp]
    \centering
    \includegraphics[width=0.95\textwidth]{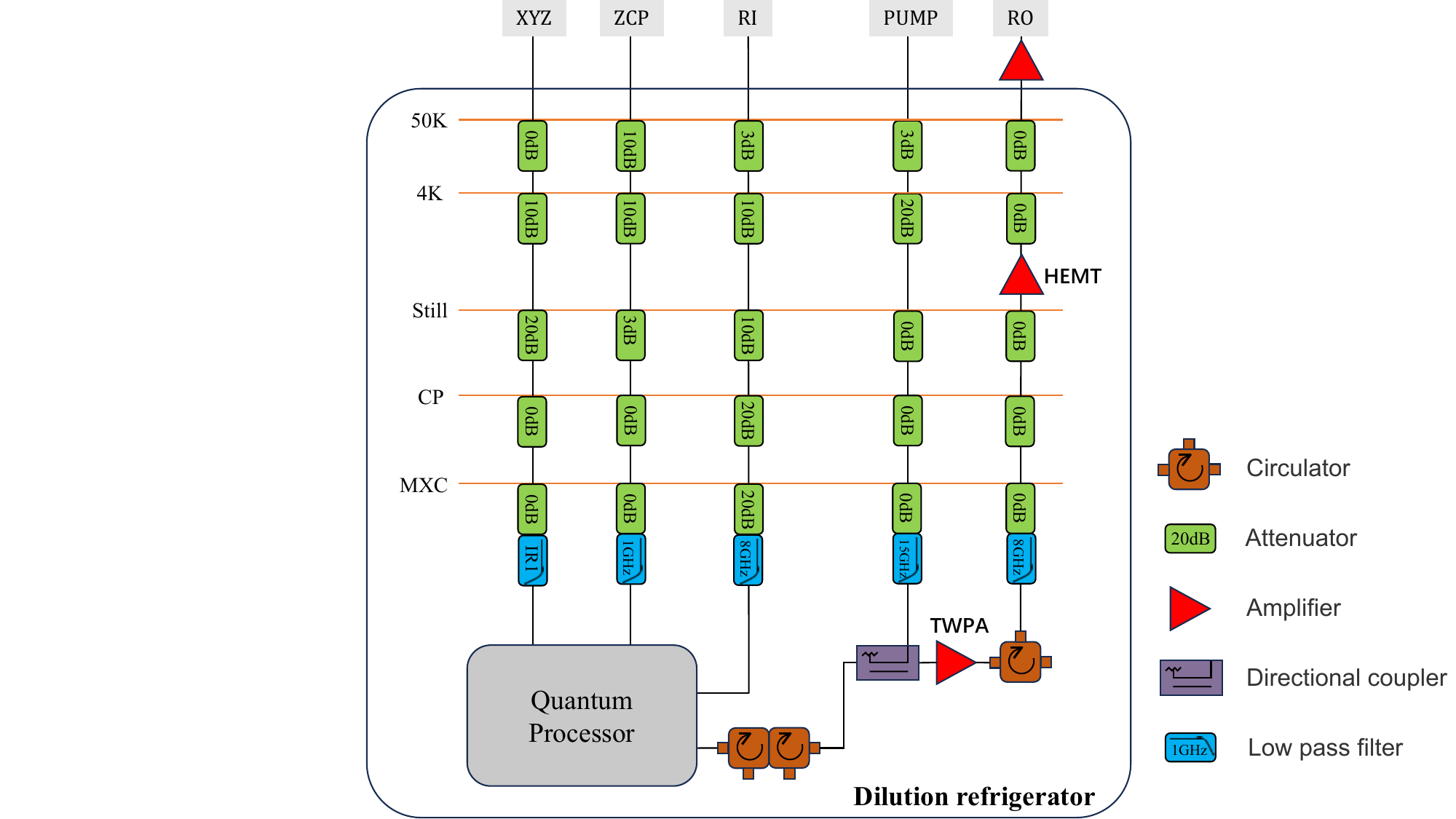}
    \caption{
        Schematic of the control electronics and wiring in the dilution refrigerator. Each qubit is equipped with independent XY and Z control lines, which are combined at room temperature through a bias tee before entering the refrigerator, while the coupler control lines are routed separately. Attenuators and filters are installed at different temperature stages to suppress noise. The readout chain consists of a TWPA, a HEMT amplifier, and room-temperature amplifiers. A directional coupler combines the pump and readout signals before the TWPA. Room-temperature electronics generate the control and readout signals, while the amplified readout signals are digitized and demodulated by ADC modules.
    }

    \label{FigureS1}
\end{figure*}

The dilution refrigerator and transmission system are illustrated in Fig.~\ref{FigureS1}. This system incorporates multiple temperature stages, including the 50\,K, 4\,K, Still, Cold Plate (CP), and Mixing Chamber (MXC) stages, each equipped with appropriately rated attenuators to suppress thermal noise. At the MXC stage, low-pass filters are used to reduce high-frequency noise, while directional couplers and circulators are incorporated to manage signal routing and to isolate the quantum processor from amplifier backaction. Signal amplification is accomplished through a combination of a High Electron Mobility Transistor (HEMT) amplifier at the 4\,K stage and a Traveling Wave Parametric Amplifier (TWPA) at the MXC stage, ensuring high-fidelity readout. This structured arrangement of attenuators, filters, and amplifiers enables precise control of signal propagation while maintaining the millikelvin temperatures essential for the operation of superconducting qubits.

\section{Calibration of AshN two-qubit gates}

During the operation of a two-qubit gate, the two qubits are tuned to resonance ($\omega_{1} = \omega_{2} = \omega$). The universal Hamiltonian for the AshN two-qubit gate in the rotating frame can be expressed as
\begin{equation}
    \begin{aligned}
        H= & \frac{g}{2}(XX+YY)+\frac{\Omega_1}{2}\left(\cos \phi_1 XI+\sin \phi_1 YI\right)+\frac{\Omega_2}{2}\left(\cos \phi_2 IX+\sin \phi_2 IY\right)+\frac{\Delta}{2}(ZI+IZ).
    \end{aligned}
    \label{Hamiltonian}
\end{equation}
Here, the parameter $g$ denotes the coupling strength of the coupler, while $\Omega_{1}$ and $\Omega_{2}$ are the XY-drive amplitudes applied to the two qubits, respectively. The parameters $\phi_{1}$ and $\phi_{2}$ denote the corresponding XY-drive phases for the two qubits, respectively. $\Delta = \omega - \omega_{d}$ represents the detuning between the qubit frequency and the XY-drive frequency.

We develop a parallel calibration framework for various types of AshN two-qubit gates. For example, we calibrate a single group of couplers for the most calibration-intensive gate family, the quantum Fourier transform ($\mathrm{QFT}$) gate. In addition, our approach supports the simultaneous calibration of different AshN two-qubit gates on different couplers. In contrast to previous work, we avoid closed-loop optimization based on XEB fidelity, owing to its substantial time overhead~\cite{chen2025efficient}. This measurement-efficient strategy substantially shortens the calibration time, making it well suited to large-scale superconducting processors that require the simultaneous deployment of many calibrated two-qubit interactions.

\begin{figure}[htb]
    \centering
    \includegraphics[width=0.48\textwidth]{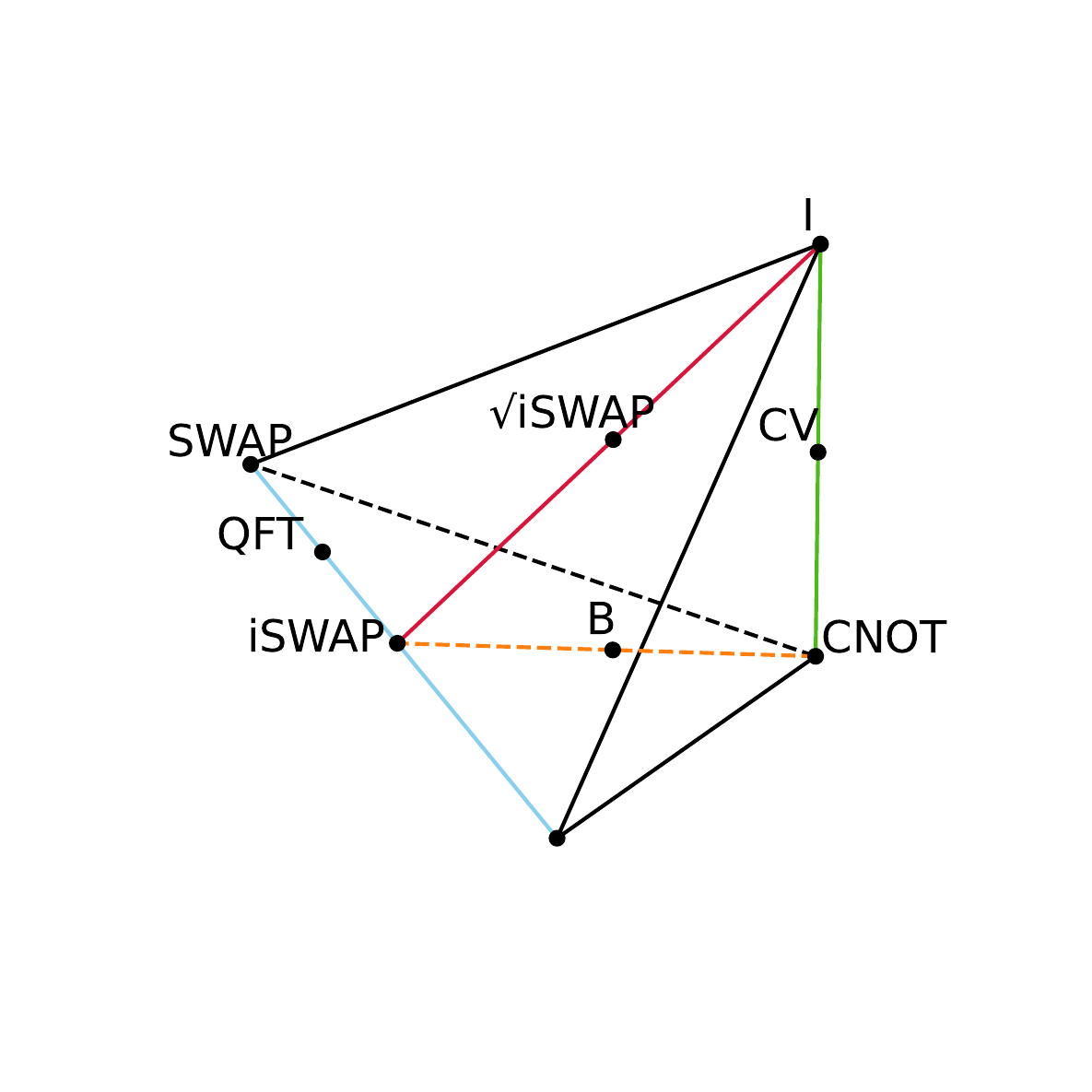}
    \caption{
        The locations of AshN two-qubit gates within the Weyl chamber are marked. Different calibration methods are employed for different regions of the Weyl chamber, as indicated by different colors.
    }
    \label{welchamber_supp}
\end{figure}

\begin{figure*}[htbp]
    \centering
    \includegraphics[width=0.7\textwidth]{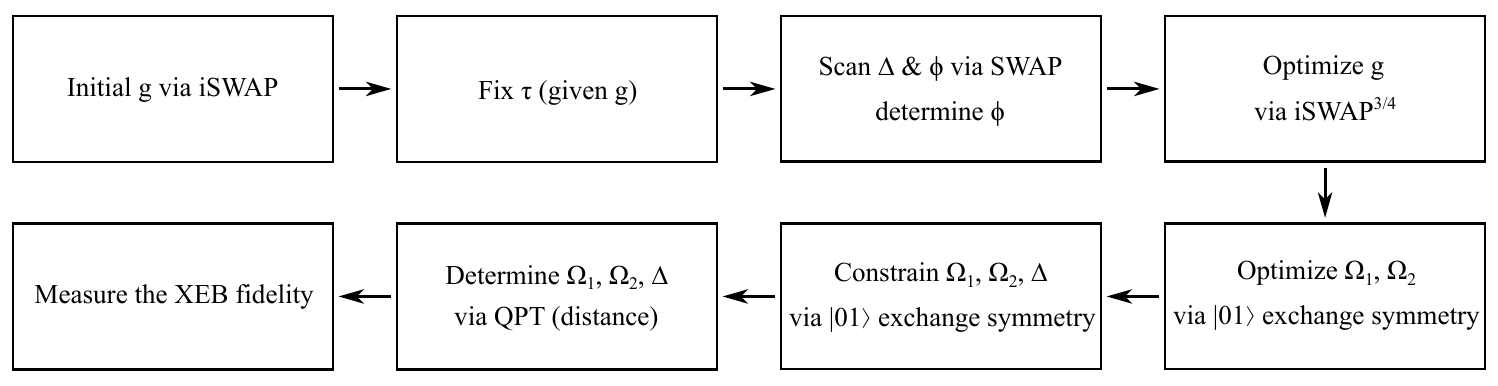}
    \caption{
        The calibration process of the $\mathrm{QFT}$ gate involves several parameters: $g$ represents the coupling strength of the couplers; $\Delta$ indicates the detuning between the qubit frequency and the XY-drive frequency; $\phi$ refers to the relative phase of the two XY-drive signals applied to the two qubits; and $\Omega_{1}$ and $\Omega_{2}$ are the XY-drive amplitudes for the two qubits, respectively.
    }
    \label{QFT_calibration_flow}
\end{figure*}

Different calibration strategies are used for gates in different regions of the Weyl chamber, as illustrated in Fig.~\ref{welchamber_supp}. As an example, we consider the $\mathrm{QFT}$ gate. Its implementation requires simultaneous XY drives on both qubits and is therefore more involved than the implementations of $\mathrm{iSWAP}$-like and controlled-$V$ ($\mathrm{CV}$) gates, where $V^2=X$, so that a controlled-$V$ gate squares to a $\mathrm{CNOT}$. These gate families require an XY drive on only one qubit. The $\mathrm{QFT}$-gate calibration workflow is shown in Fig.~\ref{QFT_calibration_flow} and described below.

\subsection{Calibration of $\mathrm{QFT}$ gate}
\subsubsection{Calibration of $\mathrm{iSWAP}$ gate parameters}

\begin{figure}[htb]
    \centering
    \includegraphics[width=0.7\textwidth]{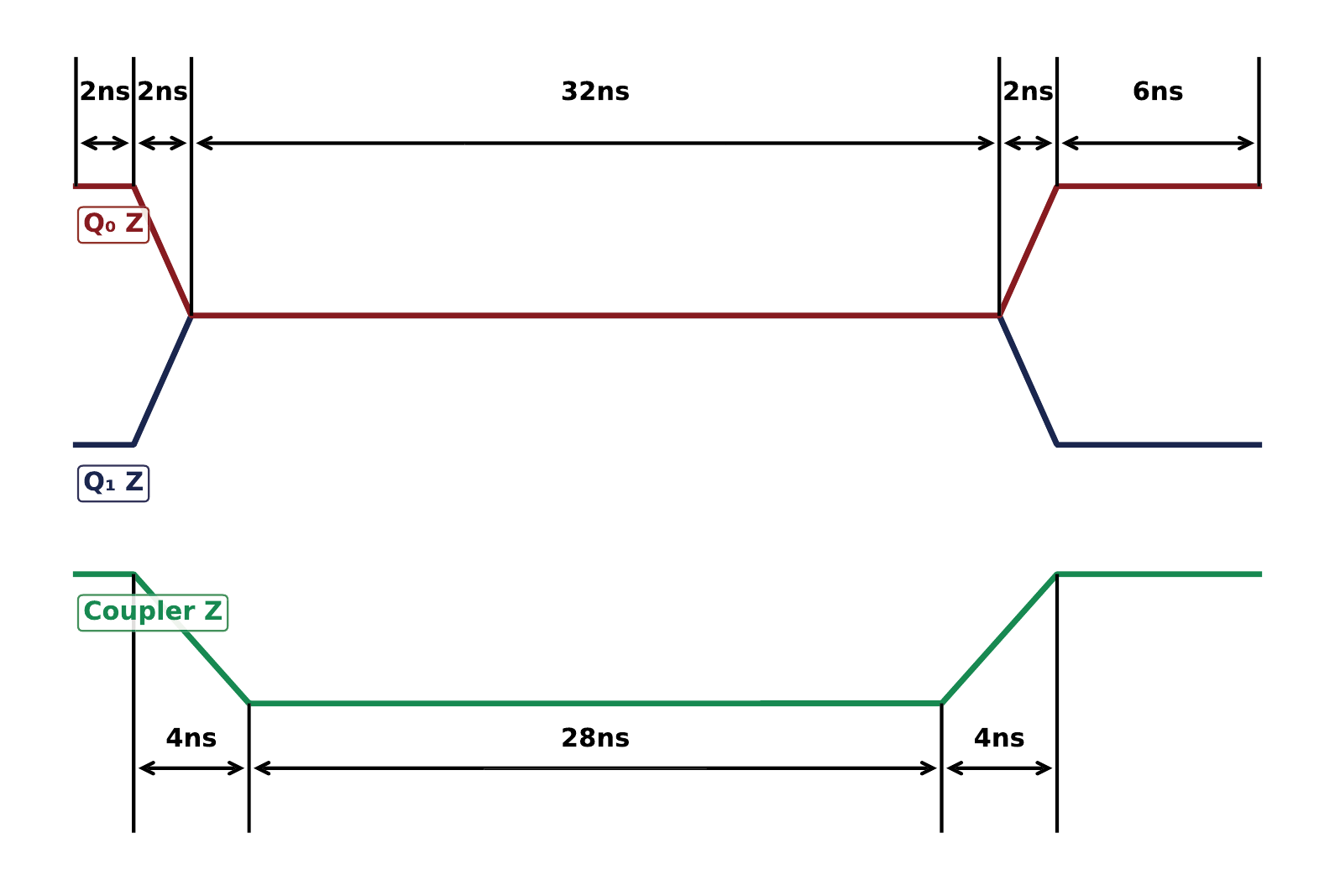}
    \caption{
        Pulse waveform of the $\mathrm{iSWAP}$ gate. From left to right, the qubit pulse consists of a $2~\mathrm{ns}$ idle segment, followed by a $2~\mathrm{ns}$ rising edge, a $30~\mathrm{ns}$ gate segment, a $2~\mathrm{ns}$ falling edge, and concludes with a $6~\mathrm{ns}$ idle segment. The coupler pulse consists of a $2~\mathrm{ns}$ idle segment, a $4~\mathrm{ns}$ rising edge, a $26~\mathrm{ns}$ gate segment, a $4~\mathrm{ns}$ falling edge, and likewise ends with a $6~\mathrm{ns}$ idle segment.
    }
    \label{ashn_wave_supp}
\end{figure}

\begin{figure*}[htbp]
    \centering
    \includegraphics[width=0.9\textwidth]{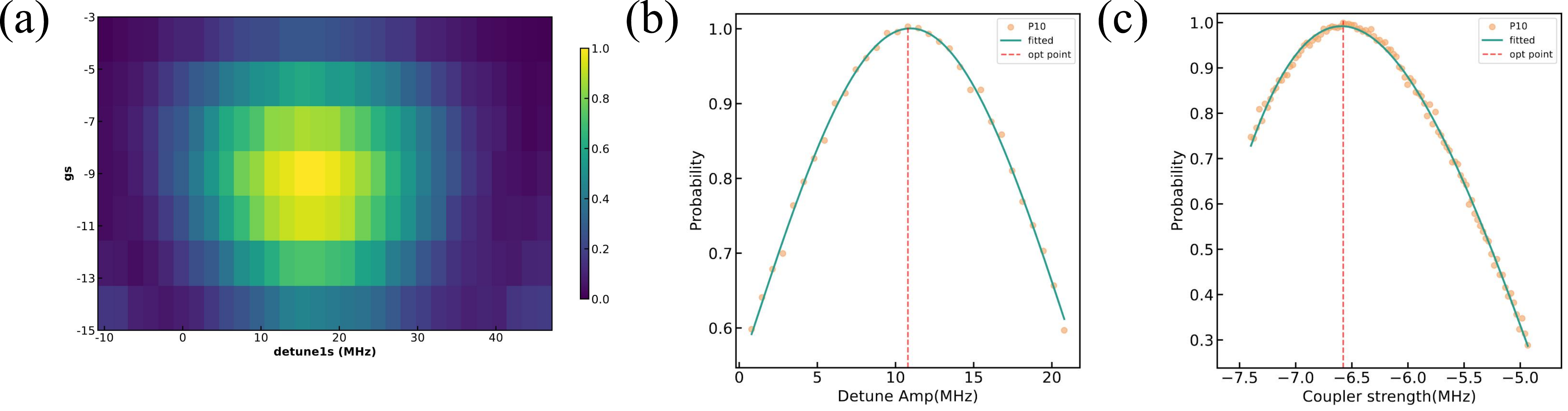}
    \caption{
        Calibration of the $\mathrm{iSWAP}$ two-qubit gate. (a) The population $P_{01}$ is measured as a function of the detuning frequency of $Q_{1}$ and the coupling strength to determine the parameter regime in which the two qubits are fully swapped, corresponding to a final-state $P_{01}$ approaching unity. (b) The population $P_{01}$ is measured as a function of the detuning frequency of $Q_{1}$ to optimize the detuning parameter of the $\mathrm{iSWAP}$ gate. The green curve represents a sinusoidal fit, while the red dotted line indicates the peak value of $P_{01}$. (c) The population $P_{01}$ is measured as a function of the coupling strength to refine the coupling parameter $g$ of the $\mathrm{iSWAP}$ gate. The green curve represents a polynomial fit, and the red dotted line marks the peak value of $P_{01}$.
    }
    \label{iswap}
\end{figure*}

Calibration of all AshN gates starts with the $\mathrm{iSWAP}$ interaction, which determines the coupling strength $g$ and the detuning of the qubits. Thus, we first calibrate the $\mathrm{iSWAP}$ operation with the XY drives turned off, setting $\Omega_{1} = \Omega_{2} = 0$. The calibration procedure is outlined as follows:
\begin{enumerate}
    \item The swap frequency is selected by taking into account the coherence time $T_{1}$ of the qubit at the operating frequency, as well as possible frequency collisions with neighboring qubits.
    \item With the total gate duration fixed at 44\,ns and the corresponding pulse waveform shown in Fig.~\ref{ashn_wave_supp}, we scan the 2D parameter space of the coupling strength $g$ and the detuning required to achieve resonance between the two qubits, as shown in Fig.~\ref{iswap}(a).
    \item We calibrate the pulse distortion of the qubits and couplers at various detuning frequencies and coupling strengths and further refine the optimal values of the coupling strength $g$ and the qubit detuning, as shown in Fig.~\ref{iswap}(b,c).
\end{enumerate}

\begin{figure}[htb]
    \centering
    \includegraphics[width=0.48\textwidth]{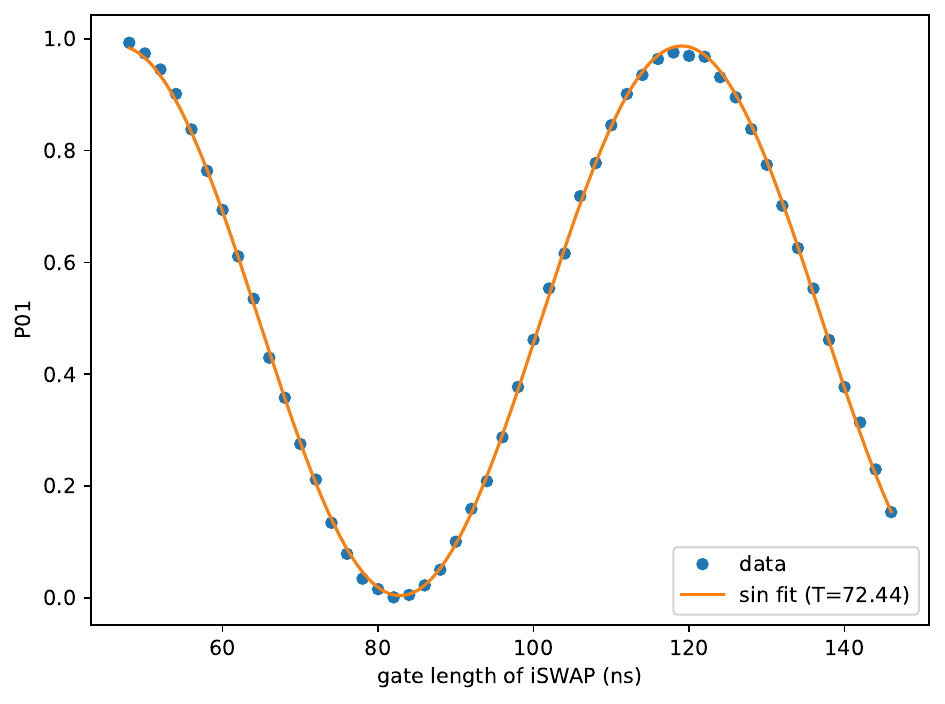}
    \caption{
        Measurement of the $P_{01}$ population as a function of the $\mathrm{iSWAP}$ gate duration. The blue dots indicate the experimental data, while the orange curve represents the sinusoidal fit.
    }
    \label{FigureS6}
\end{figure}

\subsubsection{Fixing the $\mathrm{QFT}$ Gate Duration $\tau$}

When the coupling strength $g$ is fixed, the theoretical duration of the $\mathrm{QFT}$ gate is $1.25$ times that of the $\mathrm{iSWAP}$ gate. To experimentally determine the appropriate gate duration for a given coupling strength $g$, we first turn off the XY drive and then scan the gate duration around the calibrated $\mathrm{iSWAP}$ pulse parameters. During this process, we measure the population exchange between the $\ket{01}$ and $\ket{10}$ states in the computational ($Z$) basis, which exhibits a sinusoidal oscillation as a function of the gate duration, as shown in Fig.~\ref{FigureS6}.

Ideally, the oscillation period $T$ corresponds to the period of an $\mathrm{iSWAP}$-like interaction. Therefore, compared with the calibrated $\mathrm{iSWAP}$ gate duration $\tau_{\mathrm{iSWAP}}$, the $\mathrm{QFT}$ gate duration should be extended by approximately $T/4$ under the same coupling strength. In practice, we choose the nearest even integer to
\[
    \tau_{\mathrm{iSWAP}} + \frac{T}{4}
\]
as the final duration of the $\mathrm{QFT}$ gate. This ensures a consistent total gate length for all $\mathrm{QFT}$ gates within one group by compensating for the idle time. Moreover, once the coupling strength is accurately extracted from Fig.~\ref{FigureS6}, approximate values of the remaining gate parameters, such as $\Delta$, can be obtained directly from the theoretical model, providing reliable initial values for subsequent experimental calibration~\cite{chen2024one}.

\subsubsection{Determining the Relative Phase of the XY Drive}

Only the relative phase
$\phi=\phi_1-\phi_2$
between the two microwave drives enters the AshN Hamiltonian.
Throughout this work, we choose the target relative phase
$\phi=0$,
such that the ideal operating condition is
$\phi=2n\pi$, where $n\in\mathbb Z$.
In experiments, however, a nonzero phase offset may arise from cable delays and other imperfections in the control electronics.

\begin{figure*}[htbp]
    \centering
    \includegraphics[width=0.9\textwidth]{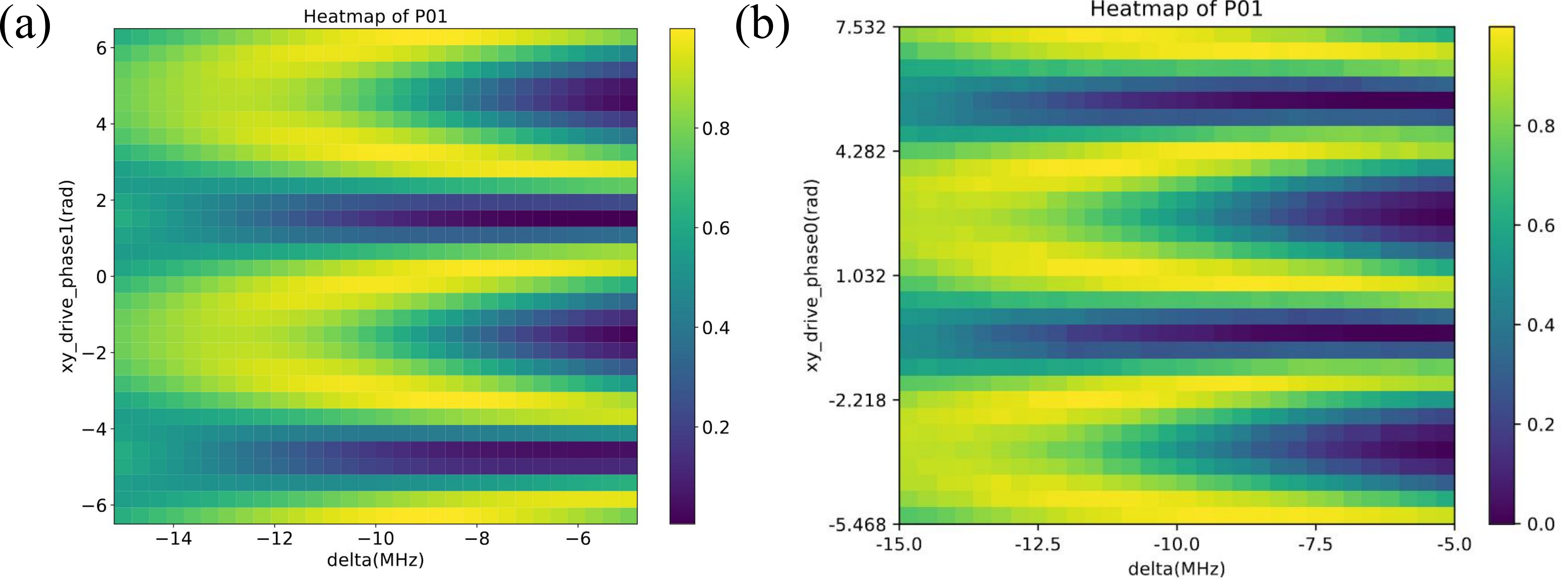}
    \caption{
        Color map of the $P_{01}$ population as a function of the relative phase between the two XY drives and the drive-frequency detuning from the qubit frequency. (a) Numerical simulation results; the ideal XY-drive phase is expected to be $0$. (b) Experimental results.
    }
    \label{Phase}
\end{figure*}

Since both $\Omega_1$ and $\Omega_2$ are nonzero for the $\mathrm{SWAP}$ gate, the $\mathrm{SWAP}$ gate and the $\mathrm{QFT}$ gate share the same XY-drive phase configuration, and the edge time of the XY-drive signal is $4~\mathrm{ns}$. Therefore, the phase calibration for the $\mathrm{QFT}$ gate can be performed via the $\mathrm{SWAP}$ interaction. Specifically, we exploit the population exchange between the $\ket{01}$ and $\ket{10}$ states in the computational ($Z$) basis. Using the estimated $\mathrm{SWAP}$ gate parameters, we simultaneously sweep the XY-drive detuning $\Delta$ and the relative phase $\phi$, while measuring the corresponding $\ket{01}$ population in the $Z$ basis. This procedure generates a 2D response surface, as shown in Fig.~\ref{Phase}.

Figure~\ref{Phase}(a) presents QuTiP numerical simulations based on the Hamiltonian in Eq.~\ref{Hamiltonian}~\cite{johansson2012qutip}. In the simulation, the remaining parameters, including $\Omega_{1}$ and $\Omega_{2}$, are fixed at their ideal theoretical values for $g = 6.25~\mathrm{MHz}$, while the relative phase $\phi$ between the two XY drives and the drive-frequency detuning $\Delta$ are varied. The ideal value of $\phi$ is $0$. When $\phi = \pi/2 + 2n\pi$, where $n \in \mathbb{Z}$, the XY drives applied to the two qubits interfere destructively, resulting in the absence of a $\mathrm{SWAP}$ interaction; consequently, the quantum state remains in its initial state.

As shown in Fig.~\ref{Phase}, the response exhibits a $2\pi$ periodicity with respect to $\phi$, evident in both the numerical simulations and the experimental results. Within a single period, we identify the phase corresponding to the minimum $P_{01}$ and subtract $\pi/2$ from it to obtain the calibrated phase,
\[
    \phi_{\mathrm{final}} = \phi_{\min(P_{01})} - \frac{\pi}{2}.
\]

\subsubsection{Optimizing the Coupling Strength $g$}

In the previous calibration steps, the $\mathrm{QFT}$ gate duration $\tau$ has already been determined. However, owing to the finite sampling rate of the Arbitrary Waveform Generator (AWG), $\tau$ is discretized and constrained to an even integer, such that its experimental value cannot always exactly match the theoretical optimum. To compensate for this discretization error, we further fine-tune the coupling strength $g$, which remains continuously adjustable in the experiment.

\begin{figure}[htb]
    \centering
    \includegraphics[width=0.48\textwidth]{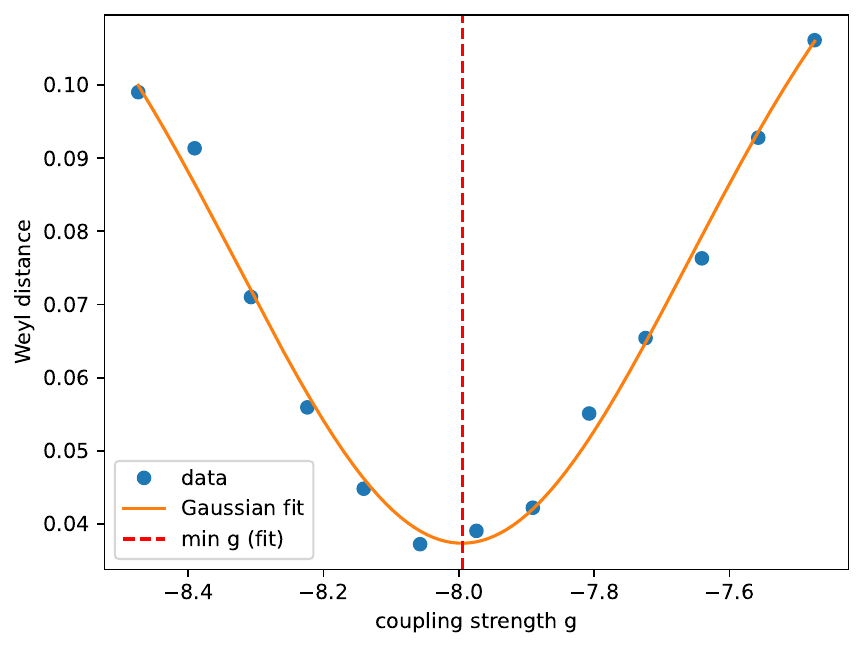}
    \caption{
        Weyl distance to the target point $\left(3\pi/16, 3\pi/16, 0\right)$ as a function of the coupling strength $g$. The blue points represent the experimental data, the orange curve shows the Gaussian fit, and the red dashed line marks the minimum of the fitted curve.
    }
    \label{FigureS7}
\end{figure}

Under the theoretical $\mathrm{QFT}$ gate duration and coupling strength, the system reduces to an $\mathrm{iSWAP}$-like interaction when the XY drives are turned off. In this case, the corresponding Weyl chamber coordinates are ideally given by
\[
    \left(\frac{3\pi}{16}, \frac{3\pi}{16}, 0\right).
\]
Therefore, with the XY drives disabled, we sweep the coupling strength $g$ and perform quantum process tomography to extract the Weyl chamber coordinates of the resulting two-qubit gate for each value of $g$. We then calculate the Weyl distance, defined as the Euclidean distance between the measured coordinates and the theoretical target point $\left(3\pi/16, 3\pi/16, 0\right)$, thereby obtaining a distance-versus-$g$ curve, as shown in Fig.~\ref{FigureS7}. The Weyl distance $D$ is defined as~\cite{chen2025efficient}
\begin{equation}
    D=\sqrt{\left(a-a_0\right)^2+\left(b-b_0\right)^2+\left(|c|-|c_0|\right)^2}.
\end{equation}
Here, $(a,b,c)$ denotes the Weyl chamber coordinates extracted from the experimentally measured gate, while $(a_0,b_0,c_0)$ denotes the coordinates of the target point.

Finally, the curve is fitted with a Gaussian function, and the value of $g$ corresponding to the minimum Weyl distance is selected as the optimized coupling strength. More generally, any $\mathrm{iSWAP}$-like gate can be calibrated using the same procedure, with only the target Weyl coordinates modified accordingly.

\subsubsection{Fixing the Ratio Between $\Omega_1$ and $\Omega_2$}

Unlike the $\mathrm{iSWAP}$ or $\mathrm{SWAP}$ gates, the physical implementation of the theoretical $\mathrm{QFT}$ gate does not exhibit a clear $\ket{01}$--$\ket{10}$ population-exchange behavior. However, in the vicinity of the theoretical values of $\Omega_1$ and $\Omega_2$, both the maximum-swap-probability points and the theoretical $\mathrm{QFT}$ gate share a common feature: they satisfy the condition $\Omega_1 = \Omega_2$.

\begin{figure*}[htbp]
    \centering
    \includegraphics[width=0.9\textwidth]{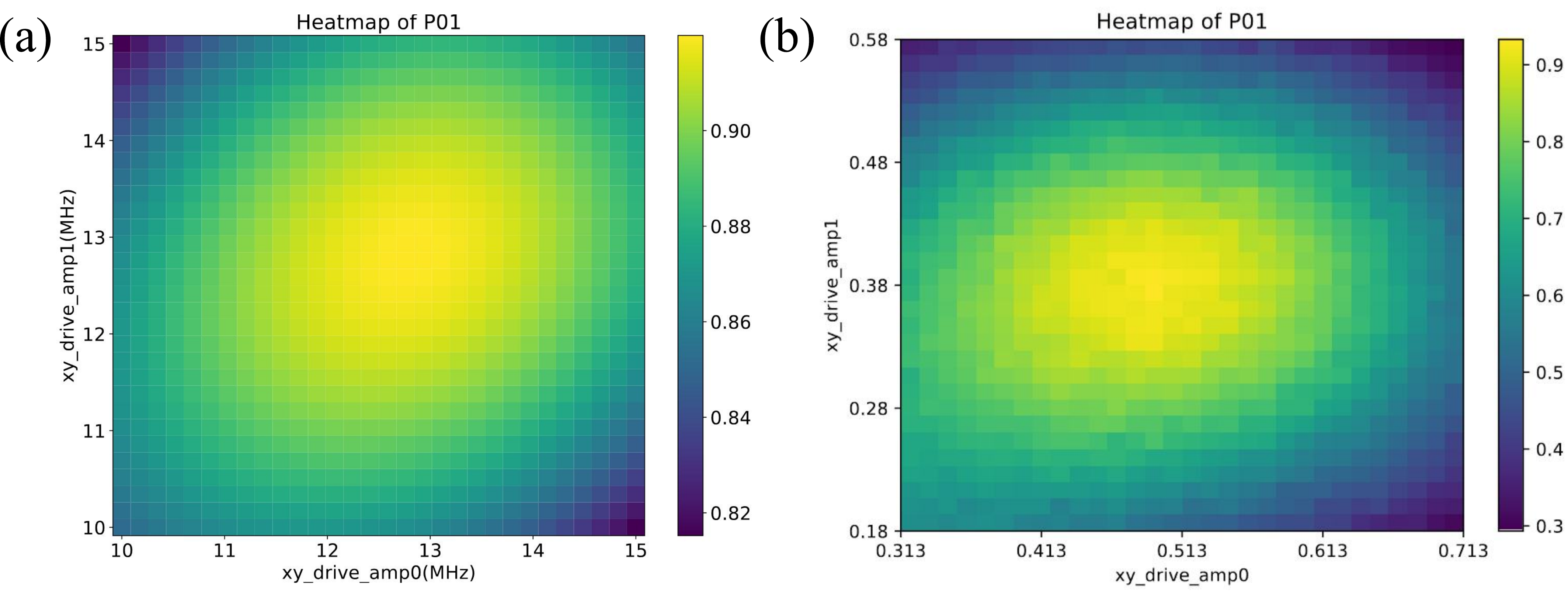}
    \caption{
        Color map of the $P_{01}$ population as a function of the amplitudes of the two XY drives, $\Omega_{1}$ and $\Omega_{2}$. (a) Numerical simulation showing that the optimal operating point corresponds to the maximum $P_{01}$ population. (b) Experimental results. The XY-drive amplitudes are given in normalized units of the control electronics output.
    }
    \label{FigureS3}
\end{figure*}

Based on this observation, we perform a 2D sweep over $\Omega_1$ and $\Omega_2$ and measure the corresponding swap probability in the computational basis, as shown in Fig.~\ref{FigureS3}(b). Although the ideal gate requires identical drive amplitudes on the two qubits, differences in the gain of the individual XY control lines result in unequal microwave amplitudes at the chip. We therefore determine the optimal operating point experimentally and fix the ratio between $\Omega_1$ and $\Omega_2$ for subsequent calibration procedures.

\subsubsection{Constraining the Relationship Between $\Omega_1$, $\Omega_2$, and $\Delta$}

Under the theoretical parameter setting of the $\mathrm{QFT}$ gate, the operating point corresponding to the maximum $\ket{01}$--$\ket{10}$ swap probability remains close to the $\mathrm{QFT}$ gate as the detuning $\Delta$ is varied. We observe that, near the $\mathrm{QFT}$ operating point, different choices of $\Omega_1$ correspond to different optimal values of $\Delta$ that maximize the swap probability. The ratio between $\Omega_1$ and $\Omega_2$ was fixed in the previous calibration step; therefore, varying $\Omega_1$ uniquely determines $\Omega_2$.

\begin{figure*}[htbp]
    \centering
    \includegraphics[width=0.9\textwidth]{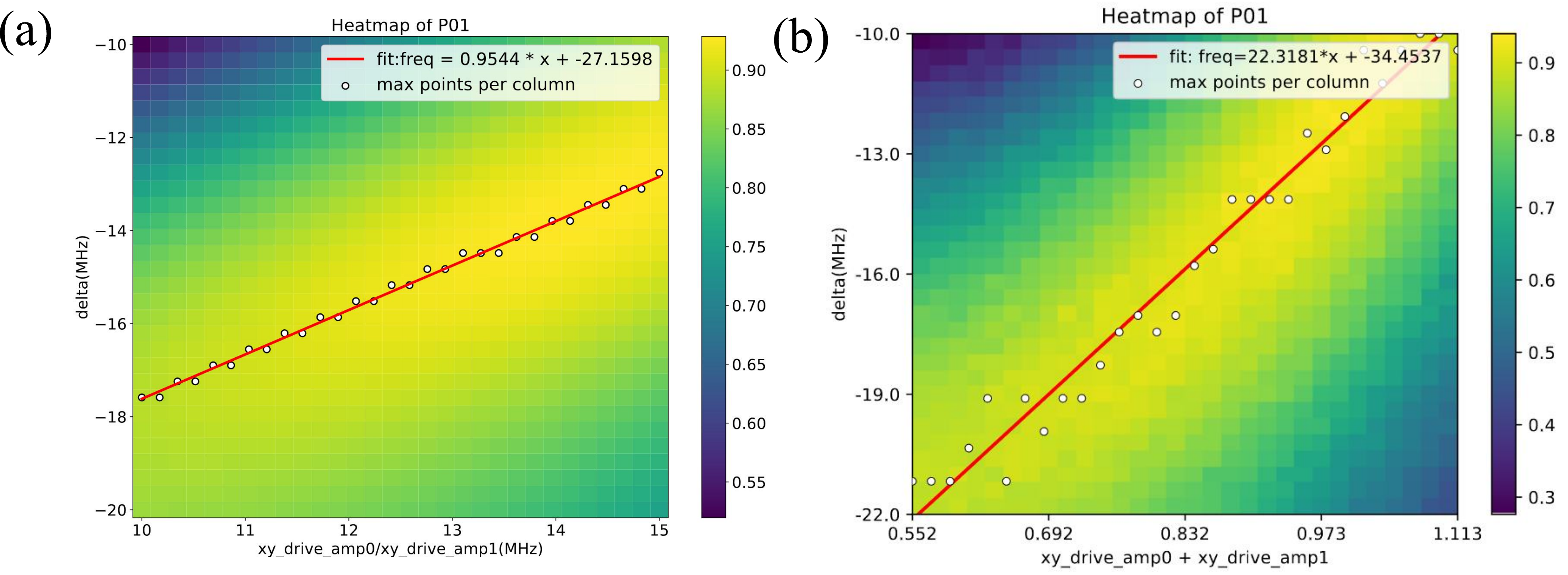}
    \caption{
        Color map of the $P_{01}$ population as a function of the XY-drive amplitude applied to one qubit ($\Omega_{1}$) and the drive-frequency detuning $\Delta$. (a) Numerical simulation results obtained under the condition $\omega_{1} = \omega_{2}$. The red line is obtained from a linear fit to the points corresponding to the maximum $P_{01}$ value in each column. (b) Experimental results, with the ratio between $\Omega_1$ and $\Omega_2$ kept fixed.
    }
    \label{FigureS4}
\end{figure*}

By extracting the optimal detuning for different values of $\Omega_1$, we find that the relationship between $\Delta$ and $\Omega_1$ is approximately linear in the vicinity of the $\mathrm{QFT}$ operating point, as shown in Fig.~\ref{FigureS4}(a). Based on this observation, we perform a 2D sweep over $\Omega_1$ and $\Delta$, identify the value of $\Delta$ that maximizes the swap probability for each $\Omega_1$, and fit the extracted optimal points using a linear function, as shown in Fig.~\ref{FigureS4}(b). The resulting linear relation between $\Delta$ and $\Omega_1$ is then imposed as a constraint in the subsequent calibration procedure.

\subsubsection{Determining $\Omega_1$, $\Omega_2$, and $\Delta$}

In the previous two calibration steps, the ratio $\Omega_1 / \Omega_2$ and the linear relationship between $\Delta$ and $\Omega_1$ have already been determined. Therefore, during the calibration procedure, $\Omega_2$ and $\Delta$ can be adjusted synchronously once a value of $\Omega_1$ is chosen.

\begin{figure*}[htbp]
    \centering
    \includegraphics[width=0.9\textwidth]{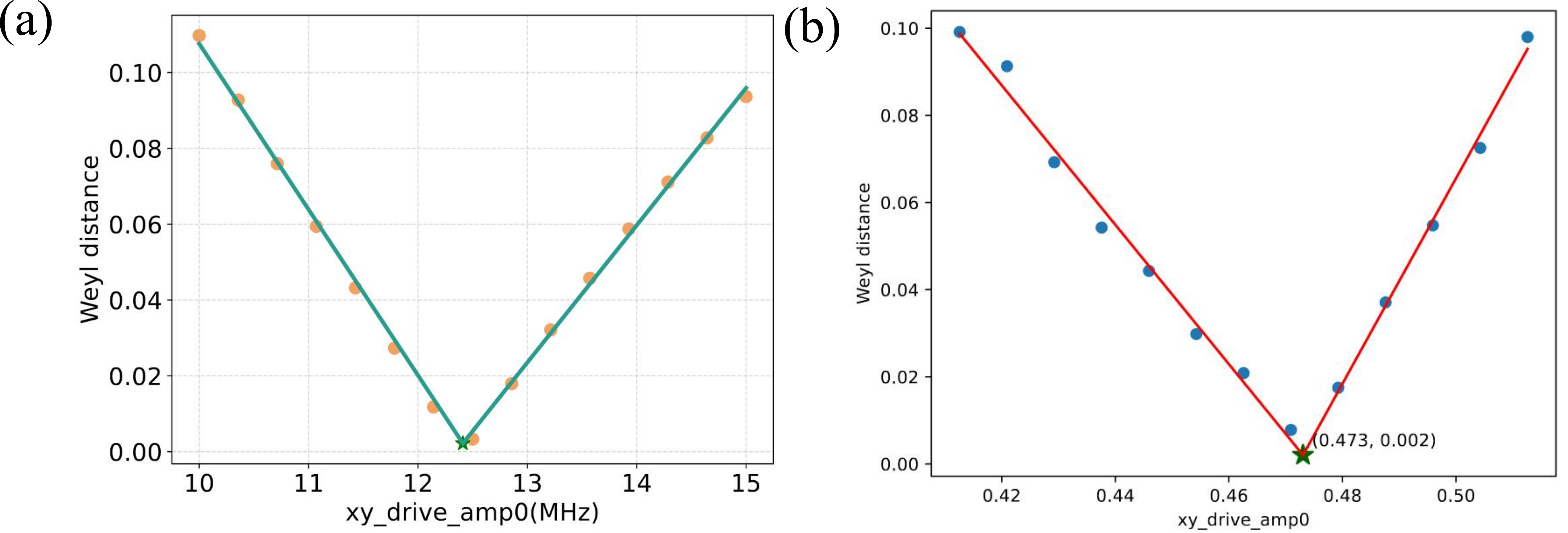}
    \caption{
        Weyl distance to the target point of the $\mathrm{QFT}$ gate, $\left(\pi/4, \pi/4, \pi/8\right)$, as a function of the XY-drive amplitude applied to one qubit, $\Omega_{1}$. (a) Numerical simulation results. The solid lines are obtained from linear fits to the data points, and the minimum point is determined by the intersection of the two fitted lines, located at $(12.41, 0.0022)$. (b) Experimental results. The solid lines are likewise obtained from linear fits to the data points, and the minimum point is determined by the intersection of the two fitted lines.
    }
    \label{FigureS5}
\end{figure*}

Based on these constraints, we directly sweep $\Omega_1$ while simultaneously updating the corresponding values of $\Omega_2$ and $\Delta$. For each value of $\Omega_1$, quantum process tomography is performed to extract the Weyl chamber coordinates of the resulting two-qubit gate, and the corresponding Weyl distance from the target $\mathrm{QFT}$ gate is calculated. This procedure produces a Weyl-distance-versus-$\Omega_1$ curve, as shown in Fig.~\ref{FigureS5}.

Finally, the curve is fitted to identify the value of $\Omega_1$ that minimizes the Weyl distance, thereby simultaneously determining the optimized values of $\Omega_2$ and $\Delta$. Figure~\ref{FigureS5}(a) presents the fitted simulation results, yielding an optimized value of $\Omega_{1}=12.41~\mathrm{MHz}$, which is very close to the theoretical value of $\Omega_{1}=12.42~\mathrm{MHz}$. This agreement demonstrates the reliability of the calibration procedure.

\subsubsection{Measuring the XEB Fidelity}

At this stage, all calibration and optimization procedures for the $\mathrm{QFT}$ gate parameters have been completed. To characterize the performance of the calibrated gate, we first perform quantum process tomography to reconstruct the actual physical operation of the gate, and subsequently carry out cross-entropy benchmarking (XEB) measurements to evaluate its XEB fidelity.

In addition to the fidelity estimation, the reconstructed process matrix also provides access to the actual unitary information of the physical gate. This information is subsequently used in experiments to calculate and apply single-qubit-gate compensation corrections.

\subsection{Calibration of Other AshN Gates}

\subsubsection{Calibration of $\sqrt{\mathrm{iSWAP}}$ Gates}

The $\sqrt{\mathrm{iSWAP}}$ gate belongs to the family of $\mathrm{iSWAP}$-like gates. Since $\mathrm{iSWAP}$-like gates do not require XY drives, their calibration procedure is relatively straightforward. In fact, the calibration procedures for all $\mathrm{iSWAP}$-like gates are essentially identical; here, we take the $\sqrt{\mathrm{iSWAP}}$ gate, which represents a relatively special case, as an example.

As introduced previously, the calibration of the $\mathrm{iSWAP}$ gate has already been completed. Based on the calibrated $\mathrm{iSWAP}$ gate parameters, we first reduce the pulse duration to one half of the original $\mathrm{iSWAP}$ gate duration. Then, with the gate duration fixed, we sweep the coupling strength $g$ and perform quantum process tomography to extract the Weyl distance for different values of $g$. By fitting the resulting Weyl-distance-versus-$g$ curve, we determine the optimized coupling strength corresponding to the minimum Weyl distance.

Finally, XEB is performed to evaluate the fidelity of the calibrated $\sqrt{\mathrm{iSWAP}}$ gate. In general, the gates located along the red line in the Weyl chamber shown in Fig.~\ref{welchamber_supp} can be calibrated using a similar procedure.

\subsubsection{Calibration of $\mathrm{CV}$ Gates}

The $\mathrm{CV}$ gate is locally equivalent to a controlled-$R_Z(\pi/2)$ gate and can therefore be regarded as a member of the controlled-$R_Z$ gate family. Under a fixed coupling strength $g$, the theoretical duration of the $\mathrm{CV}$ gate is approximately $1.5$ times that of the $\mathrm{iSWAP}$ gate. Consequently, the initial part of the calibration procedure is similar to that of the $\mathrm{QFT}$ gate.

We first initialize the coupling strength $g$ using the calibrated $\mathrm{iSWAP}$ gate parameters. By scanning the gate duration around the calibrated $\mathrm{iSWAP}$ pulse length, we measure the $\ket{01}$--$\ket{10}$ swap probability and extract the oscillation period, which is then used to determine the $\mathrm{CV}$ gate duration, as shown in Fig.~\ref{FigureS6}. With the gate duration fixed, the coupling strength $g$ is swept under the $\sqrt{\mathrm{iSWAP}}$ configuration, and quantum process tomography is performed to evaluate the corresponding Weyl distance. The optimal value of $g$ is obtained by minimizing the Weyl distance, as shown in Fig.~\ref{FigureS7}.

After calibrating $g$, we optimize $\Omega_1$ by performing a 1D scan and selecting the value that minimizes the Weyl distance. The phase $\phi$ of the XY drive is directly inherited from the calibrated $\mathrm{QFT}$ or $\mathrm{SWAP}$ gate. Finally, with $\phi$ fixed, a small-range scan of $\Omega_2$ around zero is carried out to determine its optimal value. Although the ideal $\mathrm{CV}$ gate corresponds to $\Omega_2=0$, a small nonzero drive is typically required in experiments to compensate for residual crosstalk and other systematic imperfections.

After completing all calibration procedures, the XEB fidelity of the $\mathrm{CV}$ gate is characterized experimentally. In general, the gates located along the green and orange lines in the Weyl chamber shown in Fig.~\ref{welchamber_supp} can be calibrated using a similar procedure.

\subsubsection{Calibration of $\mathrm{SWAP}$ Gates}

Under a fixed coupling strength $g$, the theoretical duration of the $\mathrm{SWAP}$ gate is likewise approximately $1.5$ times that of the $\mathrm{iSWAP}$ gate. Therefore, the initial part of the calibration procedure is identical to that of the $\mathrm{CV}$ gate.

After determining the gate duration $\tau$ and the coupling strength $g$ of the $\mathrm{SWAP}$ gate, we first estimate suitable values of $\Omega_1$ and $\Omega_2$. We then perform a 2D sweep over the XY-drive detuning $\Delta$ and the relative phase $\phi$ to determine the optimal phase, following the same procedure as in the $\mathrm{QFT}$ gate calibration.

Subsequently, the remaining calibration steps are carried out in a manner similar to the $\mathrm{QFT}$ calibration procedure by exploiting the strong $\ket{01}$--$\ket{10}$ swap dynamics of the $\mathrm{SWAP}$ gate. First, a 2D sweep over $\Omega_1$ and $\Omega_2$ is performed to determine the ratio $\Omega_1/\Omega_2$. Next, a 2D sweep over $\Omega_1$ and $\Delta$ is carried out to establish the approximately linear relationship between these two parameters.

Finally, quantum process tomography is performed while sweeping $\Omega_1$, with $\Omega_2$ and $\Delta$ synchronously updated according to the previously established constraints. The Weyl distance is calculated for each parameter set, and the optimal values of $\Omega_1$, $\Omega_2$, and $\Delta$ are obtained by minimizing the Weyl distance. After completing all calibration procedures, the XEB fidelity of the $\mathrm{SWAP}$ gate is experimentally characterized.

\section{Circuit-Level Optimization via $R_Z$ Propagation}
\label{app:rz_commutation}

A key advantage of the $\mathrm{CZ}$-based compilation framework is that the $\mathrm{CZ}$ gate commutes with single-qubit $R_Z$ rotations. As a consequence, $R_Z$ operations can be implemented virtually through phase updates (\emph{virtual-$Z$} gates), eliminating the need for additional physical pulses and substantially reducing the overall single-qubit gate count~\cite{McKay2017EfficientZGates}.

In this section, we show that an analogous reduction mechanism extends beyond the $\mathrm{CZ}$ gate to the $\mathrm{AshN}$ family of two-qubit interactions. Although a generic physical $\mathrm{AshN}$ implementation does not commute with $R_Z$ rotations, it can admit a locally equivalent representation that either (i) commutes with $R_Z$, or (ii) allows $R_Z$ rotations to propagate through the two-qubit gate and, in some cases, transfer between qubits, as for $\mathrm{iSWAP}$ gate.

\subsection*{Weyl-Chamber Condition}

Consider a two-qubit unitary characterized by Weyl chamber coordinates $(a,b,c)$.
If the coordinates satisfy either
\begin{equation}
    a = b = \frac{\pi}{4},
    \qquad \text{or} \qquad
    b = c = 0,
\end{equation}
then the unitary admits a locally equivalent form such that $R_Z$ rotations can be propagated through the two-qubit interaction up to a relabeling of qubits. We refer to this property as \emph{$R_Z$ propagation under local equivalence}.

\subsection*{Reduction by Local Equivalence}

Let $U_{\mathrm{AshN,phy}}$ denote the physical implementation of an $\mathrm{AshN}$ gate whose Weyl chamber coordinates $(a,b,c)$ satisfy either
\begin{align}
    a=b=\frac{\pi}{4},
    \qquad \text{or} \qquad
    b=c=0.
\end{align}

By the definition of local equivalence, there exist single-qubit unitaries
\begin{align}
    A_1, A_2, B_1, B_2 \in \mathrm{SU}(2)
\end{align}
such that
\begin{equation}
    U_{R_Z}
    =
    (B_1 \otimes B_2)\,
    U_{\mathrm{AshN,phy}}\,
    (A_1 \otimes A_2),
\end{equation}
where $U_{R_Z}$ satisfies
\begin{equation}
    U_{R_Z}
    (R_Z(x_1)\otimes R_Z(x_2))
    =
    (R_Z(-x_3)\otimes R_Z(-x_4))
    U_{R_Z}.
\end{equation}
Here $(-x_3,-x_4)$ may differ from $(x_1,x_2)$, reflecting the possibility that phase information can transfer between qubits during propagation.

\begin{figure*}[htbp]
    \centering
    \includegraphics[width=0.75\textwidth]{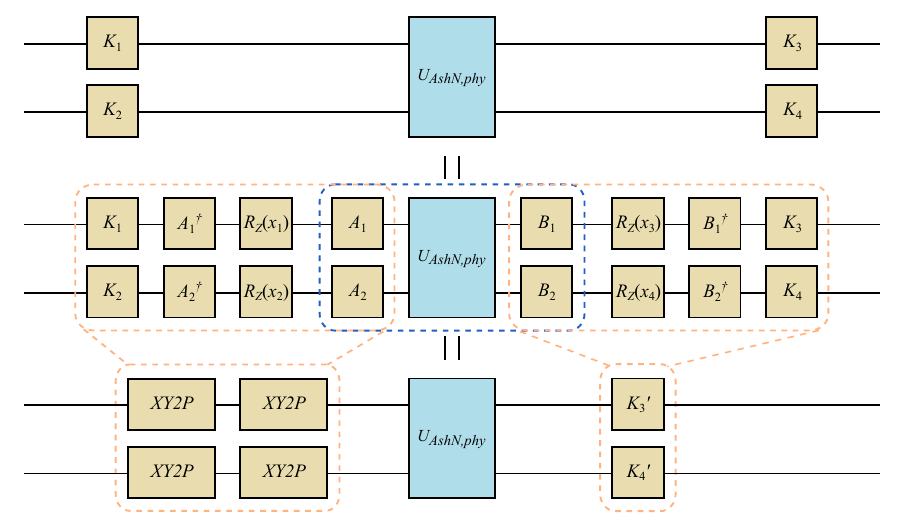}
    \caption{
        The circuit transformation used to reduce the number of single-qubit gates. Using the local-equivalence relations of the entangling gate, $R_Z$ rotations are commuted across the two-qubit interaction and absorbed into adjacent single-qubit gates. After propagating these phase rotations, redundant single-qubit operations are merged or eliminated, resulting in a functionally equivalent circuit with a lower single-qubit gate count.
    }
    \label{virtual-Z}
\end{figure*}

Now consider a compiled circuit segment of the form
\begin{equation}
    U
    =
    (K_3 \otimes K_4)\,
    U_{\mathrm{AshN,phy}}\,
    (K_1 \otimes K_2),
\end{equation}
where $K_i \in \mathrm{SU}(2)$ are arbitrary single-qubit gates.

Inserting identities induced by local equivalence gives
\begin{align}
    U
     & =
    (K_3 \otimes K_4)
    (B_1^{-1}\otimes B_2^{-1})
    \, U_{R_Z} \,
    (A_1^{-1}\otimes A_2^{-1})
    (K_1 \otimes K_2).
\end{align}

Since $U_{R_Z}$ propagates $R_Z$ rotations through the interaction, we may insert compensating rotations:
\begin{align}
    U
     & =
    (K_3 \otimes K_4)
    (B_1^{-1}\otimes B_2^{-1})
    (R_Z(x_3)\otimes R_Z(x_4))
    \nonumber      \\
     & \quad \cdot
    U_{R_Z}
    \cdot
    (R_Z(x_1)\otimes R_Z(x_2))
    (A_1^{-1}\otimes A_2^{-1})
    (K_1 \otimes K_2).
\end{align}
The corresponding circuit transformation is illustrated in Supplementary Fig.~\ref{virtual-Z}, which shows how the inserted $R_Z$ rotations are propagated through the entangling gate and subsequently absorbed into neighboring single-qubit operations.

\subsection*{Absorption into Single-Qubit Layers}

Define
\begin{equation}
    K_1' \otimes K_2'
    :=
    (A_1 \otimes A_2)
    (R_Z(x_1)\otimes R_Z(x_2))
    (A_1^{-1}\otimes A_2^{-1})
    (K_1 \otimes K_2).
\end{equation}

Because $\mathrm{SU}(2)$ is closed under composition, each $K_i'$ remains a valid single-qubit unitary and therefore admits a decomposition in the native gate basis.
By appropriately fitting the parameters $x_1$ and $x_2$, the resulting single-qubit unitaries can be constrained to satisfy
\begin{equation}
    \theta_i = \phi_i + \lambda_i,
    \qquad i=1,2,
\end{equation}
such that each unitary can be decomposed into a two-pulse hardware-efficient form:
\begin{equation}
    K_i'
    =
    U_3(\theta_i,\phi_i,\lambda_i)
    =
    XY2P(\phi_i)\,
    XY2P(-\lambda_i-\pi),
    \qquad i=1,2.
\end{equation}

The circuit can therefore be rewritten in the reduced form
\begin{align}
    U
     & =
    (K_3' \otimes K_4')
    \, U_{\mathrm{AshN,phy}}\,
    (K_1' \otimes K_2')
    \nonumber \\
     & =
    (K_3' \otimes K_4')
    \, U_{\mathrm{AshN,phy}}\,
    \Bigl(
    XY2P(\phi_1)\,XY2P(-\lambda_1-\pi) \otimes
    XY2P(\phi_2)\,XY2P(-\lambda_2-\pi)
    \Bigr).
\end{align}
Here,
\[
    XY2P(\theta)
    =
    \frac{1}{\sqrt{2}}
    \begin{pmatrix}
        1              & -i e^{-i\theta} \\
        -i e^{i\theta} & 1
    \end{pmatrix},
\]
denotes the native single-qubit gate implemented on our superconducting quantum processor. It represents a $\pi/2$ rotation about an axis lying in the equatorial ($XY$) plane of the Bloch sphere, where the parameter $\theta$ specifies the azimuthal angle of the rotation axis.

Here, the decomposition into the two-pulse form is applied only to $K_1'$ and $K_2'$, since the parameters $x_3$ and $x_4$ in the inserted $R_Z$ rotations are already determined once $x_1$ and $x_2$ are fixed, leaving no additional freedom for further fitting.
However, if the subsequent two-qubit gates associated with $K_3'$ and $K_4'$ also satisfy the above Weyl-chamber condition, then $K_3'$ and $K_4'$ can be further optimized during the reduction procedure of those subsequent two-qubit gates.

\subsection*{Implications for Circuit Complexity}

This construction demonstrates that, for any two-qubit gate satisfying the above Weyl-chamber condition, the number of single-qubit gates immediately preceding the two-qubit gate on each qubit can be reduced to at most two. In particular, the adjacent single-qubit layers before such two-qubit gates can always be compressed into a fixed two-pulse representation per qubit.

For a circuit containing $n_1$ two-qubit gates satisfying the above Weyl-chamber condition and $n_2$ generic two-qubit gates that do not satisfy the condition, acting on $m$ qubits, the total number of single-qubit gates satisfies
\begin{equation}
    N_{\mathrm{1q}}
    \le
    4n_1 + 8n_2 + 2m,
\end{equation}
where the term $2m$ accounts for the final single-qubit layers at the end of the circuit.

\subsection*{Practical Relevance}

Many experimentally relevant two-qubit gates, including $\mathrm{CNOT}$, $\mathrm{iSWAP}$, $\mathrm{SWAP}$, and $\mathrm{CV}$, satisfy the above Weyl-chamber condition. Consequently, the proposed reduction mechanism applies broadly to realistic compilation and control settings.

Compared with the $\mathrm{CZ}$ compilation framework, where exact commutation with $R_Z$ rotations is available, the $\mathrm{AshN}$ scheme achieves an analogous effect through local equivalence and parameter absorption. Combined with its capability to reduce the number of two-qubit gates, this approach can provide favorable trade-offs in both circuit depth and control complexity.



\section{Hardware-Aware Compilation on Restricted Topologies}

This section extends the circuit-level results in the main text through three successive benchmark stages, each addressing a different question. First, we use seven circuits small enough to execute on the present processor to test whether the compilation advantage survives physical implementation. Second, we broaden the compilation study to 18 application categories and larger circuits that are mostly beyond the reach of a controlled gate-set comparison on current hardware, testing whether the advantage generalizes across workloads. Third, we scale four representative algorithm families---QAOA, AQFT, Dicke-state preparation, and Grover search---over multiple problem sizes, testing how the routing advantage evolves with circuit scale rather than only across heterogeneous examples. We then complement this progression with quantum-volume random circuits as a deliberately unfavorable regime and with an equivalent-$\mathrm{CZ}$-cost analysis that relaxes the assumption of a fully calibrated AshN gate set. Together, these studies separate experimental validation, workload breadth, scalability, limitations, and native-gate-set requirements.

\subsection{$\mathrm{SWAP}$-aware routing heuristic}

All mapped AshN circuits in the following benchmarks are routed with MirrorSABRE, a gate-set-aware extension of SABRE~\cite{li2019tackling}. Standard SABRE selects $\mathrm{SWAP}$s using the physical distances of gates in the current front layer and a bounded look-ahead set. MirrorSABRE retains this criterion but rewards a candidate $\mathrm{SWAP}$ when it can be absorbed into the preceding two-qubit block on the same physical-qubit pair, as allowed by the closure of the AshN gate set under composition with $\mathrm{SWAP}$.

For a candidate $\mathrm{SWAP}$ $s=(q_i,q_j)$ and the logical-to-physical mapping $\pi_s$ that results from applying it, we use the score
\begin{align}
    H(s) = \max\{\delta(q_i),\delta(q_j)\}
    \left[
    \frac{1}{|F|}\sum_{g\in F}D(\pi_s,g)
    +W\frac{1}{|E|}\sum_{g\in E}D(\pi_s,g)
    \right] -\lambda_{\mathrm{absorb}}I_{\mathrm{absorb}}(s).
    \label{eq:mirrorsabre-score}
\end{align}
Here, $F$ is the front layer of unscheduled gates, $E$ is the bounded look-ahead set, and $D(\pi_s,g)$ is the shortest-path distance between the physical operands of gate $g$ under $\pi_s$. The factor $\delta(q)$ is the standard SABRE decay factor, which discourages repeated use of the same qubits and thereby limits depth growth, and $W$ weights the look-ahead contribution. When $E$ is empty, its contribution in Eq.~\eqref{eq:mirrorsabre-score} is defined to be zero.

The final term introduces the gate-set-aware preference. The indicator $I_{\mathrm{absorb}}(s)$ equals one when the candidate $\mathrm{SWAP}$ follows a two-qubit block on the same physical pair and is considered in the first $\mathrm{SWAP}$-search round after the scheduler has advanced; it is zero otherwise. The positive weight $\lambda_{\mathrm{absorb}}$ therefore lowers the score of an absorbable candidate. Minimizing Eq.~\eqref{eq:mirrorsabre-score} balances progress towards upcoming interactions with the opportunity to update the logical mapping without adding a separate two-qubit gate or layer. After routing, adjacent operations are partitioned and rebased into canonical $\mathrm{SU}(4)$ blocks, which realizes the selected absorptions. Related gate-set-aware routing principles have been considered previously~\cite{yang2026reconfigurable,yangUnifyingQubitRouting2026}.

\subsection{Stage I: Hardware-validated circuit benchmarks}

The starting point is the hardware experiment reported in \MainTextCircuitBenchmarkFigure{} of the main text. We benchmark seven representative circuits in the experimentally accessible four- to seven-qubit regime. For both 1D and 2D restricted topologies, each circuit is compiled using the AshN- and $\mathrm{CZ}$-based pipelines and executed on the superconducting processor using four computational-basis input states. The resulting measurements compare the computational-basis success probabilities of the two implementations. This experiment establishes that, for the circuits and hardware configurations tested, the lower two-qubit gate counts obtained with AshN compilation are accompanied by higher computational-basis success probabilities. It serves as the experimental anchor for the larger compilation-only studies below.

The AshN implementations draw from a calibrated set of five native two-qubit gates---$\mathrm{CZ}$, $\mathrm{iSWAP}$, $\mathrm{SWAP}$, $\mathrm{CV}$, and $\mathrm{QFT}$---with durations of 42, 44, 66, 66, and 58 ns, respectively. This stage tests physical execution rather than scaling; larger controlled comparisons are limited by the depth of the corresponding $\mathrm{CZ}$-based implementations.

\subsection{Stage II: Breadth across application-derived circuits}

Having established the effect experimentally at small scale, we next ask whether it persists across a substantially broader range of circuit structures. We therefore benchmark a suite drawn from 18 application categories. Most circuits are selected from RevLib~\cite{wille2008revlib} and QASMBench~\cite{li2023qasmbench}, or constructed from standard Qiskit subroutines. Unlike the seven circuits in Stage I, this stage is a compilation study: it extends to larger workloads, including circuits whose $\mathrm{CZ}$-based implementations are too deep for a meaningful controlled experiment on the present device.

We evaluate each n-qubit circuit with two hardware topologies: a 1D chain of exactly $n$ qubits, and a $\lceil\sqrt{n}\rceil \times \lceil\sqrt{n}\rceil$ square grid. The spare qubits are allowed as intermediate routing locations.

Evaluations are carried out in the same setup as \MainTextCircuitBenchmarkFigure{} of the main text.
As a preprocessing step, circuits are decomposed into one- and two-qubit gates when necessary.
For the mapped AshN pipeline, the resulting circuits are routed by $\mathrm{SWAP}$-aware MirrorSABRE, after which the routed operations are partitioned and rebased into canonical \(\mathrm{SU}(4)\) blocks. The all-to-all AshN reference is compiled and rebased separately without a device constraint.
To establish a baseline, we use Qiskit's O3 optimization level with SABRE routing for $\mathrm{CZ}$-based compilation.

\begin{figure*}[htbp]
    \centering

    \subfloat[Routing on 1D chain topology.\label{fig:routing_chain}]{
        \includegraphics[width=0.47\textwidth]
        {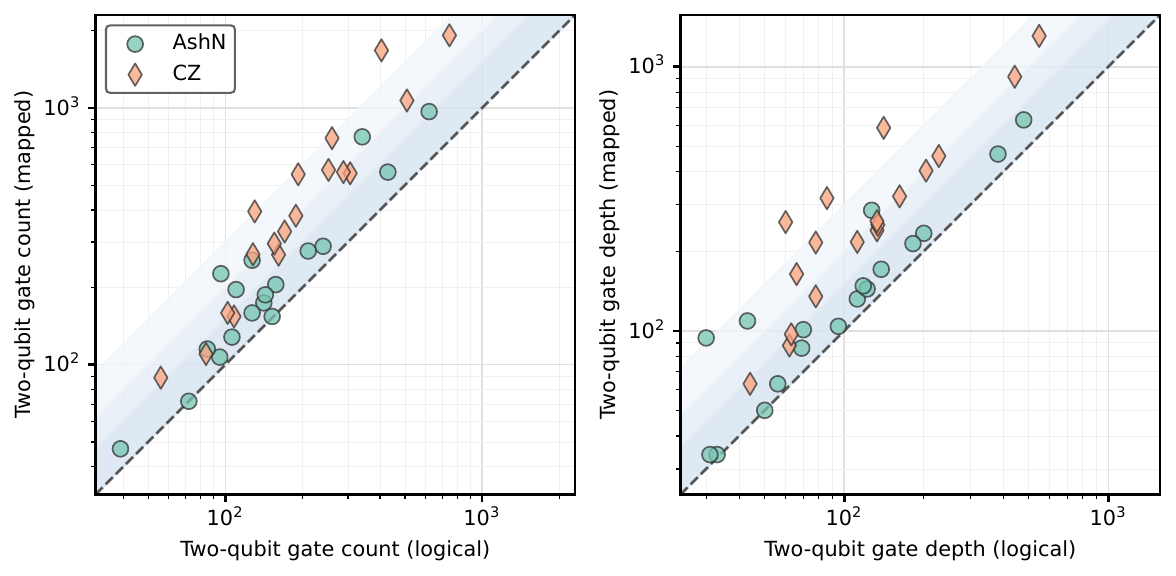}
    }
    \hfill
    \subfloat[Routing on 2D square topology.\label{fig:routing_square}]{
        \includegraphics[width=0.47\textwidth]
        {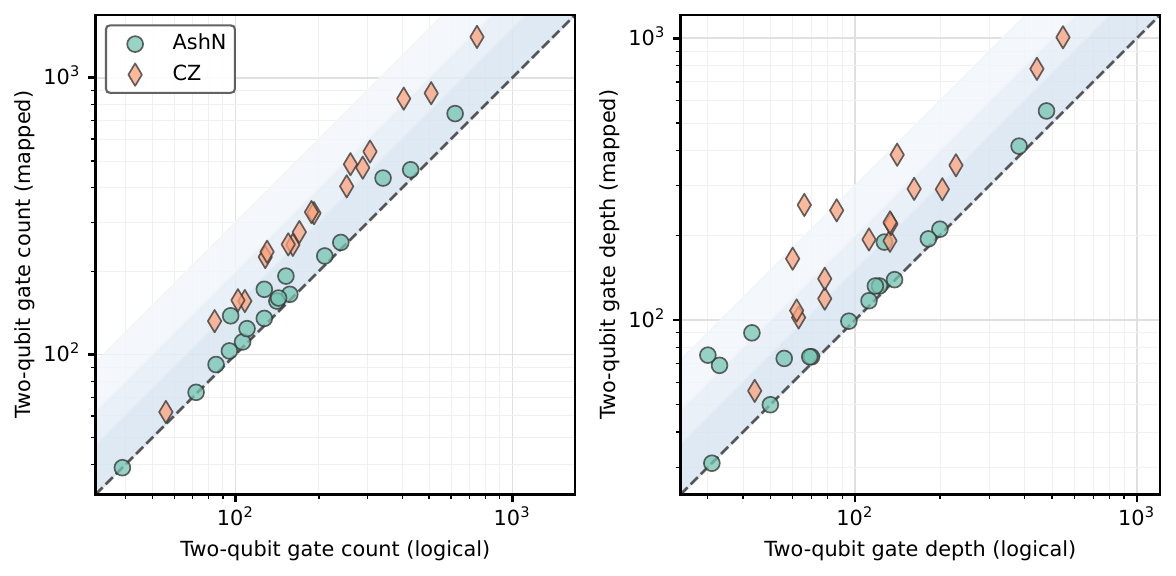}
    }

    \caption{Routing overhead for the benchmark circuits compiled into the AshN or $\mathrm{CZ}$ basis on (a) a 1D chain and (b) a 2D square lattice. The dashed diagonal line, $y=x$, denotes mapping without routing overhead. Shaded regions denote overhead factors of $1\text{--}1.5$, $1.5\text{--}2$ and $2\text{--}3$, where the overhead factor is the ratio of the mapped to logical circuit metric. Depth denotes two-qubit-gate layer depth; one-qubit operations are excluded.}
    \label{fig:routing_topologies}
\end{figure*}

The benchmark suite and compilation results are illustrated in Fig.~\ref{fig:routing_topologies} and summarized in Tables~\ref{tab:benchmark-summary-chain} and~\ref{tab:benchmark-summary-square}. Across the benchmark suite, the AshN gate set with tailored qubit routing substantially suppresses routing overhead. On the 1D chain, the geometric-mean two-qubit-gate count and depth overheads decrease from $\BenchChainCZGateGeomean$ and $\BenchChainCZDepthGeomean$ with $\mathrm{CZ}$ compilation to $\BenchChainAshNGateGeomean$ and $\BenchChainAshNDepthGeomean$ with AshN compilation, respectively. AshN therefore reduces the geometric-mean normalized gate-count and depth overheads by $\BenchChainGateReduction$ and $\BenchChainDepthReduction$, despite the heavily limited 1D-connectivity constraint of the chain. This trend is also evident from the AshN data points lying closer to the $y=x$ line in Fig.~\ref{fig:routing_topologies}.
KNN and QFT incur zero or negligible AshN routing overhead in either metric, by absorbing nearly every $\mathrm{SWAP}$ gate into aggregated $\mathrm{SU}(4)$ blocks. In contrast, QPE and QAOA retain substantial overhead, showing that the achievable reduction remains dependent on the circuit interaction pattern.

The advantage persists on the 2D square lattice, where the lower baseline routing cost leaves less room for improvement. AshN reduces the geometric-mean gate-count overhead from $\BenchSquareCZGateGeomean$ to $\BenchSquareAshNGateGeomean$ and the depth overhead from $\BenchSquareCZDepthGeomean$ to $\BenchSquareAshNDepthGeomean$, corresponding to relative reductions of $\BenchSquareGateReduction$ and $\BenchSquareDepthReduction$, respectively. Consistent with the clustering of the AshN data points near the diagonal in Fig.~\ref{fig:routing_topologies}, all benchmarks have an AshN gate-count overhead of at most $\BenchSquareAshNMaxGate$ on the square lattice. The outlier depth overheads for QFT, QPE, and QAOA demonstrate that richer native gates substantially mitigate, but do not universally eliminate, routing costs on restricted topologies; specifically, the worse result obtained with QFT on the 2D square lattice than on the 1D chain may reflect the larger search space; we examine this effect together with the QAOA scale-up results later.

\providecommand{\BenchmarkSummaryChainCaption}{\GJZ{Placeholder}}
\ProvideDocumentCommand{\BenchmarkSummaryChainTable}{g}{%
    \begin{table}[htbp]
        \centering
        \caption{\IfNoValueTF{#1}{\BenchmarkSummaryChainCaption}{#1}}
        \label{tab:benchmark-summary-chain}
        \scriptsize
        \setlength{\tabcolsep}{3pt}
        \providecommand{\BenchmarkSummaryHead}[1]{\begin{tabular}[c]{@{}c@{}}##1\end{tabular}}
        \resizebox{\textwidth}{!}{%
            \begin{tabular}{lrcrcrcrcr}
                \hline\hline
                \BenchmarkSummaryHead{Application (Circuit)}       & \BenchmarkSummaryHead{\#Qubit} &
                \BenchmarkSummaryHead{Mapped/Logical                                                                                                                      \\\#2Q ($\mathrm{CZ}$)} & \BenchmarkSummaryHead{\#2Q ratio\\($\mathrm{CZ}$)} &
                \BenchmarkSummaryHead{Mapped/Logical                                                                                                                      \\depth ($\mathrm{CZ}$)} & \BenchmarkSummaryHead{Depth ratio\\($\mathrm{CZ}$)} &
                \BenchmarkSummaryHead{Mapped/Logical                                                                                                                      \\\#2Q (AshN)} & \BenchmarkSummaryHead{\#2Q ratio\\(AshN)} &
                \BenchmarkSummaryHead{Mapped/Logical                                                                                                                      \\depth (AshN)} & \BenchmarkSummaryHead{Depth ratio\\(AshN)} \\
                \colrule
                Comparator (\texttt{4gt4-v0\_73})                  & 6                              & 296/155  & 1.91 & 252/134  & 1.88 & 159/127 & 1.25 & 132/112 & 1.18 \\
                Chemistry (\texttt{H2\_cmplt\_BK})                 & 8                              & 89/56    & 1.59 & 63/44    & 1.43 & 47/39   & 1.21 & 34/31   & 1.10 \\
                ALU (\texttt{alu-v2\_30})                          & 6                              & 380/188  & 2.02 & 323/162  & 1.99 & 205/157 & 1.31 & 171/138 & 1.24 \\
                Multiplier (\texttt{gf2\textasciicircum{}8\_mult}) & 24                             & 1674/405 & 4.13 & 587/141  & 4.16 & 771/341 & 2.26 & 286/127 & 2.25 \\
                Grover (\texttt{grover\_5})                        & 9                              & 562/288  & 1.95 & 459/228  & 2.01 & 289/240 & 1.20 & 234/200 & 1.17 \\
                Encoding (\texttt{ham7\_104})                      & 7                              & 269/128  & 2.10 & 217/112  & 1.94 & 128/106 & 1.21 & 104/95  & 1.09 \\
                HWB (\texttt{hwb5\_53})                            & 6                              & 1068/509 & 2.10 & 916/442  & 2.07 & 562/429 & 1.31 & 467/382 & 1.22 \\
                KNN (\texttt{knn\_n25})                            & 25                             & 110/84   & 1.31 & 88/62    & 1.42 & 72/72   & 1.00 & 50/50   & 1.00 \\
                QFT (\texttt{qft\_n18})                            & 18                             & 556/306  & 1.82 & 164/66   & 2.48 & 154/152 & 1.01 & 34/33   & 1.03 \\
                QPE (\texttt{qpeexact\_n16})                       & 16                             & 762/260  & 2.93 & 318/86   & 3.70 & 255/127 & 2.01 & 109/43  & 2.53 \\
                QRAM (\texttt{qram\_n20})                          & 20                             & 395/130  & 3.04 & 216/78   & 2.77 & 196/110 & 1.78 & 101/70  & 1.44 \\
                Bit Adder (\texttt{rd53\_131})                     & 7                              & 330/170  & 1.94 & 260/133  & 1.95 & 187/143 & 1.31 & 148/118 & 1.25 \\
                QAOA (\texttt{reg3\_8})                            & 8                              & 551/192  & 2.87 & 258/60   & 4.30 & 226/96  & 2.35 & 94/30   & 3.13 \\
                Ripple Adder (\texttt{ra\_10})                     & 22                             & 268/161  & 1.66 & 240/133  & 1.80 & 174/141 & 1.23 & 144/122 & 1.18 \\
                SAT (\texttt{sat\_n11})                            & 11                             & 573/252  & 2.27 & 404/204  & 1.98 & 277/210 & 1.32 & 214/182 & 1.18 \\
                Square (\texttt{squar5\_261})                      & 13                             & 1916/745 & 2.57 & 1308/547 & 2.39 & 967/621 & 1.56 & 629/478 & 1.32 \\
                SymB (\texttt{sym9\_146})                          & 12                             & 154/108  & 1.43 & 97/63    & 1.54 & 107/95  & 1.13 & 63/56   & 1.12 \\
                MCX (\texttt{tof\_10})                             & 19                             & 159/102  & 1.56 & 135/78   & 1.73 & 115/85  & 1.35 & 86/69   & 1.25 \\
                \colrule
                \multicolumn{2}{c}{Geometric mean}              & --                             & 2.09     & --   & 2.18     & --   & 1.39    & --   & 1.35           \\
                \hline\hline
            \end{tabular}%
        }
    \end{table}
}
\providecommand{\BenchmarkSummarySquareCaption}{\GJZ{Placeholder}}
\ProvideDocumentCommand{\BenchmarkSummarySquareTable}{g}{%
    \begin{table}[htbp]
        \centering
        \caption{\IfNoValueTF{#1}{\BenchmarkSummarySquareCaption}{#1}}
        \label{tab:benchmark-summary-square}
        \scriptsize
        \setlength{\tabcolsep}{3pt}
        \providecommand{\BenchmarkSummaryHead}[1]{\begin{tabular}[c]{@{}c@{}}##1\end{tabular}}
        \resizebox{\textwidth}{!}{%
            \begin{tabular}{lrcrcrcrcr}
                \hline\hline
                \BenchmarkSummaryHead{Application (Circuit)}       & \BenchmarkSummaryHead{\#Qubit} &
                \BenchmarkSummaryHead{Mapped/Logical                                                                                                                      \\\#2Q ($\mathrm{CZ}$)} & \BenchmarkSummaryHead{\#2Q ratio\\($\mathrm{CZ}$)} &
                \BenchmarkSummaryHead{Mapped/Logical                                                                                                                      \\depth ($\mathrm{CZ}$)} & \BenchmarkSummaryHead{Depth ratio\\($\mathrm{CZ}$)} &
                \BenchmarkSummaryHead{Mapped/Logical                                                                                                                      \\\#2Q (AshN)} & \BenchmarkSummaryHead{\#2Q ratio\\(AshN)} &
                \BenchmarkSummaryHead{Mapped/Logical                                                                                                                      \\depth (AshN)} & \BenchmarkSummaryHead{Depth ratio\\(AshN)} \\
                \colrule
                Comparator (\texttt{4gt4-v0\_73})                  & 6                              & 250/155  & 1.61 & 219/134  & 1.63 & 135/127 & 1.06 & 117/112 & 1.04 \\
                Chemistry (\texttt{H2\_cmplt\_BK})                 & 8                              & 62/56    & 1.11 & 56/44    & 1.27 & 39/39   & 1.00 & 31/31   & 1.00 \\
                ALU (\texttt{alu-v2\_30})                          & 6                              & 327/188  & 1.74 & 292/162  & 1.80 & 165/157 & 1.05 & 139/138 & 1.01 \\
                Multiplier (\texttt{gf2\textasciicircum{}8\_mult}) & 24                             & 840/405  & 2.07 & 386/141  & 2.74 & 434/341 & 1.27 & 189/127 & 1.49 \\
                Grover (\texttt{grover\_5})                        & 9                              & 473/288  & 1.64 & 354/228  & 1.55 & 254/240 & 1.06 & 210/200 & 1.05 \\
                Encoding (\texttt{ham7\_104})                      & 7                              & 225/128  & 1.76 & 193/112  & 1.72 & 111/106 & 1.05 & 99/95   & 1.04 \\
                HWB (\texttt{hwb5\_53})                            & 6                              & 879/509  & 1.73 & 779/442  & 1.76 & 465/429 & 1.08 & 414/382 & 1.08 \\
                KNN (\texttt{knn\_n25})                            & 25                             & 132/84   & 1.57 & 108/62   & 1.74 & 73/72   & 1.01 & 50/50   & 1.00 \\
                QFT (\texttt{qft\_n18})                            & 18                             & 541/306  & 1.77 & 256/66   & 3.88 & 192/152 & 1.26 & 69/33   & 2.09 \\
                QPE (\texttt{qpeexact\_n16})                       & 16                             & 487/260  & 1.87 & 245/86   & 2.85 & 172/127 & 1.35 & 90/43   & 2.09 \\
                QRAM (\texttt{qram\_n20})                          & 20                             & 235/130  & 1.81 & 140/78   & 1.79 & 124/110 & 1.13 & 74/70   & 1.06 \\
                Bit Adder (\texttt{rd53\_131})                     & 7                              & 277/170  & 1.63 & 222/133  & 1.67 & 160/143 & 1.12 & 132/118 & 1.12 \\
                QAOA (\texttt{reg3\_8})                            & 8                              & 324/192  & 1.69 & 165/60   & 2.75 & 138/96  & 1.44 & 75/30   & 2.50 \\
                Ripple Adder (\texttt{ra\_10})                     & 22                             & 249/161  & 1.55 & 191/133  & 1.44 & 156/141 & 1.11 & 132/122 & 1.08 \\
                SAT (\texttt{sat\_n11})                            & 11                             & 405/252  & 1.61 & 291/204  & 1.43 & 227/210 & 1.08 & 194/182 & 1.07 \\
                Square (\texttt{squar5\_261})                      & 13                             & 1404/745 & 1.88 & 1008/547 & 1.84 & 742/621 & 1.19 & 552/478 & 1.15 \\
                SymB (\texttt{sym9\_146})                          & 12                             & 156/108  & 1.44 & 102/63   & 1.62 & 103/95  & 1.08 & 73/56   & 1.30 \\
                MCX (\texttt{tof\_10})                             & 19                             & 157/102  & 1.54 & 119/78   & 1.53 & 92/85   & 1.08 & 74/69   & 1.07 \\
                \colrule
                \multicolumn{2}{c}{Geometric mean}              & --                             & 1.65     & --   & 1.86     & --   & 1.13    & --   & 1.24           \\
                \hline\hline
            \end{tabular}%
        }
    \end{table}
}
\BenchmarkSummaryChainTable{Benchmarking summary for logical circuits and mapped results on the 1D chain topology.}
\BenchmarkSummarySquareTable{Benchmarking summary for logical circuits and mapped results on the 2D square topology.}

\subsection{Stage III: Scaling within representative algorithm families}

Stage II establishes breadth across many application categories, but each category is represented by a particular circuit instance and therefore does not isolate the effect of increasing problem size. We next perform a controlled scale-up study within four circuit families. This third stage tracks the same compilation metrics as the circuits grow, allowing workload structure and circuit scale to be distinguished.

\begin{figure*}[htbp]
    \centering
    \subfloat[Routing on 1D chain topology.\label{fig:routing-scaleup-1d}]{
        \includegraphics[width=\textwidth]{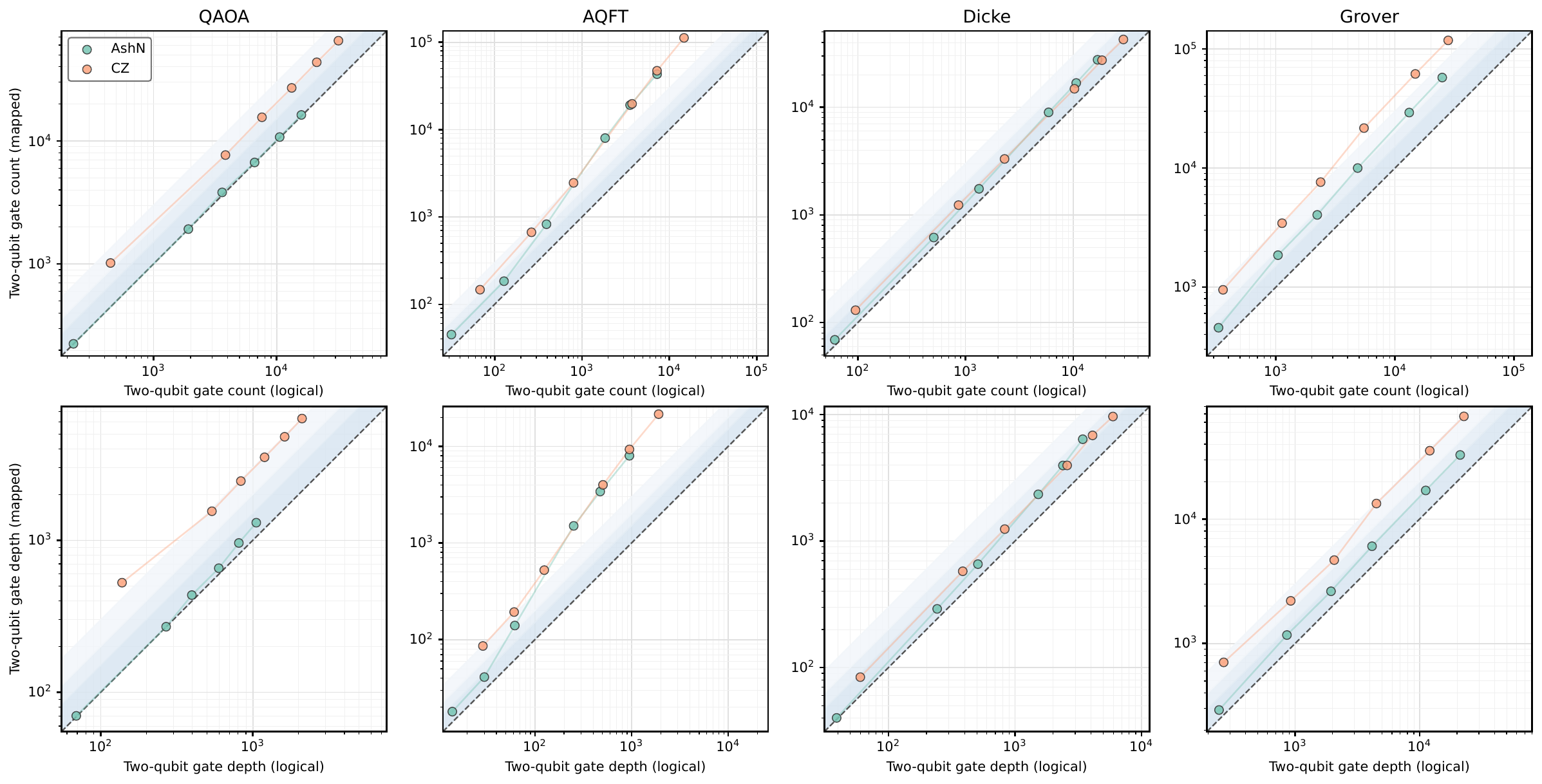}
    }\\
    \subfloat [Routing on 2D square topology.\label{fig:routing-scaleup-2d}]{
        \includegraphics[width=\textwidth]{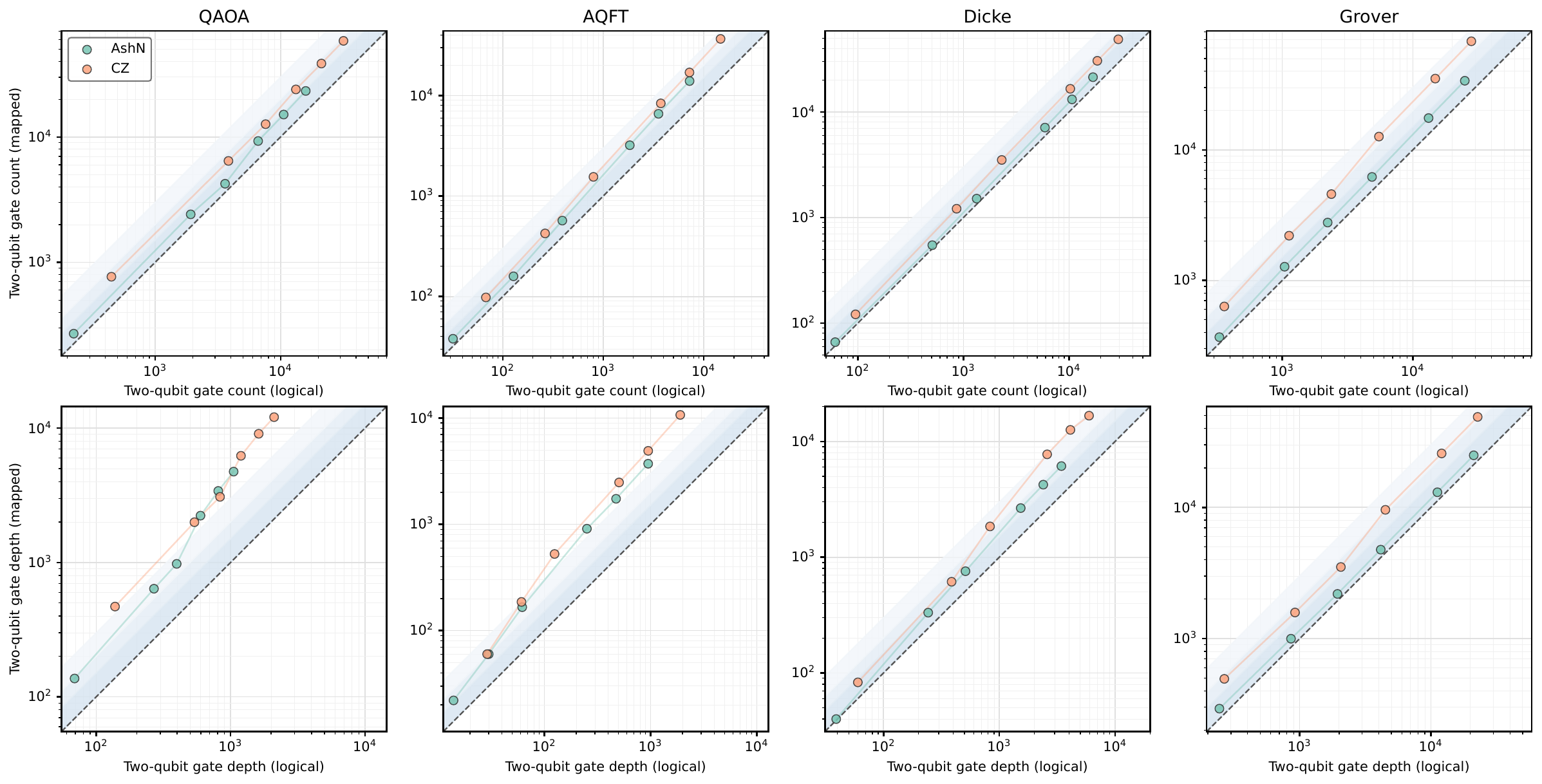}
    }
    \caption{
        Stage III routing overhead as four benchmarked quantum-algorithm families scale up. AshN- and $\mathrm{CZ}$-based routing results are shown for QAOA, approximate quantum Fourier transform (AQFT), Dicke-state preparation, and Grover search circuits on limited-connectivity topologies.
        The dashed diagonal line, $y = x$, represents zero routing overhead.
        The three shaded regions indicate routing-overhead factors of $1\text{--}1.5\times$, $1.5\text{--}2\times$, and $2\text{--}3\times$, respectively.
    }
    \label{fig:routing-scaleup}
\end{figure*}

\newcommand{\CZ}{\mathrm{CZ}}

For this controlled scaling study, we construct six progressively larger instances from each family:
\begin{enumerate}
    \item QAOA, utilizing Sherrington-Kirkpatrick problem instances generated by Cirq. Instances of qubit count $n\in\{8, 16, 20, 24, 28, 32\}$ and QAOA layer count $p=n$ are tested.
    \item AQFT, an approximate version of the QFT circuit that truncates two-qubit interactions to only those involving qubits within $m=16$ nearest neighbors. Instances of qubit count $n\in\{8, 16, 32, 64, 80, 96\}$ are tested. For $n=64, 80, 96$, to further increase quantum circuit scale, a $\times 2$, $\times 3$, and $\times 5$ repetition of the QFT circuit is applied, respectively.
    \item Dicke, the state preparation circuit from \citet{bartschi2022short}. Instances of size $n\in\{8, 16, 24, 48, 64, 80\}$ and $k=\left\lfloor \frac{n}{3}\right\rfloor$ are tested.
    \item Grover, a \texttt{GroverOperator} from Qiskit generated from a single marked-state phase oracle. Instances of $n\in\{8, 12, 16, 24, 32, 40\}$ are tested. For $n=32, 40$, a $\times 2$ and $\times 3$ repetition of the Grover circuit is applied to simulate multi-iteration search scenarios.
\end{enumerate}

Table~\ref{tab:routing-scaleup} and Fig.~\ref{fig:routing-scaleup} show how the routing overhead evolves within each algorithm family as the instances scale up.

Across the evaluated benchmark families and topologies, AshN routing with $\mathrm{SWAP}$ absorption generally incurs less routing overhead than $\mathrm{CZ}$ routing:
On the 1D chain, the two-qubit gate-count and depth overheads
decrease from \BenchScaleupChainCZGateGeomean{} and
\BenchScaleupChainCZDepthGeomean{} with $\mathrm{CZ}$ routing to
\BenchScaleupChainAshNGateGeomean{} and
\BenchScaleupChainAshNDepthGeomean{} with AshN routing, respectively.
On the 2D square lattice, the corresponding overheads decrease
from \BenchScaleupSquareCZGateGeomean{} and
\BenchScaleupSquareCZDepthGeomean{} to
\BenchScaleupSquareAshNGateGeomean{} and
\BenchScaleupSquareAshNDepthGeomean{}, respectively.
At the aggregate level, AshN achieves lower geometric-mean overhead in both two-qubit gate count and circuit depth. The advantage is nevertheless workload- and topology-dependent rather than universal: for example, the largest Dicke circuits on the 1D chain do not retain the same depth advantage, and the QAOA family has higher AshN routing overhead on the 2D square lattice than on the 1D chain.

The QAOA trend is similar to that of the \texttt{qft\_n18} circuit in Stage II. Both have dense, approximately all-to-all interaction patterns. On the 1D chain, MirrorSABRE can identify the well-known interact-then-$\mathrm{SWAP}$ pattern~\cite{yangUnifyingQubitRouting2026}, which is particularly effective for such interactions. On the 2D square lattice, the larger search space for both the initial mapping and routing-$\mathrm{SWAP}$ insertion makes this structure harder for the heuristic to identify. Thus, a less restrictive physical topology need not yield a lower compiled overhead for every heuristic run; this observation separates compiler-search effects from the native-gate advantage itself.

\providecommand{\RoutingScaleupCaption}{\GJZ{Placeholder}}
\ProvideDocumentCommand{\RoutingScaleupTable}{g}{%
    \begin{table*}[htbp]
        \centering
        \caption{\IfNoValueTF{#1}{\RoutingScaleupCaption}{#1}}
        \label{tab:routing-scaleup}
        \scriptsize
        \setlength{\tabcolsep}{3pt}
        \resizebox{\textwidth}{!}{%
            \begin{tabular}{llrrrrrrrrrrrrrrrr}
                \hline\hline
                Circuit family                        & Circuit scale                 & \multicolumn{8}{c}{1D chain}
                                                      & \multicolumn{8}{c}{2D square}                                                                                                                                                                                                                      \\
                \cline{3-10}\cline{11-18}
                                                      &                               & \multicolumn{4}{c}{$\mathrm{CZ}$}       & \multicolumn{4}{c}{AshN}
                                                      & \multicolumn{4}{c}{$\mathrm{CZ}$}        & \multicolumn{4}{c}{AshN}                                                                                                                                                                                           \\
                                                      &                               & \#2Q                         & $R_{\#2Q}$               & Depth       & $R_D$ & \#2Q        & $R_{\#2Q}$ & Depth       & $R_D$
                                                      & \#2Q                          & $R_{\#2Q}$                   & Depth                    & $R_D$       & \#2Q  & $R_{\#2Q}$  & Depth      & $R_D$                                                                                                   \\
                \colrule
                QAOA                                  & $n=8,\ p=8$                   & 1020/448                     & 2.28                     & 526/138     & 3.81  & 225/224     & 1.00       & 70/69       & 1.01  & 768/448     & 1.71 & 470/138     & 3.41 & 270/224     & 1.21 & 137/69      & 1.99 \\
                QAOA                                  & $n=16,\ p=16$                 & 7676/3840                    & 2.00                     & 1552/538    & 2.88  & 1923/1920   & 1.00       & 270/269     & 1.00  & 6443/3840   & 1.68 & 1995/538    & 3.71 & 2416/1920   & 1.26 & 638/269     & 2.37 \\
                QAOA                                  & $n=20,\ p=20$                 & 15558/7600                   & 2.05                     & 2451/834    & 2.94  & 3819/3610   & 1.06       & 436/397     & 1.10  & 12661/7600  & 1.67 & 3080/834    & 3.69 & 4238/3610   & 1.17 & 977/397     & 2.46 \\
                QAOA                                  & $n=24,\ p=24$                 & 26954/13248                  & 2.03                     & 3514/1194   & 2.94  & 6684/6624   & 1.01       & 655/597     & 1.10  & 23962/13248 & 1.81 & 6224/1194   & 5.21 & 9292/6624   & 1.40 & 2232/597    & 3.74 \\
                QAOA                                  & $n=28,\ p=28$                 & 43522/21168                  & 2.06                     & 4795/1618   & 2.96  & 10746/10584 & 1.02       & 960/809     & 1.19  & 38540/21168 & 1.82 & 9080/1618   & 5.61 & 15103/10584 & 1.43 & 3409/809    & 4.21 \\
                QAOA                                  & $n=32,\ p=32$                 & 65166/31744                  & 2.05                     & 6318/2106   & 3.00  & 16253/15872 & 1.02       & 1305/1053   & 1.24  & 58441/31744 & 1.84 & 12071/2106  & 5.73 & 23300/15872 & 1.47 & 4747/1053   & 4.51 \\
                AQFT                                  & $n=8,\ m=16$                  & 147/68                       & 2.16                     & 86/29       & 2.97  & 45/32       & 1.41       & 18/14       & 1.29  & 98/68       & 1.44 & 60/29       & 2.07 & 38/32       & 1.19 & 22/14       & 1.57 \\
                AQFT                                  & $n=16,\ m=16$                 & 668/264                      & 2.53                     & 193/61      & 3.16  & 184/128     & 1.44       & 41/30       & 1.37  & 425/264     & 1.61 & 186/61      & 3.05 & 159/128     & 1.24 & 60/30       & 2.00 \\
                AQFT                                  & $n=32,\ m=16$                 & 2459/800                     & 3.07                     & 524/125     & 4.19  & 826/392     & 2.11       & 140/62      & 2.26  & 1553/800    & 1.94 & 526/125     & 4.21 & 570/392     & 1.45 & 166/62      & 2.68 \\
                AQFT                                  & $n=64,\ m=16,\times 2$        & 19743/3744                   & 5.27                     & 3992/506    & 7.89  & 8006/1840   & 4.35       & 1503/252    & 5.96  & 8376/3744   & 2.24 & 2480/506    & 4.90 & 3208/1840   & 1.74 & 909/252     & 3.61 \\
                AQFT                                  & $n=80,\ m=16,\times 3$        & 47243/7224                   & 6.54                     & 9293/951    & 9.77  & 19112/3552  & 5.38       & 3405/474    & 7.18  & 17013/7224  & 2.36 & 4918/951    & 5.17 & 6592/3552   & 1.86 & 1740/474    & 3.67 \\
                AQFT                                  & $n=96,\ m=16,\times 5$        & 112459/14720                 & 7.64                     & 21510/1905  & 11.29 & 43158/7240  & 5.96       & 7987/950    & 8.41  & 36700/14720 & 2.49 & 10716/1905  & 5.63 & 13965/7240  & 1.93 & 3719/950    & 3.91 \\
                Dicke                                 & $n=8,\ k=2$                   & 130/95                       & 1.37                     & 84/60       & 1.40  & 69/61       & 1.13       & 40/39       & 1.03  & 121/95      & 1.27 & 83/60       & 1.38 & 66/61       & 1.08 & 40/39       & 1.03 \\
                Dicke                                 & $n=16,\ k=5$                  & 1232/861                     & 1.43                     & 577/387     & 1.49  & 616/507     & 1.21       & 291/243     & 1.20  & 1212/861    & 1.41 & 614/387     & 1.59 & 547/507     & 1.08 & 332/243     & 1.37 \\
                Dicke                                 & $n=24,\ k=8$                  & 3310/2305                    & 1.44                     & 1243/831    & 1.50  & 1746/1334   & 1.31       & 657/510     & 1.29  & 3512/2305   & 1.52 & 1845/831    & 2.22 & 1510/1334   & 1.13 & 757/510     & 1.48 \\
                Dicke                                 & $n=48,\ k=16$                 & 14846/10277                  & 1.44                     & 3962/2587   & 1.53  & 8970/5914   & 1.52       & 2341/1530   & 1.53  & 16584/10277 & 1.61 & 7740/2587   & 2.99 & 7111/5914   & 1.20 & 2656/1530   & 1.74 \\
                Dicke                                 & $n=64,\ k=21$                 & 27390/18533                  & 1.48                     & 6842/4107   & 1.67  & 16809/10675 & 1.57       & 3957/2395   & 1.65  & 30533/18533 & 1.65 & 12583/4107  & 3.06 & 13164/10675 & 1.23 & 4233/2395   & 1.77 \\
                Dicke                                 & $n=80,\ k=26$                 & 42672/29283                  & 1.46                     & 9647/5952   & 1.62  & 27633/16861 & 1.64       & 6381/3435   & 1.86  & 48807/29283 & 1.67 & 16695/5952  & 2.80 & 21365/16861 & 1.27 & 6125/3435   & 1.78 \\
                Grover                                & $n=8$                         & 949/360                      & 2.64                     & 705/267     & 2.64  & 456/330     & 1.38       & 292/245     & 1.19  & 632/360     & 1.76 & 494/267     & 1.85 & 368/330     & 1.12 & 293/245     & 1.20 \\
                Grover                                & $n=12$                        & 3445/1128                    & 3.05                     & 2201/923    & 2.38  & 1855/1042   & 1.78       & 1169/860    & 1.36  & 2201/1128   & 1.95 & 1583/923    & 1.72 & 1274/1042   & 1.22 & 999/860     & 1.16 \\
                Grover                                & $n=16$                        & 7623/2376                    & 3.21                     & 4678/2063   & 2.27  & 4043/2226   & 1.82       & 2627/1944   & 1.35  & 4582/2376   & 1.93 & 3516/2063   & 1.70 & 2775/2226   & 1.25 & 2193/1944   & 1.13 \\
                Grover                                & $n=24$                        & 21671/5496                   & 3.94                     & 13326/4504  & 2.96  & 10006/4864  & 2.06       & 6048/4152   & 1.46  & 12631/5496  & 2.30 & 9589/4504   & 2.13 & 6200/4864   & 1.27 & 4766/4152   & 1.15 \\
                Grover                                & $n=32,\times 2$               & 61794/14832                  & 4.17                     & 35419/12080 & 2.93  & 29282/13184 & 2.22       & 16981/11216 & 1.51  & 35101/14832 & 2.37 & 25793/12080 & 2.14 & 17503/13184 & 1.33 & 13048/11216 & 1.16 \\
                Grover                                & $n=40,\times 3$               & 118239/28008                 & 4.22                     & 66871/22728 & 2.94  & 57602/24960 & 2.31       & 32740/21205 & 1.54  & 67760/28008 & 2.42 & 48961/22728 & 2.15 & 33786/24960 & 1.35 & 24954/21205 & 1.18 \\
                \colrule
                \multicolumn{2}{c}{Geometric mean} & --                            & 2.55                         & --                       & 2.91        & --    & 1.67        & --         & 1.64        & --    & 1.82        & --   & 2.95        & --   & 1.31        & --   & 1.96               \\
                \hline\hline
            \end{tabular}%
        }
    \end{table*}
}

\RoutingScaleupTable{Routing scalability across circuit families and scales. Each entry is the mapped-to-all-to-all overhead ratio for native two-qubit gate count ($R_{\#2Q}$) or two-qubit depth ($R_{D}$).}

\subsection{Boundary case: random interaction patterns}

\begin{figure*}[htbp]
    \centering
    \includegraphics[width=\textwidth]{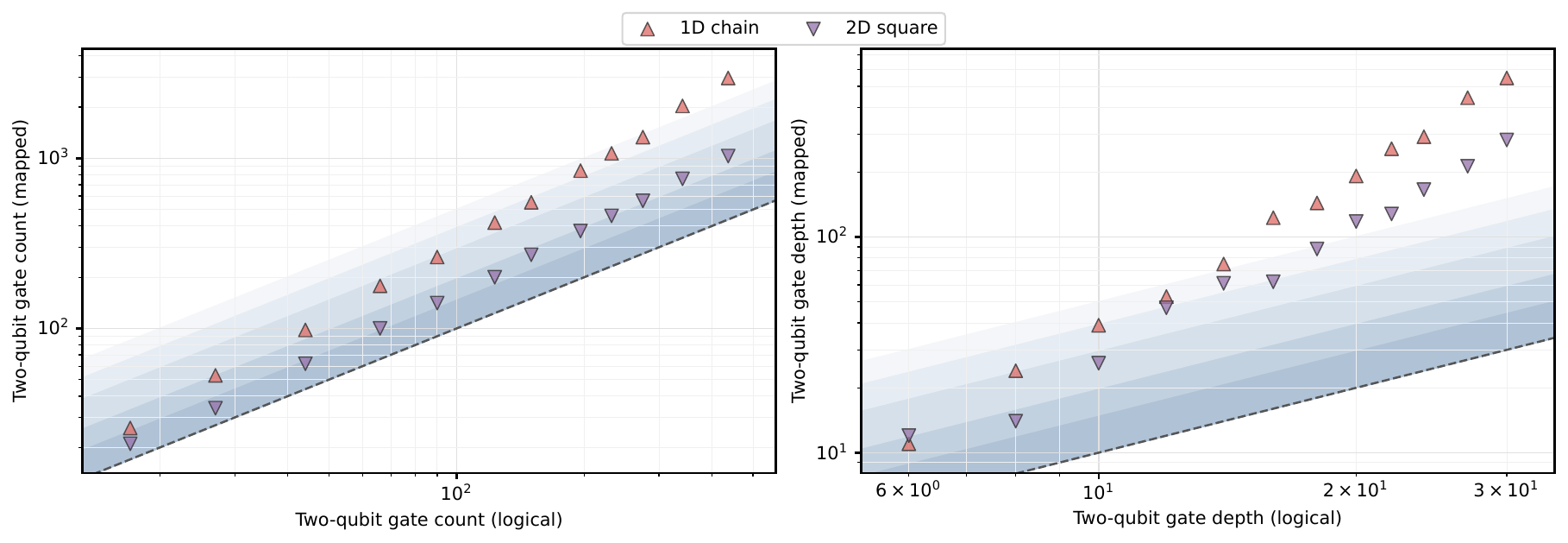}
    \caption{Qubit routing benchmarking for random sampling circuits (quantum volume) on limited-connectivity topologies. Only AshN-based routing results are shown since both AshN- and $\mathrm{CZ}$-based results show similar scaling trends, as reported in Table~\ref{tab:qv-routing}. The dashed diagonal line, $y=x$, represents an ideal mapping with no routing overhead. The five shaded regions indicate routing-overhead factors of $1\text{--}1.5\times$, $1.5\text{--}2\times$, $2\text{--}3\times$, $3\text{--}4\times$, and $4\text{--}5\times$, respectively, where the overhead factor is defined as the ratio between the mapped and logical circuit metrics.}
    \label{FigureS_qv_routing}
\end{figure*}

The three benchmark stages above concern structured application circuits and show where $\mathrm{SWAP}$ absorption is useful in experiment, across workloads, and with increasing scale. To delineate the opposite regime, we benchmark quantum-volume (QV) random circuits~\cite{cross2019validating} on both 1D chains and 2D square lattices. For each circuit, the QV width and the number of random $\mathrm{SU}(4)$ layers are both set to $n$, with $n\in\{6,8,10,12,14,16,18,20,22,24,27,30\}$. As shown in Fig.~\ref{FigureS_qv_routing}, the routing overhead grows rapidly with circuit size. On the chain, the two-qubit-gate overhead
increases from \BenchQVChainMinGate{} at $n=\BenchQVMinWidth{}$ to
\BenchQVChainMaxGate{} at $n=\BenchQVMaxWidth{}$, while the depth overhead
increases from \BenchQVChainMinDepth{} to \BenchQVChainMaxDepth{}. The square
lattice alleviates, but does not remove, this scaling: the gate-count overhead
ranges from \BenchQVSquareMinGate{} to \BenchQVSquareMaxGate{}, and the depth overhead
ranges from \BenchQVSquareMinDepth{} to \BenchQVSquareMaxDepth{}.

Unlike the structured application circuits considered above, QV circuits contain random two-qubit interactions whose partners change from layer to layer. Consequently, a required $\mathrm{SWAP}$ rarely coincides with a preceding two-qubit operation on the same physical-qubit pair, leaving few opportunities for systematic $\mathrm{SWAP}$ absorption. Routing overheads for quantum volume circuits for both $\CZ$- and AshN-based pipelines are given in Table~\ref{tab:qv-routing}, and AshN-based data are plotted in Fig.~\ref{FigureS_qv_routing}. These results identify random circuits as a fundamental limitation of the approach: enriched native two-qubit gates can substantially reduce routing costs for structured workloads, but cannot in general overcome the growing connectivity penalty induced by unstructured, all-to-all interaction patterns.

\providecommand{\QVRoutingCaption}{\GJZ{Placeholder}}
\ProvideDocumentCommand{\QVRoutingTable}{g}{%
    \begin{table*}[htbp]
        \centering
        \caption{\IfNoValueTF{#1}{\QVRoutingCaption}{#1}}
        \label{tab:qv-routing}
        \scriptsize
        \resizebox{\textwidth}{!}{%
            \begin{tabular}{rrrrrrrrrrrrrrrrr}
                \hline\hline
                Qubits & \multicolumn{8}{c}{1D chain} & \multicolumn{8}{c}{2D square}                                                                                                                           \\
                       & \multicolumn{4}{c}{$\mathrm{CZ}$}       & \multicolumn{4}{c}{AshN}
                       & \multicolumn{4}{c}{$\mathrm{CZ}$}       & \multicolumn{4}{c}{AshN}                                                                                                                                \\
                       & \#2Q                         & $R$                           & Depth   & $R$   & \#2Q     & $R$  & Depth  & $R$
                       & \#2Q                         & $R$                           & Depth   & $R$   & \#2Q     & $R$  & Depth  & $R$                                                                        \\
                \colrule
                6      & 81/51                        & 1.59                          & 36/18   & 2.00  & 26/17    & 1.53 & 11/6   & 1.83  & 57/51     & 1.12 & 24/18  & 1.33 & 21/17    & 1.24 & 12/6   & 2.00 \\
                8      & 150/81                       & 1.85                          & 57/24   & 2.38  & 53/27    & 1.96 & 24/8   & 3.00  & 99/81     & 1.22 & 45/24  & 1.88 & 34/27    & 1.26 & 14/8   & 1.75 \\
                10     & 285/132                      & 2.16                          & 126/30  & 4.20  & 98/44    & 2.23 & 39/10  & 3.90  & 183/132   & 1.39 & 75/30  & 2.50 & 62/44    & 1.41 & 26/10  & 2.60 \\
                12     & 510/198                      & 2.58                          & 141/36  & 3.92  & 178/66   & 2.70 & 53/12  & 4.42  & 282/198   & 1.42 & 90/36  & 2.50 & 100/66   & 1.52 & 47/12  & 3.92 \\
                14     & 732/270                      & 2.71                          & 219/42  & 5.21  & 263/90   & 2.92 & 75/14  & 5.36  & 414/270   & 1.53 & 144/42 & 3.43 & 141/90   & 1.57 & 61/14  & 4.36 \\
                16     & 1188/369                     & 3.22                          & 288/48  & 6.00  & 419/123  & 3.41 & 123/16 & 7.69  & 594/369   & 1.61 & 186/48 & 3.88 & 200/123  & 1.63 & 62/16  & 3.88 \\
                18     & 1518/450                     & 3.37                          & 414/54  & 7.67  & 551/150  & 3.67 & 144/18 & 8.00  & 777/450   & 1.73 & 285/54 & 5.28 & 271/150  & 1.81 & 88/18  & 4.89 \\
                20     & 2274/588                     & 3.87                          & 483/60  & 8.05  & 846/196  & 4.32 & 192/20 & 9.60  & 1065/588  & 1.81 & 306/60 & 5.10 & 374/196  & 1.91 & 118/20 & 5.90 \\
                22     & 3006/696                     & 4.32                          & 591/66  & 8.95  & 1069/232 & 4.61 & 257/22 & 11.68 & 1323/696  & 1.90 & 447/66 & 6.77 & 459/232  & 1.98 & 128/22 & 5.82 \\
                24     & 3735/825                     & 4.53                          & 732/72  & 10.17 & 1333/275 & 4.85 & 292/24 & 12.17 & 1620/825  & 1.96 & 471/72 & 6.54 & 562/275  & 2.04 & 166/24 & 6.92 \\
                27     & 5412/1023                    & 5.29                          & 972/81  & 12.00 & 2032/341 & 5.96 & 443/27 & 16.41 & 2145/1023 & 2.10 & 630/81 & 7.78 & 757/341  & 2.22 & 213/27 & 7.89 \\
                30     & 7836/1311                    & 5.98                          & 1245/90 & 13.83 & 2966/437 & 6.79 & 547/30 & 18.23 & 2928/1311 & 2.23 & 816/90 & 9.07 & 1031/437 & 2.36 & 282/30 & 9.40 \\
                \hline\hline
            \end{tabular}%
        }
    \end{table*}
}

\QVRoutingTable{Routing overhead for Quantum Volume circuits compiled with $\mathrm{CZ}$ and AshN gate sets.}

\subsection{Native-gate-set requirements and equivalent $\mathrm{CZ}$ cost}

\begin{figure*}[htbp]
    \centering
    \subfloat[Real-world applications.\label{subfig:su4decomp-real}]{
        \includegraphics[width=\textwidth]{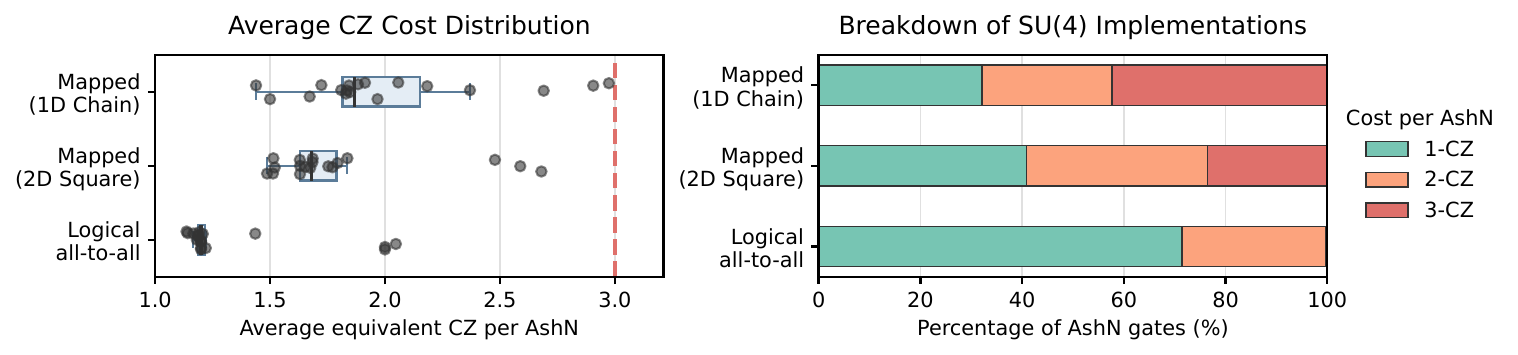}
    }\\
    \subfloat [Quantum volume.\label{subfig:su4decomp-qv}]{
        \includegraphics[width=\textwidth]{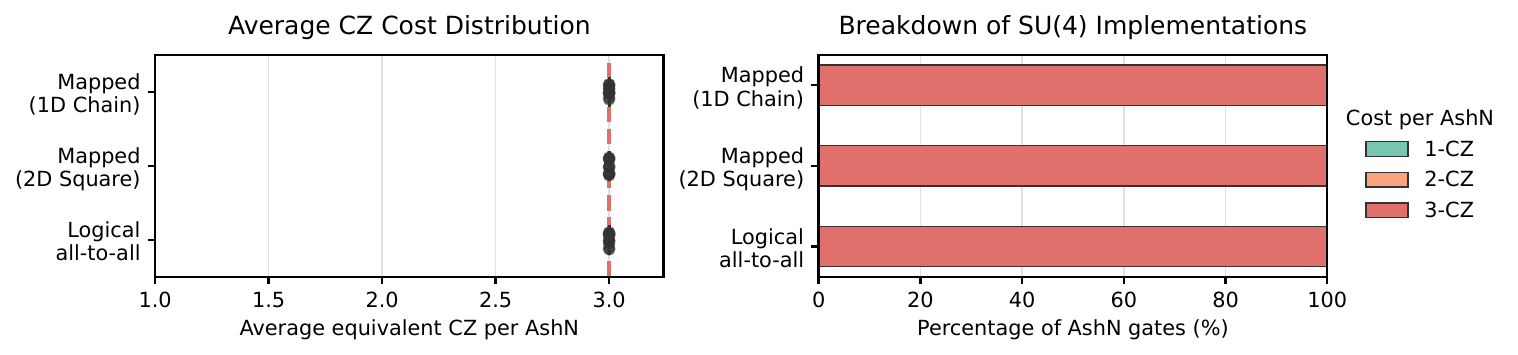}
    }\\
    \subfloat[Weyl chamber coverage of 1 to 3 $\CZ$ gates.]{
        \includegraphics[width=0.8\textwidth]{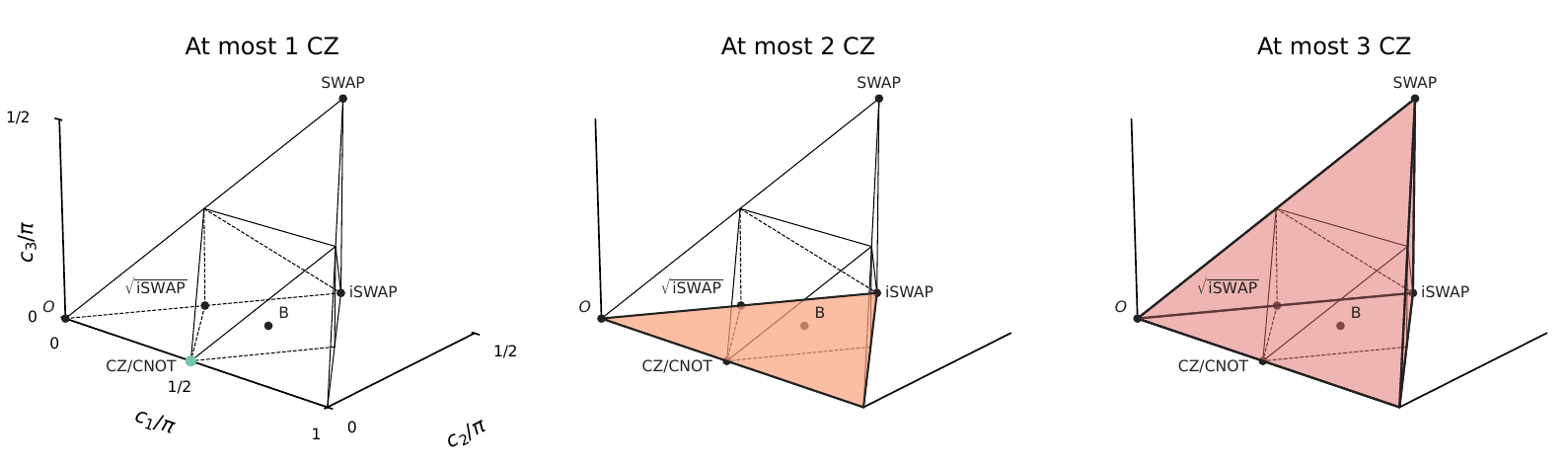}
        \label{subfig:su4decomp-weyl}
    }
    \caption{
        Categories of $\mathrm{SU}(4)$ blocks in logical and mapped $\mathrm{SU}(4)$ circuits of (a) real-world applications and (b) quantum volume circuits. In (c), the Weyl chamber coverage of 1 to 3 $\CZ$ gates is shown.
    }
    \label{fig:su4decomp}
\end{figure*}

The preceding comparisons treat each calibrated AshN operation as one native two-qubit gate. We finally test how strongly the conclusions depend on that hardware assumption. In particular, if only a restricted subset of the AshN family were calibrated, a general $\mathrm{SU}(4)$ block could require as many as three $\CZ$ gates, as shown by the Weyl-chamber coverage in Fig.~\ref{subfig:su4decomp-weyl}. An equivalent-$\mathrm{CZ}$-cost analysis therefore provides a conservative bridge from the full-AshN results to a restricted native gate set.

We collected all $\mathrm{SU}(4)$ blocks from logical and mapped circuits for both real-world applications and quantum volume circuits, and plotted the distribution of $\mathrm{SU}(4)$ blocks in Fig.~\ref{fig:su4decomp}.

For real-world circuits, a large fraction of $\mathrm{SU}(4)$ blocks are simply locally equivalent to a $\mathrm{CZ}$ gate, and among the remainder, a large fraction lie on the floor of the Weyl chamber (the $c=0$ plane) and can be decomposed into $2$ $\CZ$ gates. Absorbing $\mathrm{SWAP}$ gates into a $\mathrm{SU}(4)$ block would reduce the overall equivalent $\CZ$-cost to at most $3$.
In fact, as shown in Table~\ref{tab:legacy-ashn-cz-cost}, the gate-count-weighted average $\CZ$-cost of $\mathrm{SU}(4)$ blocks in real-world logical circuits is
\BenchLegacyAshNCZCostLogicalWeightedOverall{}, and increases to \BenchLegacyAshNCZCostMappedOneDWeightedOverall{} on 1D topology and \BenchLegacyAshNCZCostMappedTwoDWeightedOverall{} on 2D topology, each of which is still much lower than the theoretical bound $3$.

The distribution of $\mathrm{SU}(4)$ in quantum volume circuits is much simpler: as shown in Fig.~\ref{subfig:su4decomp-qv}, the decomposition procedure assigns a cost of three $\CZ$ gates to essentially all sampled generic $\mathrm{SU}(4)$ blocks, demonstrating that in worst-case scenarios one $\mathrm{SU}(4)$ block should almost always be counted as $3$ $\CZ$ gates.

We conclude this section by showing the MirrorSABRE flow on a potential target only supporting $\CZ$ gate, by converting all $\mathrm{SU}(4)$ blocks back to local $\CZ$ implementations.
As shown in Table~\ref{tab:legacy-can-intermediate-cz-comparison}, the equivalent $\mathrm{CZ}$ implementation cost measured in terms of equivalent $\CZ$ count relies heavily on circuit type and topology.
On the 1D chain, the geometric-mean cost ratio relative to the direct $\CZ$-based pipeline is \BenchLegacyCANChainRatioGeomean{}, with substantial variation across circuits, indicating that the limited connectivity cannot be fully mitigated by AshN gate set and MirrorSABRE.
On the 2D square lattice, the equivalent $\CZ$ implementation cost shows a clear overall advantage, where the geometric-mean cost ratio is \BenchLegacyCANSquareRatioGeomean{}, and most circuits benefit from $\mathrm{SWAP}$ absorption.

\providecommand{\LegacyAshNCZCostTableCaption}{\GJZ{Placeholder}}
\ProvideDocumentCommand{\LegacyAshNCZCostTable}{g}{%
    \begin{table*}[htbp]
        \centering
        \caption{\IfNoValueTF{#1}{\LegacyAshNCZCostTableCaption}{#1}}
        \label{tab:legacy-ashn-cz-cost}
        \scriptsize
        \setlength{\tabcolsep}{3pt}
        \providecommand{\benchhead}[1]{\begin{tabular}[c]{@{}c@{}}##1\end{tabular}}
        \resizebox{\textwidth}{!}{%
            \begin{tabular}{lrrrrrrrrrrrrrrr}
                \hline\hline
                \benchhead{Application                                                                                                                                                                                        \\(Circuit)} & \multicolumn{5}{c}{Logical}
                                                                   & \multicolumn{5}{c}{Mapped 1D} & \multicolumn{5}{c}{Mapped 2D}                                                                                            \\
                \cline{2-6}\cline{7-11}\cline{12-16}
                                                                   & \benchhead{\#AshN}            & \benchhead{1-$\mathrm{CZ}$                                                                                               \\(\%)}
                                                                   & \benchhead{2-$\mathrm{CZ}$                                                                                                                               \\(\%)}
                                                                   & \benchhead{3-$\mathrm{CZ}$                                                                                                                               \\(\%)} & \benchhead{Avg.\\$\mathrm{CZ}$/AshN}
                                                                   & \benchhead{\#AshN}            & \benchhead{1-$\mathrm{CZ}$                                                                                               \\(\%)}
                                                                   & \benchhead{2-$\mathrm{CZ}$                                                                                                                               \\(\%)}
                                                                   & \benchhead{3-$\mathrm{CZ}$                                                                                                                               \\(\%)} & \benchhead{Avg.\\$\mathrm{CZ}$/AshN}
                                                                   & \benchhead{\#AshN}            & \benchhead{1-$\mathrm{CZ}$                                                                                               \\(\%)}
                                                                   & \benchhead{2-$\mathrm{CZ}$                                                                                                                               \\(\%)}
                                                                   & \benchhead{3-$\mathrm{CZ}$                                                                                                                               \\(\%)} & \benchhead{Avg.\\$\mathrm{CZ}$/AshN} \\
                \colrule
                Comparator (\texttt{4gt4-v0\_73})                  & 127                           & 78.7                          & 20.5  & 0.8 & 1.22 & 159  & 41.5 & 34.0 & 24.5 & 1.83 & 135  & 50.4 & 36.3 & 13.3 & 1.63 \\
                Chemistry (\texttt{H2\_cmplt\_BK})                 & 39                            & 56.4                          & 43.6  & 0.0 & 1.44 & 47   & 44.7 & 38.3 & 17.0 & 1.72 & 39   & 51.3 & 48.7 & 0.0  & 1.49 \\
                ALU (\texttt{alu-v2\_30})                          & 157                           & 80.3                          & 19.7  & 0.0 & 1.20 & 205  & 41.5 & 32.2 & 26.3 & 1.85 & 165  & 48.5 & 40.0 & 11.5 & 1.63 \\
                Multiplier (\texttt{gf2\textasciicircum{}8\_mult}) & 341                           & 81.2                          & 18.8  & 0.0 & 1.19 & 771  & 21.8 & 19.5 & 58.8 & 2.37 & 434  & 42.9 & 30.6 & 26.5 & 1.84 \\
                Grover (\texttt{grover\_5})                        & 240                           & 80.0                          & 20.0  & 0.0 & 1.20 & 289  & 41.2 & 36.7 & 22.1 & 1.81 & 254  & 40.6 & 41.7 & 17.7 & 1.77 \\
                Encoding (\texttt{ham7\_104})                      & 106                           & 79.2                          & 20.8  & 0.0 & 1.21 & 128  & 40.6 & 34.4 & 25.0 & 1.84 & 111  & 47.7 & 41.4 & 10.8 & 1.63 \\
                HWB (\texttt{hwb5\_53})                            & 429                           & 80.9                          & 19.1  & 0.0 & 1.19 & 562  & 41.5 & 28.8 & 29.7 & 1.88 & 465  & 48.6 & 37.4 & 14.0 & 1.65 \\
                KNN (\texttt{knn\_n25})                            & 72                            & 83.3                          & 16.7  & 0.0 & 1.17 & 72   & 66.7 & 16.7 & 16.7 & 1.50 & 73   & 63.0 & 21.9 & 15.1 & 1.52 \\
                QFT (\texttt{qft\_n18})                            & 152                           & 0.0                           & 100.0 & 0.0 & 2.00 & 154  & 0.0  & 2.6  & 97.4 & 2.97 & 192  & 0.0  & 41.1 & 58.9 & 2.59 \\
                QPE (\texttt{qpeexact\_n16})                       & 127                           & 0.8                           & 93.7  & 5.5 & 2.05 & 255  & 0.0  & 9.4  & 90.6 & 2.91 & 172  & 0.6  & 30.8 & 68.6 & 2.68 \\
                QRAM (\texttt{qram\_n20})                          & 110                           & 81.8                          & 18.2  & 0.0 & 1.18 & 196  & 31.6 & 18.4 & 50.0 & 2.18 & 124  & 50.8 & 29.8 & 19.4 & 1.69 \\
                Bit Adder (\texttt{rd53\_131})                     & 143                           & 81.1                          & 18.9  & 0.0 & 1.19 & 187  & 43.3 & 29.9 & 26.7 & 1.83 & 160  & 50.0 & 32.5 & 17.5 & 1.68 \\
                QAOA (\texttt{reg3\_8})                            & 96                            & 0.0                           & 100.0 & 0.0 & 2.00 & 226  & 0.0  & 31.0 & 69.0 & 2.69 & 138  & 0.0  & 52.2 & 47.8 & 2.48 \\
                Ripple Adder (\texttt{ra\_10})                     & 141                           & 85.8                          & 14.2  & 0.0 & 1.14 & 174  & 52.9 & 27.0 & 20.1 & 1.67 & 156  & 60.3 & 28.2 & 11.5 & 1.51 \\
                SAT (\texttt{sat\_n11})                            & 210                           & 80.0                          & 20.0  & 0.0 & 1.20 & 277  & 37.5 & 28.2 & 34.3 & 1.97 & 227  & 43.2 & 38.3 & 18.5 & 1.75 \\
                Square (\texttt{squar5\_261})                      & 621                           & 80.0                          & 20.0  & 0.0 & 1.20 & 967  & 33.6 & 27.0 & 39.4 & 2.06 & 742  & 43.3 & 34.1 & 22.6 & 1.79 \\
                SymB (\texttt{sym9\_146})                          & 95                            & 86.3                          & 13.7  & 0.0 & 1.14 & 107  & 67.3 & 21.5 & 11.2 & 1.44 & 103  & 61.2 & 26.2 & 12.6 & 1.51 \\
                MCX (\texttt{tof\_10})                             & 85                            & 80.0                          & 20.0  & 0.0 & 1.20 & 115  & 36.5 & 35.7 & 27.8 & 1.91 & 92   & 46.7 & 38.0 & 15.2 & 1.68 \\
                \colrule
                \multicolumn{1}{c}{Weighted overall}               & 3291                          & 71.4                          & 28.3  & 0.2 & 1.29 & 4891 & 32.1 & 25.6 & 42.3 & 2.10 & 3782 & 40.9 & 35.6 & 23.5 & 1.83 \\
                \hline\hline
            \end{tabular}%
        }
    \end{table*}
}

\LegacyAshNCZCostTable{Distribution of AshN gates in Weyl chamber by $\mathrm{CZ}$ implementation cost. Avg. $\mathrm{CZ}$/AshN is the equivalent number of $\mathrm{CZ}$ gates per AshN gate. Weighted overall aggregates the underlying AshN gate counts across all circuits before computing the cost percentages and average.}

\providecommand{\LegacyCANIntermediateCZComparisonCaption}{\GJZ{Placeholder}}
\ProvideDocumentCommand{\LegacyCANIntermediateCZComparisonTable}{g}{%
    \begin{table*}[htbp]
        \centering
        \caption{\IfNoValueTF{#1}{\LegacyCANIntermediateCZComparisonCaption}{#1}}
        \label{tab:legacy-can-intermediate-cz-comparison}
        \scriptsize
        \setlength{\tabcolsep}{3pt}
        \providecommand{\LegacyCANComparisonHead}[1]{\begin{tabular}[c]{@{}c@{}}##1\end{tabular}}
        \resizebox{\textwidth}{!}{%
            \begin{tabular}{lrrrrrcrrrrc}
                \hline\hline
                \LegacyCANComparisonHead{Application                                                                                                                                                                                                  \\(Circuit)} & \LegacyCANComparisonHead{Unmapped\\$\mathrm{CZ}$}
                                                                   & \multicolumn{5}{c}{1D chain} & \multicolumn{5}{c}{2D square}                                                                                                                     \\
                \cline{3-7}\cline{8-12}
                                                                   &                              & $\mathrm{CZ}$                            & SU(4)                          & \LegacyCANComparisonHead{SU(4)                                                   \\equiv. $\mathrm{CZ}$} & Ratio & Result
                                                                   & $\mathrm{CZ}$                           & SU(4)                         & \LegacyCANComparisonHead{SU(4)                                                                                    \\equiv. $\mathrm{CZ}$} & Ratio & Result \\
                \colrule
                Comparator (\texttt{4gt4-v0\_73})                  & 155                          & 296                           & 159                            & 291                            & 0.983 & Win  & 250  & 135 & 220  & 0.880 & Win  \\
                Chemistry (\texttt{H2\_cmplt\_BK})                 & 56                           & 89                            & 47                             & 81                             & 0.910 & Win  & 62   & 39  & 58   & 0.935 & Win  \\
                ALU (\texttt{alu-v2\_30})                          & 188                          & 380                           & 205                            & 379                            & 0.997 & Win  & 327  & 165 & 269  & 0.823 & Win  \\
                Multiplier (\texttt{gf2\textasciicircum{}8\_mult}) & 405                          & 1674                          & 771                            & 1827                           & 1.091 & Loss & 840  & 434 & 797  & 0.949 & Win  \\
                Grover (\texttt{grover\_5})                        & 288                          & 562                           & 289                            & 523                            & 0.931 & Win  & 473  & 254 & 450  & 0.951 & Win  \\
                Encoding (\texttt{ham7\_104})                      & 128                          & 269                           & 128                            & 236                            & 0.877 & Win  & 225  & 111 & 181  & 0.804 & Win  \\
                HWB (\texttt{hwb5\_53})                            & 509                          & 1068                          & 562                            & 1058                           & 0.991 & Win  & 879  & 465 & 769  & 0.875 & Win  \\
                KNN (\texttt{knn\_n25})                            & 84                           & 110                           & 72                             & 108                            & 0.982 & Win  & 132  & 73  & 111  & 0.841 & Win  \\
                QFT (\texttt{qft\_n18})                            & 306                          & 556                           & 154                            & 458                            & 0.824 & Win  & 541  & 192 & 497  & 0.919 & Win  \\
                QPE (\texttt{qpeexact\_n16})                       & 260                          & 762                           & 255                            & 741                            & 0.972 & Win  & 487  & 172 & 461  & 0.947 & Win  \\
                QRAM (\texttt{qram\_n20})                          & 130                          & 395                           & 196                            & 428                            & 1.084 & Loss & 235  & 124 & 209  & 0.889 & Win  \\
                Bit Adder (\texttt{rd53\_131})                     & 170                          & 330                           & 187                            & 343                            & 1.039 & Loss & 277  & 160 & 268  & 0.968 & Win  \\
                QAOA (\texttt{reg3\_8})                            & 192                          & 551                           & 226                            & 608                            & 1.103 & Loss & 324  & 138 & 342  & 1.056 & Loss \\
                Ripple Adder (\texttt{ra\_10})                     & 161                          & 268                           & 174                            & 291                            & 1.086 & Loss & 249  & 156 & 236  & 0.948 & Win  \\
                SAT (\texttt{sat\_n11})                            & 252                          & 573                           & 277                            & 545                            & 0.951 & Win  & 405  & 227 & 398  & 0.983 & Win  \\
                Square (\texttt{squar5\_261})                      & 745                          & 1916                          & 967                            & 1990                           & 1.039 & Loss & 1404 & 742 & 1331 & 0.948 & Win  \\
                SymB (\texttt{sym9\_146})                          & 108                          & 154                           & 107                            & 154                            & 1.000 & Tie  & 156  & 103 & 156  & 1.000 & Tie  \\
                MCX (\texttt{tof\_10})                             & 102                          & 159                           & 115                            & 220                            & 1.384 & Loss & 157  & 92  & 155  & 0.987 & Win  \\
                \colrule
                \multicolumn{2}{c}{Geometric mean}              & --                           & --                            & --
                                                                   & 1.007
                                                                   & --
                                                                   & --                           & --                            & --                             & 0.926
                                                                   & --                                                                                                                                                                               \\
                \multicolumn{2}{c}{Win/Tie/Loss}                   & --                           & --                            & --                             & --
                                                                   & 10/1/7
                                                                   & --                           & --                            & --                             & --
                                                                   & 16/1/1                                                                                                                                                                           \\
                \hline\hline
            \end{tabular}
        }
    \end{table*}
}

\LegacyCANIntermediateCZComparisonTable{$\mathrm{CZ}$ implementation cost of real-world application circuits, computed from circuit statistics from Table~\ref{tab:benchmark-summary-chain} and Table~\ref{tab:benchmark-summary-square}, and $\mathrm{CZ}$ implementation cost from Table~\ref{tab:legacy-ashn-cz-cost}.}

Taken together, the evidence has a nested interpretation. Stage I shows on hardware that the compiler-level reduction survives execution for the seven experimentally accessible circuits. Stage II supports breadth across application-derived workloads, and Stage III supports persistence of the advantage as several representative families grow, while also exposing topology- and heuristic-dependent exceptions. The QV study then identifies rapidly changing random interactions as a boundary of $\mathrm{SWAP}$ absorption, and the equivalent-$\mathrm{CZ}$ analysis shows how the conclusion weakens when arbitrary AshN operations are not available natively. The resulting claim is therefore not that enriched control removes connectivity constraints universally, but that it can substantially reduce them for structured workloads when the native gate set and compiler are co-designed for $\mathrm{SWAP}$ absorption.

\newpage
\section{State Preparation of Dicke State}

\begin{figure*}[htbp]
    \centering
    \includegraphics[width=0.6\textwidth]{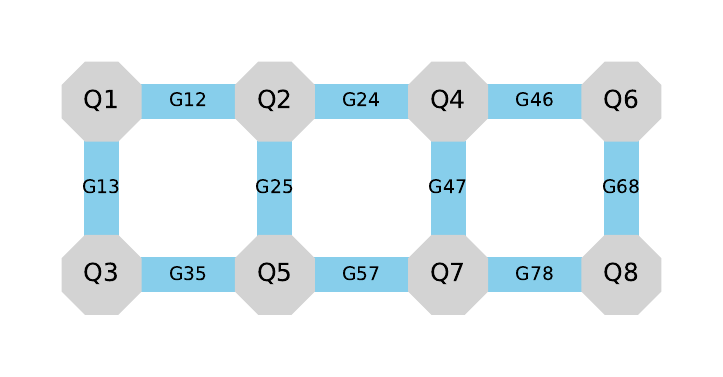}
    \caption{Schematic of the defect-free $2\times4$ qubit topology. Nodes represent qubits and edges denote available two-qubit couplings.
    }
    \label{topo_24}
\end{figure*}

Figure~\ref{topo_24} shows the system topology used for preparing the Dicke state on the $2\times4$ array. Single-qubit parameters are listed in Table~\ref{tab:qubit_params}, and the errors of the different two-qubit gates used in the AshN and $\mathrm{CZ}$ schemes are summarized in Table~\ref{tab:2q_gate_errors}; all gate errors were obtained via parallel XEB. In the AshN implementation, logical $\mathrm{CNOT}$ operations are implemented using locally equivalent $\mathrm{CZ}$ gates and single-qubit corrections.

The preparation circuit for the eight-qubit, two-excitation Dicke state, $D(8,2)$, is based on the $\mathrm{CZ}$ circuit proposed in Ref.~\cite{bartschi2022short}, which gives a short-depth construction for a perfect $2\times4$ rectangular lattice optimized for nearest-neighbor interactions on an ideal grid.

We considered eight hardware topologies in total: five variants of the $2\times4$ array and three variants of the $3\times3$ array. Because the original circuit is tailored to a perfect $2\times4$ lattice, it cannot run directly on topologies with defective couplers, nor is it compatible with the $3\times3$ geometry, whose connectivity differs from the rectangular lattice.

We therefore recompiled the circuit for each hardware topology using the Mirror-SABRE framework~\cite{li2019tackling,zou2024lightsabre}, which jointly optimizes qubit mapping and routing under the connectivity constraints of the target topology while minimizing the $\mathrm{SWAP}$ overhead introduced during routing. This procedure was applied uniformly across all topologies, yielding executable circuits that preserve the target Dicke state.

For the AshN scheme, circuits for all topologies were synthesized independently through the same Mirror-SABRE procedure. Depending on the topology and the resulting compilation, the output circuits use a native gate set comprising $\mathrm{CZ}$, $\mathrm{iSWAP}$, $\sqrt{\mathrm{iSWAP}}$, $\mathrm{SWAP}$, and a B-like entangling gate, with gate durations of 48, 44, 28, 60, and 44 ns, respectively.

\begin{figure}[htbp]
    \centering
    \includegraphics[width=0.9\textwidth]{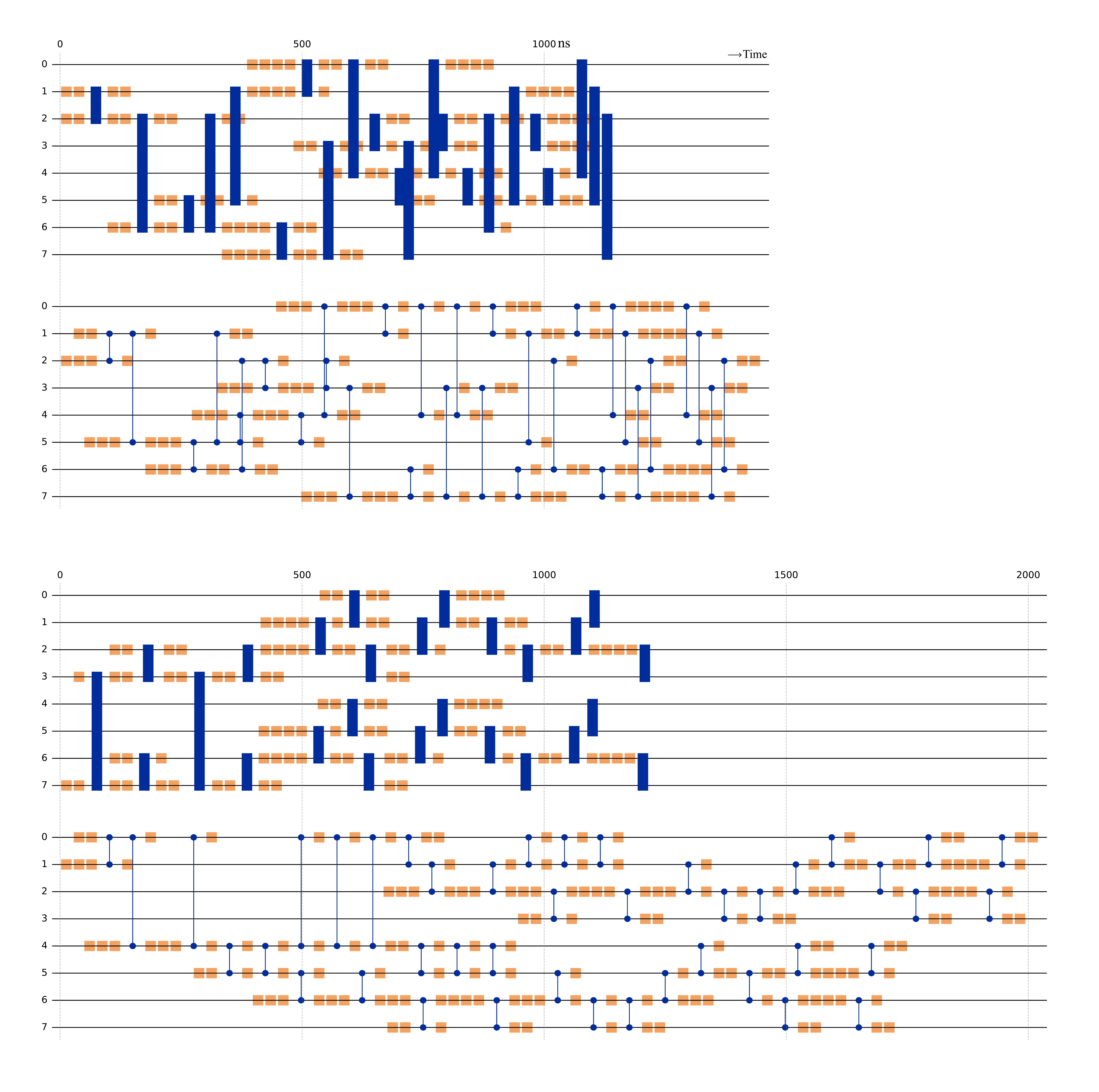}
    \caption{
        Quantum circuits for preparing the eight-qubit two-excitation Dicke state under different gate sets and connectivity configurations. From top to bottom, the panels show the AshN-based implementation on the defect-free $2\times4$ topology, the corresponding $\mathrm{CZ}$-based implementation, the AshN-based implementation on the Defect$_{247}$ topology [\MainTextTopologyFigure{}(a) of the main text], and the corresponding $\mathrm{CZ}$-based implementation. The comparison illustrates that the AshN gate set substantially reduces circuit recompilation and routing overhead while maintaining efficient state preparation in the presence of connectivity defects.
    }
    \label{topo_24_circuit}
\end{figure}

The B-like gate has Weyl chamber coordinates
\begin{equation}
    \left(
    \frac{\pi}{4},
    \frac{\pi}{4}
    -\frac{1}{2}\arccos\sqrt{\frac{5}{6}},
    0
    \right),
\end{equation}
placing it between the $\mathrm{CZ}$ and $\mathrm{iSWAP}$ families. As an additional native entangling gate, it enlarges the available gate set and gives the compiler more flexibility when adapting circuits to a given topology, particularly around connectivity defects. All circuits reported here were compiled subject to the connectivity constraints of their respective hardware topologies; representative AshN- and $\mathrm{CZ}$-based circuits for both the defect-free and defective $2\times4$ topologies are shown in Fig.~\ref{topo_24_circuit}.

All experimental results reported in this work are corrected for single-qubit readout error using a tensor-product noise model: the readout error matrix for each qubit is characterized independently, and the resulting correction is applied to the measured probability distributions via matrix inversion~\cite{bravyi2021mitigating}.

To evaluate the robustness of Dicke-state preparation against variations in hardware connectivity, we prepared the Dicke state on eight hardware topologies using both the AshN and $\mathrm{CZ}$ schemes. The resulting PPT-mixture entanglement witness values, $\mathrm{Tr}(W_{8, 2}^{\textrm{opti}}\cdot\rho)$, for the defect-free and defective configurations of the $2\times4$ topology are summarized in Table~\ref{tab:PPT_witness_result}. Negative values of $\mathrm{Tr}(W_{8, 2}^{\textrm{opti}}\cdot\rho)$ certify genuine multipartite entanglement (GME), providing a quantitative comparison of the two compilation schemes across different hardware topologies.

\begin{table}[htbp]
    \centering
    \caption{Device parameters for the eight-qubit system.}
    \label{tab:qubit_params}
    \setlength{\tabcolsep}{6pt}
    \begin{tabular}{c c c c c c c c c}
        \hline\hline
        Qubit                            & Q1   & Q2   & Q3   & Q4   & Q5   & Q6   & Q7   & Q8   \\
        \colrule
        Frequency (GHz)                  & 4.97 & 4.88 & 4.75 & 4.85 & 4.81 & 4.75 & 4.72 & 4.91 \\
        $\alpha$ (MHz)                   & 224  & 229  & 227  & 228  & 228  & 231  & 230  & 227  \\
        $f_{\mathrm{readout}}$ (GHz)     & 6.51 & 6.58 & 6.24 & 6.43 & 6.51 & 6.56 & 6.24 & 6.43 \\
        $T_1$ ($\mu$s)                   & 42.3 & 49.7 & 46.5 & 50.5 & 51.7 & 57.1 & 33.1 & 48.5 \\
        $T_2^{\mathrm{Ramsey}}$ ($\mu$s) & 6.8  & 21.5 & 8.8  & 5.2  & 11.7 & 14.7 & 9.6  & 8.3  \\
        Readout error $|0\rangle$ (\%)   & 0.6  & 0.9  & 0.8  & 0.7  & 1.9  & 0.6  & 1.2  & 0.5  \\
        Readout error $|1\rangle$ (\%)   & 2.1  & 2.0  & 1.6  & 2.2  & 1.8  & 0.9  & 2.2  & 1.5  \\
        1Q gate error (\textperthousand) & 0.9  & 1.0  & 1.0  & 1.3  & 1.2  & 0.9  & 0.8  & 1.6  \\
        \hline\hline
    \end{tabular}
\end{table}

\begin{table}[htbp]
    \centering
    \caption{Two-qubit gate errors for different native gate realizations on each coupler.}
    \label{tab:2q_gate_errors}

    \setlength{\tabcolsep}{4.5pt}
    \renewcommand{\arraystretch}{1.2}

    \small

    \begin{tabular}{lcccccccccc}
        \hline\hline
        Coupler &
        G12     & G13   & G25   & G35   & G24   &
        G57     & G47   & G46   & G78   & G68                   \\
        \colrule
        $\mathrm{CZ}$ gate error (\%)
                & 0.413 & 0.353 & 0.410 & 0.554 & 0.860
                & 0.565 & 0.566 & 0.539 & 0.540 & 0.696         \\

        $\mathrm{iSWAP}$ gate error (\%)
                & --    & 0.522 & 0.319 & --
                & --    & --    & 0.744 & 0.563 & 0.469 & 0.615 \\

        $\sqrt{\mathrm{iSWAP}}$ gate error (\%)
                & --    & 0.170 & 0.126 & 0.238
                & --    & --    & 0.362 & 0.280 & 0.199 & 0.294 \\

        $\mathrm{SWAP}$ gate error (\%)
                & --    & --    & 1.256 & --
                & --    & --    & 1.480 & 0.901 & 1.154 & --    \\

        B-like gate error (\%)
                & --    & 0.538 & --    & 1.197
                & --    & --    & --    & 0.379 & 1.037 & 0.690 \\
        \hline\hline
    \end{tabular}
\end{table}

\begin{table}[htbp]
    \centering
    \caption{Values of the PPT-mixture entanglement witness for the states prepared using the AshN and $\mathrm{CZ}$ gate sets on the $2\times4$ qubit topology with different defect configurations, including the defect-free case. All errors, after rounding, are 0.0008.}
    \label{tab:PPT_witness_result}
    \setlength{\tabcolsep}{6pt}
    \renewcommand{\arraystretch}{1.2}
    \begin{tabular}{c c c c c c}
        \hline\hline
         & Perfect & Defect(G25) & Defect(G12/G46) & Defect(G13/G25/G47) & Defect(G13/G25/G68) \\
        \hline
        AshN
         & -0.0344 & -0.0140     & -0.0360         & -0.0089             & -0.0119             \\
        \hline
        $\mathrm{CZ}$
         & 0.0593  & 0.0997      & 0.1383          & 0.1441              & 0.1945              \\
        \hline\hline
    \end{tabular}
\end{table}

\newpage
\subsection{Evaluating Dicke-state quality}

To rigorously evaluate and compare Dicke states obtained from two different compilation schemes, we adopt two important metrics: state fidelity and a genuine multipartite entanglement (GME) witness, by performing full state tomography and obtaining unbiased estimators for several observable expectations under the sampling model described below~\cite{schwemmerSystematicErrorsCurrent2015}. Both the readout-error correction, which inverts the independently calibrated single-qubit response matrices, and the linear-inversion tomographic reconstruction are linear maps of the measured frequencies; they therefore preserve the unbiasedness of these estimators, provided the calibrated readout response matrices are exact.

Suppose the prepared state is $\rho$ with $N=8$ qubits.
We carried out full QST (quantum state tomography) to obtain data as follows: given every Pauli basis string $\mathbf{b}\in \{X, Y, Z\}^{\otimes N}$, QST gives us samples $s \in \{0, 1\}^{\otimes N}$ from probability distribution $P(\mathbf{s}\mid \mathbf{b})$:
\begin{equation}
    P(\mathbf{s} \mid \mathbf{b}) = \text{Tr}\left( \Pi_{\mathbf{s}}^{(\mathbf{b})} \rho  \right),
\end{equation}
where $\Pi_{\mathbf{s}}^{(\mathbf{b})} = \bigotimes_{i=1}^{N} \Pi_{\mathbf{s}_i}^{(\mathbf{b}_i)}$ is the projector onto measurement outcome $\mathbf{s}$ under basis $\mathbf{b}$, and $ \Pi_{\mathbf{s}_i}^{(\mathbf{b}_i)} = \frac{I + (-1)^{\mathbf{s}_i}\sigma_{\mathbf{b}_i}}{2}$ is defined as single-qubit projector onto measurement outcome $\mathbf{s}_i$ under basis $\mathbf{b}_i$; $\sigma_{\mathbf{b}_i}$ is defined as the Pauli operator corresponding to $\mathbf{b}_i$. Experimentally, each measurement basis $\mathbf{b}$ was sampled with $N_{\mathrm{shots}}=3584$ repetitions, yielding empirical estimates of the probability distribution $P(\mathbf{s}|\mathbf{b})$.

Given $N_{\mathbf{b}}$ samples $\left\{\mathbf{s}_{\mathbf{b}}^{(1)}, \mathbf{s}_{\mathbf{b}}^{(2)}, \dots, \mathbf{s}_{\mathbf{b}}^{(N_{\mathbf{b}})}\right\}$ obtained by measuring in a given basis $\mathbf{b}$, we construct the estimator:
\begin{equation}
    \tilde{P}(\mathbf{s} \mid \mathbf{b}) = \frac{1}{N_{\mathbf{b}}} \sum_{i=1}^{N_{\mathbf{b}}} \mathbb{I} \left(\mathbf{s}_{\mathbf{b}}^{(i)} = \mathbf{s}\right),
\end{equation}
where $\mathbb{I}(\cdot)$ is the indicator function that evaluates to $1$ when the condition is satisfied and $0$ otherwise.

We focus on the following problem: given an observable $O$, construct an estimator $\widetilde{\braket{O}}$ from samples $\{\mathbf{s}_{\mathbf{b}}^{(i)}\}$ of observable expectation:
\begin{equation}
    \braket{O} = \text{Tr}(O\rho).
\end{equation}

Since observable $O$ is Hermitian, it can be expanded in the Pauli basis:
\begin{equation}
    O = \sum_{P\in \{I, X, Y, Z\}^{\otimes N}} c_P P,
\end{equation}
where $c_P \in \mathbb{R}$ is the coefficient of Pauli string $P$ satisfying:
\begin{equation}
    c_P = \frac{1}{2^N} \text{Tr}(O P).
\end{equation}
Under the decomposition, $\braket{O}$ can be expressed as:
\begin{equation}
    \label{eqn:o_pauli_decomp}
    \braket{O} = \sum_{P\in \{I, X, Y, Z\}^{\otimes N}} c_P \braket{P}, \quad \text{where} \quad \braket{P} = \text{Tr}(P\rho).
\end{equation}
To rewrite $\braket{O}$ in terms of $P(\mathbf{s} \mid \mathbf{b})$, we first define the set of Pauli basis strings that are compatible with $P$.
We define the support of $P$ as $\text{supp}(P) = \{i \in [N] : P_i \neq I\}$; we say Pauli string $P$ is compatible with Pauli basis string $\mathbf{b}$ if $\mathbf{b}_i = P_i$ for all $i \in \text{supp}(P)$, denoted as $\mathbf{b} \triangleright P$.
For every Pauli string $P$, there are $m_P = 3^{N - |\text{supp}(P)|}$ Pauli basis strings such that $\mathbf{b} \triangleright P$.

For a given Pauli string $P$, every compatible Pauli basis string $\mathbf{b} \triangleright P$ can be used to estimate $\braket{P}$ by defining $\braket{P}_{\mathbf{b}}$ as:
\begin{equation}
    \braket{P}_{\mathbf{b}} = \sum_{\mathbf{s}}P(\mathbf{s} \mid \mathbf{b}) V_P(\mathbf{s}), \quad \text{where} \quad V_P(\mathbf{s}) = (-1)^{\sum_{i\in \text{supp}(P)}\mathbf{s}_i} \quad \text{is the parity function}.
\end{equation}
It can be shown that $\braket{P}_{\mathbf{b}}=\braket{P}$:
\begin{align}
    \braket{P}_{\mathbf{b}}
     & = \sum_{\mathbf{s}}P(\mathbf{s} \mid \mathbf{b}) V_P(\mathbf{s}) \nonumber                                  \\
     & = \sum_{\mathbf{s}} \text{Tr}\left( \Pi_{\mathbf{s}}^{(\mathbf{b})} \rho  \right) V_P(\mathbf{s}) \nonumber \\
     & = \sum_{\mathbf{s}} \text{Tr}\left( \Pi_{\mathbf{s}}^{(\mathbf{b})} V_P(\mathbf{s})\rho  \right) \nonumber  \\
     & = \text{Tr}\left( \sum_{\mathbf{s}} V_P(\mathbf{s})\Pi_{\mathbf{s}}^{(\mathbf{b})}  \rho  \right)
\end{align}
It suffices to show that $\sum_{\mathbf{s}} V_P(\mathbf{s})\Pi_{\mathbf{s}}^{(\mathbf{b})} = P$. Define sign function $c_i(x)$ as:
\begin{equation}
    c_i(x) = \begin{cases}
        (-1)^{x}, & \text{if } i \in \text{supp}(P), \\
        1,        & \text{otherwise}.
    \end{cases}
\end{equation}
Then $V_P$ can be expressed as $V_P(\mathbf{s}) = \prod_{i=1}^{N} c_i(\mathbf{s}_i)$. It follows that:
\begin{align}
    \sum_{\mathbf{s}} V_P(\mathbf{s})\Pi_{\mathbf{s}}^{(\mathbf{b})}
     & = \sum_{\mathbf{s}} \prod_{i=1}^{N} c_i(\mathbf{s}_i) \bigotimes_{i=1}^{N} \Pi_{\mathbf{s}_i}^{(\mathbf{b}_i)} \nonumber \\
     & = \bigotimes_{i=1}^{N} \sum_{\mathbf{s}_i=0}^{1} c_i(\mathbf{s}_i) \Pi_{\mathbf{s}_i}^{(\mathbf{b}_i)} \nonumber         \\
     & = \bigotimes_{i=1}^{N} P_i = P.
\end{align}
The step $\sum_{\mathbf{s}_i=0}^{1} c_i(\mathbf{s}_i) \Pi_{\mathbf{s}_i}^{(\mathbf{b}_i)} = P_i$ follows from the fact that:
\begin{itemize}
    \item For $i \in \text{supp}(P)$, it can be shown that $\sum_{\mathbf{s}_i=0}^{1} c_i(\mathbf{s}_i) \Pi_{\mathbf{s}_i}^{(\mathbf{b}_i)} = \Pi_{0}^{(\mathbf{b}_i)} - \Pi_{1}^{(\mathbf{b}_i)} = \sigma_{\mathbf{b}_i} = P_i$.
    \item For $i \notin \text{supp}(P)$, it can be shown that $\sum_{\mathbf{s}_i=0}^{1} c_i(\mathbf{s}_i) \Pi_{\mathbf{s}_i}^{(\mathbf{b}_i)} = \Pi_{0}^{(\mathbf{b}_i)} + \Pi_{1}^{(\mathbf{b}_i)} = I = P_i$.
\end{itemize}

By expressing $\braket{P}$ as a weighted sum of  $\braket{P}_{\mathbf{b}}$ over compatible $\mathbf{b}$, \eqref{eqn:o_pauli_decomp} can be rewritten as:
\begin{equation}
    \braket{O} = \sum_{P} \sum_{\mathbf{b}: \mathbf{b} \triangleright P} c_{P, \mathbf{b}} \braket{P}_{\mathbf{b}} = \sum_{\mathbf{b}}  \sum_{P: \mathbf{b} \triangleright P} c_{P, \mathbf{b}} \braket{P}_{\mathbf{b}}.
\end{equation}
The weights $c_{P, \mathbf{b}}$ should be chosen s.t. $\sum_{\mathbf{b}: \mathbf{b} \triangleright P} c_{P, \mathbf{b}} = c_P$ for every Pauli string $P$. One way to construct such $c_{P, \mathbf{b}}$ is to set $c_{P, \mathbf{b}} = \frac{c_P}{m_P}$.
By expanding $\braket{P}_{\mathbf{b}}$ in terms of $P(\mathbf{s} \mid \mathbf{b})$, we have:
\begin{align}
    \braket{O} & = \sum_{\mathbf{b}}
    \sum_{P: \mathbf{b} \triangleright P} c_{P, \mathbf{b}} \sum_{\mathbf{s}}P(\mathbf{s} \mid \mathbf{b}) V_P(\mathbf{s}) \nonumber                                         \\
               & = \sum_{\mathbf{b}}  \sum_{\mathbf{s}} P(\mathbf{s} \mid \mathbf{b}) \left(\sum_{P: \mathbf{b} \triangleright P} c_{P, \mathbf{b}} V_P(\mathbf{s}) \right).
\end{align}
By grouping $C(\mathbf{b}, \mathbf{s}) = \sum_{P: \mathbf{b} \triangleright P} c_{P, \mathbf{b}} V_P(\mathbf{s})$, we have:
\begin{equation}
    \braket{O} = \sum_{\mathbf{b}}  \sum_{\mathbf{s}} P(\mathbf{s} \mid \mathbf{b}) C(\mathbf{b}, \mathbf{s}).
\end{equation}
The derivation above allows us to construct an unbiased estimator $\widetilde{\braket{O}}$ from empirical probability distribution $\tilde{P}(\mathbf{s} \mid \mathbf{b})$:
\begin{equation}
    \widetilde{\braket{O}} = \sum_{\mathbf{b}}  \sum_{\mathbf{s}} \tilde{P}(\mathbf{s} \mid \mathbf{b}) C(\mathbf{b}, \mathbf{s}).
\end{equation}

Since all Pauli string bases $\textbf{b}$ are measured independently, the variance of the estimator $\widetilde{\braket{O}}$ is defined as:
\begin{equation}
    \mathrm{Var}\left(\widetilde{\braket{O}}\right) = \sum_{\mathbf{b}} \frac{1}{N_{\mathbf{b}}}  \left[ \sum_{\mathbf{s}} P(\mathbf{s} \mid \mathbf{b}) C(\mathbf{b}, \mathbf{s})^2 - \left(\sum_{\mathbf{s}}P(\mathbf{s} \mid \mathbf{b}) C(\mathbf{b}, \mathbf{s})\right)^2\right].
\end{equation}
The unbiased estimated variance can be constructed from empirical probability distribution $\tilde{P}(\mathbf{s} \mid \mathbf{b})$:
\begin{equation}
    \widetilde{\mathrm{Var}}\left(\widetilde{\braket{O}}\right) = \sum_{\mathbf{b}} \frac{1}{N_{\mathbf{b}} - 1}  \left[ \sum_{\mathbf{s}} \tilde{P}(\mathbf{s} \mid \mathbf{b}) C(\mathbf{b}, \mathbf{s})^2 - \left(\sum_{\mathbf{s}}\tilde{P}(\mathbf{s} \mid \mathbf{b}) C(\mathbf{b}, \mathbf{s})\right)^2\right].
\end{equation}

By using different observables $O$, we can estimate both the fidelity and the GME witness expectation, each with an estimated variance.
For instance, for the fidelity of the prepared state $\rho$ with respect to the target Dicke state $\ket{\psi}$, we use the squared Uhlmann fidelity, which reduces to $\bra{\psi}\rho\ket{\psi}$ for a pure target state:
\begin{equation}
    F(\rho, \ket{\psi}) = \left(\text{Tr}\sqrt{\sqrt{\ket{\psi}\bra{\psi}}\rho \sqrt{\ket{\psi}\bra{\psi}}}\right)^2 = \text{Tr}\left(\ket{\psi}\bra{\psi}\rho\right).
\end{equation}
By defining $O_{\ket{\psi}} = \ket{\psi}\bra{\psi}$, the state fidelity can be estimated as:
\begin{equation}
    \widetilde{F}(\rho, \ket{\psi}) = \widetilde{\braket{O_{\ket{\psi}}}},
\end{equation}
with estimated variance $\widetilde{\mathrm{Var}}\left(\widetilde{\braket{O_{\ket{\psi}}}}\right)$.

The task of detecting genuine multipartite entanglement can be formulated as finding an entanglement witness $W$ such that for all biseparable states $\sigma$, $\text{Tr}(W\sigma) \ge 0$, while for the target state $\rho$, $\text{Tr}(W\rho) < 0$. Assuming $k \le N/2$ without loss of generality, we adopt the fully positive partial transpose (PPT) witness for the Dicke state $\ket{D(N, k)}$ from Corollary 7 of \citet{bergmann2013entanglement}:
\begin{equation}
    W_{N,k} = \sum_{i=0}^{2k} \omega_i \Pi_i - \ket{D(N,k)}\bra{D(N,k)},
\end{equation}
where:
\begin{itemize}
    \item $\Pi_i$ is the projector onto the $i$-excitation subspace:
          \begin{equation}
              \Pi_i = \sum_{\mathbf{s}: |\mathbf{s}|=i} \ket{\mathbf{s}}\bra{\mathbf{s}},
          \end{equation}
          where $|\mathbf{s}|$ is the Hamming weight of $\mathbf{s}$.
    \item For $i\neq k$, define $\omega_{k+\delta}$ and $\omega_{k-\delta}$ in pairs:
          \begin{equation}
              \omega_{k-\delta} = \sqrt{\frac{\binom{N}{k+\delta}}{\binom{N}{k-\delta}}}\Lambda_\delta, \quad
              \omega_{k+\delta} =\sqrt{\frac{\binom{N}{k-\delta}}{\binom{N}{k+\delta}}}\Lambda_\delta,
          \end{equation}
          where $\delta = 1, 2, \dots, \min\{k, N-k\}$, and $\Lambda_\delta$ is defined as:
          \begin{equation}
              \Lambda_\delta = \max_{x_1\geq \delta, x_1 + x_2 \leq \frac{N}{2}} \sqrt{\binom{N-x_1-x_2}{k-x_2}
                  \binom{x_1+x_2}{x_2}
                  \binom{N-x_1-x_2}{k-\delta-x_2}
                  \binom{x_1+x_2}{x_2+\delta}
                  \bigg/
                  \binom{N}{k}^{2}}.
          \end{equation}
    \item For $i=k$, \begin{equation}
              \omega_k =
              \begin{cases}
                  \dfrac{N-k}{N},    & k<N/2, \\[4pt]
                  \dfrac{N}{2(N-1)}, & k=N/2.
              \end{cases}
          \end{equation}
\end{itemize}
By using observable $W_{N,k}$, the entanglement witness expectation $\text{Tr}(W_{N,k}\rho)$ can be estimated with variance $\widetilde{\mathrm{Var}}\left(\widetilde{\braket{W_{N,k}}}\right)$.

We end this section by using the above framework for estimating $\tilde{\rho}$ with linear inversion QST, by splitting every matrix entry of $\rho$ into real and imaginary parts:
\begin{equation}
    \rho_{u, v} = {\mathrm{Re}}(\rho_{u, v}) + i{\mathrm{Im}}(\rho_{u, v}).
\end{equation}
By defining two new observables $O_{u, v}^{\mathrm{Re}} = \frac{1}{2}(\ket{u}\bra{v} + \ket{v}\bra{u})$ and $O_{u, v}^{\mathrm{Im}} = \frac{i}{2}(\ket{u}\bra{v} - \ket{v}\bra{u})$, we have:
\begin{equation}
    {\mathrm{Re}}(\rho_{u, v}) = \braket{O_{u, v}^{\mathrm{Re}}}, \quad
    {\mathrm{Im}}(\rho_{u, v}) = \braket{O_{u, v}^{\mathrm{Im}}}.
\end{equation}
Therefore, the matrix entry $\rho_{u, v}$ can be estimated as:
\begin{equation}
    \tilde{\rho}_{u, v} = \widetilde{\braket{O_{u, v}^{\mathrm{Re}}}} + i\widetilde{\braket{O_{u, v}^{\mathrm{Im}}}}.
\end{equation}
The variances of these estimators quantify the estimated mean-squared Frobenius error of the reconstructed density matrix:
\begin{align}
    \mathbb{E}\left[\|\tilde{\rho} - \rho\|_F^2\right]
     & = \mathbb{E}\left[\sum_{u, v} |\rho_{u, v} - \tilde{\rho}_{u, v}|^2 \right] \nonumber                                                                                                                                              \\
     & = \mathbb{E}\left[\sum_{u, v}\left( |\braket{O_{u, v}^{\mathrm{Re}}} - \widetilde{\braket{O_{u, v}^{\mathrm{Re}}}}|^2 + |\braket{O_{u, v}^{\mathrm{Im}}} - \widetilde{\braket{O_{u, v}^{\mathrm{Im}}}}|^2\right) \right] \nonumber \\
     & = \sum_{u, v} \left(\mathrm{Var}\left(\widetilde{\braket{O_{u, v}^{\mathrm{Re}}}}\right) + \mathrm{Var}\left(\widetilde{\braket{O_{u, v}^{\mathrm{Im}}}}\right)\right).
\end{align}
The equalities $\mathbb{E}\left[|\braket{O_{u, v}^{\mathrm{Re}}} - \widetilde{\braket{O_{u, v}^{\mathrm{Re}}}}|^2\right] = \mathrm{Var}\left(\widetilde{\braket{O_{u, v}^{\mathrm{Re}}}}\right)$ and $\mathbb{E}\left[|\braket{O_{u, v}^{\mathrm{Im}}} - \widetilde{\braket{O_{u, v}^{\mathrm{Im}}}}|^2\right] = \mathrm{Var}\left(\widetilde{\braket{O_{u, v}^{\mathrm{Im}}}}\right)$ hold because $\widetilde{\braket{O_{u, v}^{\mathrm{Re}}}}$ and $\widetilde{\braket{O_{u, v}^{\mathrm{Im}}}}$ are unbiased estimators and satisfy $\mathbb{E}\left[|X - \mathbb{E}[X]|^2\right] = \mathrm{Var}(X)$ for any random variable $X$.

\bibliography{supp}